\documentclass[a4paper,11pt]{article}
\pdfoutput=1
\usepackage{jheppub}
\usepackage[T1]{fontenc}
\usepackage[english]{babel}
\usepackage{amsthm,amsfonts}
\allowdisplaybreaks
\usepackage[mathscr]{euscript}
\hyphenchar\font=\string"7F
\usepackage{tikz}
\usetikzlibrary{arrows}
\usetikzlibrary{calc}
\usetikzlibrary{decorations.pathreplacing}
\usepackage{dsfont}
\usepackage{bbm}
\usepackage{upgreek}
\usepackage{contour}
\usepackage{enumitem}
\usepackage{tabls}
\usepackage{cite}
\usepackage{mcite}

\DeclareMathAlphabet{\mathdsl}{U}{bbm}{m}{sl}

\newcommand{\DD}{\mathcal{D}}
\newcommand{\mfd}{\mathfrak{d}}
\renewcommand{\b}{\beta}
\newcommand{\dd}{\mathrm{d}}

\newcommand{\Pb}{\overline{P}}
\newcommand{\dg}{\delta g}
\newcommand{\dB}{\delta B}
\newcommand{\dphi}{\delta \phi}
\newcommand{\bh}{\widehat{\beta}}
\newcommand{\bb}{\overline{\beta}}
\newcommand{\B}{\upbeta}
\newcommand{\Bb}{\overline{\upbeta}}
\newcommand{\Bh}{\widehat{\upbeta}}
\newcommand{\Eh}{\widehat{E}}
\newcommand{\dEh}{\delta\Eh}
\newcommand{\dEgf}{\delta\Eh^{\mathrm{gf}}}
\newcommand{\Fh}{\widehat{F}}
\newcommand{\Dh}{\widehat{D}}
\newcommand{\A}{A}
\newcommand{\Al}{A_\lambda}
\newcommand{\intvol}{\int \dd^D x e^{-2d}}
\newcommand{\BhEgf}{\Bh^{E\mathrm{gf}}}
\newcommand{\AlE}{\Al^{(1)} \begin{pmatrix} \widehat E & d \end{pmatrix}}
\newcommand{\DL}{\Delta_\Lambda}
\newcommand{\Lh}{\widehat{L}}
\newcommand{\Sh}{\widehat{S}}
\newcommand{\Kh}{\widehat{K}}

\newcommand{\scR}{\mathcal{R}}
\newcommand{\scRh}{\widehat{\mathcal{R}}}
\newcommand{\scGh}{\widehat{\mathcal{G}}}
\newcommand{\qh}{\widehat{q}}
\newcommand{\jb}{\overline{j}}
\newcommand{\kh}{\widehat{k}}
\newcommand{\zb}{\overline{z}}
\newcommand{\wb}{\overline{w}}
\newcommand{\ci}{\mathrm{i}}
\newcommand{\notebook}{{\tt PLtwoloop.nb}}
\newcommand{\appnotebook}{\hyperref[app:notebook]{\notebook}}

\newcommand{\co}[1]{\parbox{1.7em}{\centering$\displaystyle #1$}}

\newcommand{\reflambdaetaPL}{Hoare:2015gda,Sfetsos:2015nya,Klimcik:2015gba}
\newcommand{\reflongDFTPLlist}{Hassler:2017yza,Demulder:2018lmj,Sakatani:2019jgu,*Catal-Ozer:2019hxw,*Hassler:2019wvn}
\newcommand{\refonelooprenorm}{Valent:2009nv,*Sfetsos:2009dj,Sfetsos:2009vt,*Avramis:2009xi,*Severa:2016lwc,*Pulmann:2020omk}
\newcommand{\refPLalphaprime}{Hassler:2020tvz,*Borsato:2020wwk,*Codina:2020yma}
\newcommand{\refoneLRGintegr}{Delduc:2020vxy,*Hoare:2020fye,*Georgiou:2018gpe,*Georgiou:2017jfi}

\usetikzlibrary{arrows}
\usetikzlibrary{decorations.markings}
\newcommand{\yshift}{.6cm}
\tikzstyle{small}=[scale=0.6, every node/.style={scale=0.6}]
\tikzstyle{resize}=[scale=0.8, every node/.style={scale=0.8}]
\tikzstyle{A}=[fill=black, draw=black, shape=circle, inner sep=2pt,anchor=center]
\tikzstyle{BETA}=[draw=gray!60!black,shape=circle,inner sep=10pt,{label={center:{\huge $\beta$}}}]
\tikzstyle{FA}=[fill=black, draw=black, shape=rectangle, inner sep=3pt,anchor=center]
\tikzstyle{none}=[]
\tikzstyle{arrow}=[style={-,decoration={markings,mark=at position 0.6 with {\arrow[scale=1.,black]{>}}}, postaction={decorate}, line width=1pt, shorten <=0, shorten >=0}]
\tikzstyle{dashed}=[dash pattern=on 3pt off 3pt, -, draw=black]
\tikzstyle{der}=[-triangle 60]
\tikzstyle{dashder}=[dash pattern=on 3pt off 3pt, -triangle 60]
\tikzstyle{derI}=[triangle 60-]
\tikzstyle{dashderI}=[dash pattern=on 3pt off 3pt, triangle 60-]
\tikzstyle{derder}=[triangle 60-triangle 60]
\tikzstyle{dashderder}=[dash pattern=on 3pt off 3pt, triangle 60-triangle 60]
\tikzstyle{derloop}=[loop, -triangle 60, looseness=20]
\tikzstyle{derloopback}=[style={-,decoration={markings,mark=at position 0.9 with {\arrow[scale=1.,black]{triangle 60}}}, postaction={decorate}, loop, looseness=20}]
\tikzstyle{dashderloop}=[dash pattern=on 3pt off 3pt, -triangle 60, looseness=20]
\tikzstyle{dashderbackI}=[style={-,decoration={markings,mark=at position 0.2 with {\arrow[scale=-1.,black]{triangle 60}}}, postaction={decorate}, looseness=20, dash pattern=on 3pt off 3pt}]
\tikzstyle{derbackI}=[style={-,decoration={markings,mark=at position 0.2 with {\arrow[scale=-1.,black]{triangle 60}}}, postaction={decorate}, looseness=20}]
\tikzstyle{dashderloopback}=[style={-,decoration={markings,mark=at position 0.9 with {\arrow[scale=1.,black]{triangle 60}}}, postaction={decorate}, loop, looseness=20, dash pattern=on 3pt off 3pt}]
\tikzstyle{derback}=[style={-,decoration={markings,mark=at position 0.98 with {\arrow[scale=1.,black]{triangle 60}}}, postaction={decorate}, loop, looseness=20}]
\tikzstyle{dashderback}=[style={-,decoration={markings,mark=at position 0.98 with {\arrow[scale=1.,black]{triangle 60}}}, postaction={decorate}, loop, looseness=20, dash pattern=on 3pt off 3pt}]
\pgfdeclarelayer{nodelayer}
\pgfdeclarelayer{edgelayer}
\pgfsetlayers{nodelayer,edgelayer,main}
\newcommand{\width}{.7}
\newcommand{\height}{.7}
\newcommand{\FAFA}[1]
{\scalebox{\width}[\height]{
\begin{tikzpicture}[baseline={([yshift=-.5ex]current bounding box.center)},small]
	\begin{pgfonlayer}{nodelayer}
		\node [style=FA] (0) at (-1.75, 0) {};
		\node [style=FA] (1) at (1.75, 0) {};
	\end{pgfonlayer}
	\begin{pgfonlayer}{edgelayer}
          \ifnum #1 >0 {\draw [style=dashed] (0) to (1);} \else {\draw (0) to (1);}\fi
	\end{pgfonlayer}
\end{tikzpicture}
}}
\newcommand{\FADA}[1]
{\scalebox{\width}[\height]{
\begin{tikzpicture}[baseline={([yshift=-.5ex-\yshift]current bounding box.center)},small]
	\begin{pgfonlayer}{nodelayer}
		\node [style=FA] (0) at (0, 0) {};
	\end{pgfonlayer}
	\begin{pgfonlayer}{edgelayer}
	  \ifnum #1 >0 {\draw [style=dashderback] (0) arc (-90:270:1cm);}
          \else {\draw [style=derback] (0) arc (-90:270:1cm);}\fi
	\end{pgfonlayer}
\end{tikzpicture}
}}
\newcommand{\Fsquare}[3]
{\scalebox{\width}[\height]{
\begin{tikzpicture}[baseline={([yshift=-.5ex]current bounding box.center)},small]
	\begin{pgfonlayer}{nodelayer}
		\node [style=A] (0) at (-2, 0) {};
		\node [style=A] (1) at (2, 0) {};
	\end{pgfonlayer}
	\begin{pgfonlayer}{edgelayer}
		\ifnum #1 >0 {\draw [style=dashed] (0) arc (180:0:2);}\else {\draw (0) arc (180:0:2);}\fi
		\ifnum #2 >0 {\draw [style=dashed] (0) to (1);}\else {\draw (0) to (1);}\fi
		\ifnum #3 >0 {\draw [style=dashed] (0) arc (-180:0:2);}\else {\draw (0) arc (-180:0:2);}\fi
	\end{pgfonlayer}
\end{tikzpicture}
}}
\newcommand{\dDAFA}
{\scalebox{\width}[\height]{
\begin{tikzpicture}[baseline={([yshift=-.5ex]current bounding box.center)},small]
	\begin{pgfonlayer}{nodelayer}
		\node [style=FA] (0) at (0, 0) {};
		\node [style=none] (1) at (2, 0) {};
		\node [style=none] (2) at (-2, 0) {};
	\end{pgfonlayer}
	\begin{pgfonlayer}{edgelayer}
		\draw (2.center) to (0);
		\draw [style=dashder] (1.center) to (0);
	\end{pgfonlayer}
\end{tikzpicture}
}}
\newcommand{\dFsquare}[2]
{\scalebox{\width}[\height]{
\begin{tikzpicture}[baseline={([yshift=-.5ex]current bounding box.center)},small]
  \begin{pgfonlayer}{nodelayer}
    \node [style=A] (0) at (0.75, 0) {};
    \node [style=A] (1) at (-.75, 0) {};
    \node [style=none] (2) at (2, 0) {};
    \node [style=none] (3) at (-2, 0) {};
  \end{pgfonlayer}
  \begin{pgfonlayer}{edgelayer}
    \draw (3.center) to (1);
    \draw [style=dashed] (2.center) to (0);
    \ifnum #1 >0 {\draw [style=dashed] (0) arc (0:-180:0.75);}\else {\draw (0) arc (0:-180:0.75);}\fi
    \ifnum #2 >0 {\draw [style=dashed] (0) arc (0:180:0.75);} \else {\draw (0) arc (0:180:0.75);}\fi
  \end{pgfonlayer}
\end{tikzpicture}
}}
\newcommand{\dFADA}[1]
{\scalebox{\width}[\height]{
\begin{tikzpicture}[baseline={([yshift=-.5ex-\yshift]current bounding box.center)},small]
  \begin{pgfonlayer}{nodelayer}
    \node [style=A] (0) at (0, 0) {};
    \node [style=none] (1) at (2, 0) {};
    \node [style=none] (2) at (-2, 0) {};
  \end{pgfonlayer}
  \begin{pgfonlayer}{edgelayer}
    \draw (2.center) to (0);
    \draw [style=dashed] (1.center) to (0);
    \ifnum #1 >0 {\draw [style=dashderloopback] (0) arc (-90:270:1cm);}\else {\draw [style=derloopback] (0) arc (-90:270:1cm);}\fi
  \end{pgfonlayer}
\end{tikzpicture}
}}
\newcommand{\dFFA}[1]
{\scalebox{\width}[\height]{
\begin{tikzpicture}[baseline={([yshift=-\yshift]current bounding box.center)},small]
  \begin{pgfonlayer}{nodelayer}
    \node [style=A] (0) at (0, 0) {};
    \node [style=none] (1) at (2, 0) {};
    \node [style=none] (2) at (-2, 0) {};
    \node [style=FA] (3) at (0, 1.75) {};
  \end{pgfonlayer}
  \begin{pgfonlayer}{edgelayer}
    \draw (2.center) to (0);
    \draw [style=dashed] (1.center) to (0);
    \ifnum #1 >0 {\draw [style=dashed] (0) to (3);}\else {\draw (0) to (3);}\fi
  \end{pgfonlayer}
\end{tikzpicture}
}}
\newcommand{\HCNaT}
{\scalebox{\width}[\height]{
  \begin{tikzpicture}[baseline={([yshift=-.5ex]current bounding box.center)},small]
    \begin{pgfonlayer}{nodelayer}
      \node [style=none] (0) at (-2, 0) {};
      \node [style=A] (1) at (-1.25, 0) {};
      \node [style=A] (2) at (-0.375, 0) {};
      \node [style=none] (3) at (2, 0) {};
      \node [style=A] (4) at (0.375, 0) {};
      \node [style=A] (5) at (1.25, 0) {};
    \end{pgfonlayer}
    \begin{pgfonlayer}{edgelayer}
      \draw (0.center) to (1);
      \draw (2) arc (0:180:0.875/2);
      \draw (1) arc (-180:0:0.875/2);
      \draw (2) to (4);
      \draw (5) arc (0:180:0.875/2);
      \draw (4) arc (-180:0:0.875/2);
      \draw (5) to (3.center);
    \end{pgfonlayer}
\end{tikzpicture}
}}
\newcommand{\HCNa}[5]
{\scalebox{\width}[\height]{
  \begin{tikzpicture}[baseline={([yshift=-.5ex]current bounding box.center)},small]
    \begin{pgfonlayer}{nodelayer}
      \node [style=none] (0) at (-2, 0) {};
      \node [style=A] (1) at (-1.25, 0) {};
      \node [style=A] (2) at (-0.375, 0) {};
      \node [style=none] (3) at (2, 0) {};
      \node [style=A] (4) at (0.375, 0) {};
      \node [style=A] (5) at (1.25, 0) {};
    \end{pgfonlayer}
    \begin{pgfonlayer}{edgelayer}
      \draw (0.center) to (1);
      \ifnum #1 >0 {\draw [style=dashed] (2) arc (0:180:0.875/2);}\else {\draw (2) arc (0:180:0.875/2);}\fi
      \ifnum #2 >0 {\draw [style=dashed] (1) arc (-180:0:0.875/2);}\else {\draw (1) arc (-180:0:0.875/2);}\fi
      \ifnum #3 >0 {\draw [style=dashed] (2) to (4);}\else {\draw (2) to (4);}\fi
      \ifnum #4 >0 {\draw [style=dashed] (5) arc (0:180:0.875/2);}\else {\draw (5) arc (0:180:0.875/2);}\fi
      \ifnum #5 >0 {\draw [style=dashed] (4) arc (-180:0:0.875/2);}\else {\draw (4) arc (-180:0:0.875/2);}\fi
      \draw [style=dashed] (5) to (3.center);
    \end{pgfonlayer}
\end{tikzpicture}
}}
\newcommand{\HCNbT}
{\scalebox{\width}[\height]{
  \begin{tikzpicture}[baseline={([yshift=-.5ex]current bounding box.center)},small]
	\begin{pgfonlayer}{nodelayer}
		\node [style=none] (0) at (-2, 0) {};
		\node [style=A] (1) at (-1, 0) {};
		\node [style=A] (2) at (0, 1) {};
		\node [style=A] (3) at (1, 0) {};
		\node [style=none] (4) at (2, 0) {};
		\node [style=A] (5) at (0, -1) {};
	\end{pgfonlayer}
	\begin{pgfonlayer}{edgelayer}
		\draw (0.center) to (1);
		\draw (3) to (4.center);
                \draw (1) arc (0:-90:-1);
                \draw (1) arc (0:90:-1);
                \draw (2) to (5);
                \draw (3) arc (0:90:1);
                \draw (3) arc (0:-90:1);
	\end{pgfonlayer}
  \end{tikzpicture}
}  }
\newcommand{\HCNb}[5]
{\scalebox{\width}[\height]{
  \begin{tikzpicture}[baseline={([yshift=-.5ex]current bounding box.center)},small]
	\begin{pgfonlayer}{nodelayer}
		\node [style=none] (0) at (-2, 0) {};
		\node [style=A] (1) at (-1, 0) {};
		\node [style=A] (2) at (0, 1) {};
		\node [style=A] (3) at (1, 0) {};
		\node [style=none] (4) at (2, 0) {};
		\node [style=A] (5) at (0, -1) {};
	\end{pgfonlayer}
	\begin{pgfonlayer}{edgelayer}
		\draw (0.center) to (1);
		\draw [style=dashed] (3) to (4.center);
                \ifnum #1 >0 {\draw [style=dashed] (1) arc (0:-90:-1);}\else {\draw (1) arc (0:-90:-1);}\fi
                \ifnum #2 >0 {\draw [style=dashed] (1) arc (0:90:-1);}\else {\draw (1) arc (0:90:-1);}\fi
                \ifnum #3 >0 {\draw [style=dashed](2) to (5);}\else {\draw (2) to (5);}\fi
                \ifnum #4 >0 {\draw [style=dashed] (3) arc (0:90:1);}\else {\draw (3) arc (0:90:1);}\fi
                \ifnum #5 >0 {\draw [style=dashed] (3) arc (0:-90:1);}\else {\draw (3) arc (0:-90:1);}\fi
	\end{pgfonlayer}
  \end{tikzpicture}
}  }
\newcommand{\HCNcT}
{\scalebox{\width}[\height]{
  \begin{tikzpicture}[baseline={([yshift=-.5ex]current bounding box.center)},small]
	\begin{pgfonlayer}{nodelayer}
		\node [style=none] (0) at (-2, 0) {};
		\node [style=A] (1) at (-1, 0) {};
		\node [style=A,anchor=north] (2) at (-0.5, 1) {};
		\node [style=A] (3) at (1, 0) {};
		\node [style=none] (4) at (2, 0) {};
		\node [style=A,anchor=north] (5) at (0.5, 1) {};
	\end{pgfonlayer}
	\begin{pgfonlayer}{edgelayer}
		\draw (0.center) to (1);
		\draw (3) to (4.center);
                \draw (1) arc (0:-atan2(1,.5):-1);
                \draw (3) arc (0:-180:1);
                \draw (5) arc (0:180:.5);
                \draw (5) arc (0:-180:.5);
                \draw (3) arc (0:atan2(1,.5):1);
	\end{pgfonlayer}
  \end{tikzpicture}
}  }
\newcommand{\HCNc}[5]
{\scalebox{\width}[\height]{
  \begin{tikzpicture}[baseline={([yshift=-.5ex]current bounding box.center)},small]
	\begin{pgfonlayer}{nodelayer}
		\node [style=none] (0) at (-2, 0) {};
		\node [style=A] (1) at (-1, 0) {};
		\node [style=A,anchor=north] (2) at (-0.5, 1) {};
		\node [style=A] (3) at (1, 0) {};
		\node [style=none] (4) at (2, 0) {};
		\node [style=A,anchor=north] (5) at (0.5, 1) {};
	\end{pgfonlayer}
	\begin{pgfonlayer}{edgelayer}
		\draw (0.center) to (1);
		\draw [style=dashed] (3) to (4.center);
                \ifnum #1 >0 {\draw [style=dashed] (1) arc (0:-atan2(1,.5):-1);}\else {\draw (1) arc (0:-atan2(1,.5):-1);}\fi
                \ifnum #2 >0 {\draw [style=dashed] (3) arc (0:-180:1);}\else {\draw (3) arc (0:-180:1);}\fi
                \ifnum #3 >0 {\draw [style=dashed] (5) arc (0:180:.5);}\else {\draw (5) arc (0:180:.5);}\fi
                \ifnum #4 >0 {\draw [style=dashed] (5) arc (0:-180:.5);}\else {\draw (5) arc (0:-180:.5);}\fi
                \ifnum #5 >0 {\draw [style=dashed] (3) arc (0:atan2(1,.5):1);}\else {\draw (3) arc (0:atan2(1,.5):1);}\fi
	\end{pgfonlayer}
  \end{tikzpicture}
}   }         
\newcommand{\betaLoop}[1]{(#1.north east) let \p1 = ($(#1.north)-(#1)$) in arc (-45:225:{veclen(\x1,\y1)})}
\newcommand{\DADA}[1]
{\scalebox{\width}[\height]{
  \begin{tikzpicture}[baseline={([yshift=-14pt]current bounding box.center)},resize]
    \begin{pgfonlayer}{nodelayer}
      \node [style=none] (0) at (2, 0) {};
      \node [style=BETA] (1) at (0, 0) {};
      \node [style=none] (2) at (-2, 0) {};
      \node [style=none] (3) at (1.center) {};
      \node [style=none] (4) at (1.north east) {};
    \end{pgfonlayer}
    \begin{pgfonlayer}{edgelayer}
      \ifnum #1 >0 {\draw [style=dashderder] \betaLoop1;}
      \else {\draw [style=derder] \betaLoop1;}\fi
      \draw [style=dashed] (0.center) to (1);
      \draw (2.center) to (1);
    \end{pgfonlayer}
  \end{tikzpicture}
}}
\newcommand{\DDl}[1]
{\scalebox{\width}[\height]{
  \begin{tikzpicture}[baseline={([yshift=-14pt]current bounding box.center)},resize]
    \begin{pgfonlayer}{nodelayer}
	\node [style=none] (0) at (2, 0) {};
	\node [style=BETA] (1) at (0, 0) {};
	\node [style=none] (2) at (-2, 0) {};
    \end{pgfonlayer}
    \begin{pgfonlayer}{edgelayer}
	\draw [style=dashed] (0.center) to (1);
	\draw [style=der] (2.center) to (1);
        \ifnum #1 >0 {\draw [style=dashder] \betaLoop1 ;}
        \else {\draw [style=derloop] \betaLoop1 ;}\fi
	\end{pgfonlayer}
  \end{tikzpicture}
}   }
\newcommand{\FDla}[2]
{\scalebox{\width}[\height]{
  \begin{tikzpicture}[baseline={([yshift=1pt]current bounding box.south)},resize]
    \begin{pgfonlayer}{nodelayer}
      \node [style=none] (0) at (2, 0) {};
      \node [style=A] (1) at (0, 0) {};
      \node [style=none] (2) at (-2, 0) {};
      \node [style=BETA] (3) at (0, 1) {};
    \end{pgfonlayer}
    \begin{pgfonlayer}{edgelayer}
      \draw [style=dashed] (0.center) to (1);
      \draw (2.center) to (1);
      \ifnum #1 >0 {\draw [style=dashed] (1) to (3);} \else {\draw (1) to (3);}\fi
      \ifnum #2 >0 {\draw [style=dashder] \betaLoop3;} \else {\draw [style=der] \betaLoop3 ;}\fi
    \end{pgfonlayer}
  \end{tikzpicture}
}       }
\newcommand{\FDlb}[2]
{\scalebox{\width}[\height]{
  \begin{tikzpicture}[baseline={([yshift=-10pt]current bounding box.center)},resize]
    \begin{pgfonlayer}{nodelayer}
      \node [style=none] (0) at (2, 0) {};
      \node [style=none] (1) at (-2, 0) {};
      \node [style=BETA] (2) at (.5, 0) {};
      \node [style=A] (3) at (-1, 0) {};
    \end{pgfonlayer}
    \begin{pgfonlayer}{edgelayer}
      \draw [style=dashed] (0.center) to (2);
      \draw (1.center) to (3);
      \ifnum #1 >0 {\draw [style=dashed] (3) to (2);} \else {\draw (3) to (2);}\fi
      \ifnum #2 >0 {\draw [style=dashderloopback] (3) arc (-90:270:.5);} \else {\draw [style=derloopback] (3) arc (-90:270:.5);}\fi
    \end{pgfonlayer}
  \end{tikzpicture}
}  }                       
\newcommand{\DFa}[2]
{\scalebox{\width}[\height]{
  \begin{tikzpicture}[baseline={([yshift=1pt]current bounding box.south)},resize]
	\begin{pgfonlayer}{nodelayer}
		\node [style=none] (0) at (2, 0) {};
		\node [style=A] (1) at (0, 0) {};
		\node [style=none] (2) at (-2, 0) {};
		\node [style=BETA] (3) at (0, 1.5) {};
	\end{pgfonlayer}
	\begin{pgfonlayer}{edgelayer}
		\draw [style=dashed] (0.center) to (1);
		\draw [style=der] (2.center) to (1);
                \ifnum #1 >0 {\draw [style=dashed] (1) arc (-90:-230:.75);}\else {\draw (1) arc (-90:-230:.75);}\fi
                \ifnum #2 >0 {\draw [style=dashed] (1) arc (-90:50:.75);}\else {\draw (1) arc (-90:50:.75);}\fi
	\end{pgfonlayer}
  \end{tikzpicture}
}  }
\newcommand{\DFb}[2]
{\scalebox{\width}[\height]{
  \begin{tikzpicture}[baseline={([yshift=1pt]current bounding box.south)},resize]
	\begin{pgfonlayer}{nodelayer}
		\node [style=none] (0) at (2, 0) {};
		\node [style=A] (1) at (0, 0) {};
		\node [style=none] (2) at (-2, 0) {};
		\node [style=BETA] (3) at (0,1.5) {};
	\end{pgfonlayer}
	\begin{pgfonlayer}{edgelayer}
		\draw [style=dashed] (0.center) to (1);
		\draw  (2.center) to (1);
                \ifnum #1 >0 {\draw [style=dashderbackI] (1) arc (-90:-230:.75);}\else {\draw [style=derbackI] (1) arc (-90:-230:.75);}\fi
                \ifnum #2 >0 {\draw [style=dashed] (1) arc (-90:50:.75);}\else {\draw (1) arc (-90:50:.75);}\fi
	\end{pgfonlayer}
  \end{tikzpicture}
}}
\newcommand{\DFc}[2]
{\scalebox{\width}[\height]{
  \begin{tikzpicture}[baseline={([yshift=-2pt]current bounding box.center)},resize]
    \begin{pgfonlayer}{nodelayer}
      \node [style=none] (0) at (2, 0) {};
      \node [style=A] (1) at (1, 0) {};
      \node [style=none] (2) at (-2, 0) {};
      \node [style=BETA] (3) at (-.5, 0) {};
    \end{pgfonlayer}
    \begin{pgfonlayer}{edgelayer}
      \draw [style=dashed] (0.center) to (1);
      \draw [style=der] (2.center) to (3);
      \ifnum #1 >0 {\draw [style=dashed] (3.south east) arc (30:180:-.6);}\else {\draw (3.south east) arc (30:180:-.6);}\fi
      \ifnum #2 >0 {\draw [style=dashed] (3.north east) arc (-30:-180:-.6);}\else {\draw (3.north east) arc (-30:-180:-.6);}\fi
    \end{pgfonlayer}
  \end{tikzpicture}
}}
\newcommand{\DFd}[2]
{\scalebox{\width}[\height]{
  \begin{tikzpicture}[baseline={([yshift=-3pt]current bounding box.center)},resize]
    \begin{pgfonlayer}{nodelayer}
      \node [style=none] (0) at (2, 0) {};
      \node [style=A] (1) at (1, 0) {};
      \node [style=none] (2) at (-2, 0) {};
      \node [style=BETA] (3) at (-.5, 0) {};
    \end{pgfonlayer}
    \begin{pgfonlayer}{edgelayer}
      \draw [style=dashed] (0.center) to (1);
      \draw [style=none] (2.center) to (3);
      \ifnum #1 >0 {\draw [style=dashderI] (3.south east) arc (30:180:-.6);}\else {\draw [style=derI] (3.south east) arc (30:180:-.6);}\fi
      \ifnum #2 >0 {\draw [style=dashed] (3.north east) arc (-30:-180:-.6);}\else {\draw (3.north east) arc (-30:-180:-.6);}\fi
    \end{pgfonlayer}
  \end{tikzpicture}
}}
\newcommand{\FADla}[1]
{\scalebox{\width}[\height]{
  \begin{tikzpicture}[baseline={([yshift=11pt]current bounding box.south)},resize]
    \begin{pgfonlayer}{nodelayer}
      \node [style=none] (0) at (2, 0) {};
      \node [style=BETA] (1) at (0, 0) {};
      \node [style=none] (2) at (-2, 0) {};
      \node [style=FA] (3) at (-1, 0) {};
    \end{pgfonlayer}
    \begin{pgfonlayer}{edgelayer}
      \draw [style=dashed] (0.center) to (1);
      \ifnum #1 >0 {\draw [style=dashderloop] \betaLoop1;}\else {\draw [style=derloop] \betaLoop1;}\fi
      \draw (2.center) to (3);
    \end{pgfonlayer}
  \end{tikzpicture}
}}
\newcommand{\FADlb}[1]
{\scalebox{\width}[\height]{
  \begin{tikzpicture}[baseline={([yshift=11pt]current bounding box.south)},resize]
    \begin{pgfonlayer}{nodelayer}
      \node [style=none] (0) at (2, 0) {};
      \node [style=BETA] (1) at (0, 0) {};
      \node [style=none] (2) at (-2, 0) {};
      \node [style=FA] (3) at (0, 1.75) {};
    \end{pgfonlayer}
    \begin{pgfonlayer}{edgelayer}
      \draw [style=dashed] (0.center) to (1);
      \draw (2.center) to (1);
      \ifnum #1 >0 {\draw [style=derloopback] (3) arc (-270:90:.5);}\else {\draw [style=derloopback] (3) arc (-270:90:.5);}\fi
    \end{pgfonlayer}
  \end{tikzpicture}
}}
\newcommand{\DFAa}[1]
{\scalebox{\width}[\height]{
  \begin{tikzpicture}[baseline={([yshift=11pt]current bounding box.south)},resize]
	\begin{pgfonlayer}{nodelayer}
		\node [style=none] (0) at (2, 0) {};
		\node [style=BETA] (1) at (0, 0) {};
		\node [style=none] (2) at (-2, 0) {};
		\node [style=FA] (3) at (0, 1.25) {};
	\end{pgfonlayer}
	\begin{pgfonlayer}{edgelayer}
		\draw [style=dashed] (0.center) to (1);
		\draw [style=der] (2.center) to (1);
		\ifnum #1 >0 {\draw [style=dashed] (1) to (3);}\else {\draw (1) to (3);}\fi
	\end{pgfonlayer}
  \end{tikzpicture}
}}
\newcommand{\DFAb}[1]
{\scalebox{\width}[\height]{
  \begin{tikzpicture}[baseline={([yshift=-.5ex]current bounding box.center)},resize]
	\begin{pgfonlayer}{nodelayer}
		\node [style=none] (0) at (2, 0) {};
		\node [style=none] (1) at (-2, 0) {};
		\node [style=BETA] (2) at (.5, 0) {};
		\node [style=FA] (3) at (-1, 0) {};
	\end{pgfonlayer}
	\begin{pgfonlayer}{edgelayer}
		\draw [style=dashed] (0.center) to (2);
		\draw (1.center) to (3);
		\ifnum #1 >0 {\draw [style=dashder] (2) to (3);}\else {\draw [style=der] (2) to (3);}\fi
	\end{pgfonlayer}
  \end{tikzpicture}
}}
\newcommand{\DFAc}[1]
{\scalebox{\width}[\height]{
  \begin{tikzpicture}[baseline={([yshift=-.5ex]current bounding box.center)},resize]
	\begin{pgfonlayer}{nodelayer}
		\node [style=none] (0) at (2, 0) {};
		\node [style=none] (1) at (-2, 0) {};
		\node [style=BETA] (2) at (0.5, 0) {};
		\node [style=FA] (3) at (-1, 0) {};
	\end{pgfonlayer}
	\begin{pgfonlayer}{edgelayer}
		\draw [style=dashed] (0.center) to (2);
		\draw [style=der] (1.center) to (3);
		\ifnum #1 >0 {\draw [style=dash] (2) to (3);}\else {\draw (2) to (3);}\fi
	\end{pgfonlayer}
  \end{tikzpicture}
}}
\newcommand{\DFAd}[1]
{\scalebox{\width}[\height]{
  \begin{tikzpicture}[baseline={([yshift=11pt]current bounding box.south)},resize]
	\begin{pgfonlayer}{nodelayer}
		\node [style=none] (0) at (2, 0) {};
		\node [style=BETA] (1) at (0, 0) {};
		\node [style=none] (2) at (-2, 0) {};
		\node [style=FA] (3) at (0, 1.75) {};
	\end{pgfonlayer}
	\begin{pgfonlayer}{edgelayer}
		\draw [style=dashed] (0.center) to (1);
		\draw (2.center) to (1);
                \ifnum #1 >0 {\draw [style=dashder] (3) to (1);}\else {\draw [style=der] (3) to (1);}\fi
	\end{pgfonlayer}
  \end{tikzpicture}
}}
\newcommand{\FFAa}[2]
{\scalebox{\width}[\height]{
  \begin{tikzpicture}[baseline={([yshift=-.5ex]current bounding box.center)},resize]
	\begin{pgfonlayer}{nodelayer}
		\node [style=none] (0) at (2, 0) {};
		\node [style=FA] (1) at (1.25, 0) {};
		\node [style=none] (2) at (-2, 0) {};
		\node [style=BETA] (3) at (0.25, 0) {};
		\node [style=A] (4) at (-1.25, 0) {};
	\end{pgfonlayer}
	\begin{pgfonlayer}{edgelayer}
		\draw [style=dashed] (0.center) to (1);
		\draw (2.center) to (4);
                \ifnum #1 >0 {\draw [style=dashed] (4) arc (0:-141:-.75);}\else {\draw (4) arc (0:-141:-.75);}\fi
                \ifnum #2 >0 {\draw [style=dashed] (4) arc (0:141:-.75);}\else {\draw (4) arc (0:141:-.75);}\fi
	\end{pgfonlayer}
  \end{tikzpicture}
}}
\newcommand{\FFAb}[2]
{\scalebox{\width}[\height]{
  \begin{tikzpicture}[baseline={([yshift=1pt]current bounding box.south)},resize]
	\begin{pgfonlayer}{nodelayer}
		\node [style=none] (0) at (2, 0) {};
		\node [style=A] (1) at (0, 0) {};
		\node [style=none] (2) at (-2, 0) {};
		\node [style=BETA] (3) at (0,1) {};
		\node [style=FA] (4) at (0,2) {};
	\end{pgfonlayer}
	\begin{pgfonlayer}{edgelayer}
		\draw [style=dashed] (0.center) to (1);
		\draw (2.center) to (1);
                \ifnum #1 >0 {\draw [style=dashed] (1) to (3);}\else {\draw (1) to (3);}\fi
                \ifnum #2 >0 {\draw [style=dashed] (3) to (4);}\else {\draw (3) to (4);}\fi
	\end{pgfonlayer}
  \end{tikzpicture}
}}
\newcommand{\FFAc}[2]
{\scalebox{\width}[\height]{
  \begin{tikzpicture}[baseline={([yshift=8pt]current bounding box.south)},resize]
	\begin{pgfonlayer}{nodelayer}
		\node [style=none] (0) at (2, 0) {};
		\node [style=none] (1) at (-2, 0) {};
		\node [style=BETA] (2) at (.5, 0) {};
		\node [style=A] (3) at (-1, 0) {};
		\node [style=FA] (4) at (-1, 1.25) {};
	\end{pgfonlayer}
	\begin{pgfonlayer}{edgelayer}
		\draw [style=dashed] (0.center) to (2);
		\draw (1.center) to (3);
                \ifnum #1 >0 {\draw [style=dashed] (3) to (4);}\else {\draw (3) to (4);}\fi
                \ifnum #2 >0 {\draw [style=dashed] (2) to (3);}\else {\draw (2) to (3);}\fi
	\end{pgfonlayer}
  \end{tikzpicture}
}}
\newcommand{\FFa}[3]
{\scalebox{\width}[\height]{
  \begin{tikzpicture}[baseline={([yshift=1pt]current bounding box.south)},resize]
    \begin{pgfonlayer}{nodelayer}
      \node [style=none] (0) at (2, 0) {};
      \node [style=A] (1) at (0, .75) {};
      \node [style=none] (2) at (-2, 0) {};
      \node [style=BETA] (3) at (0, 1.75) {};
      \node [style=A] (4) at (0, 0) {};
    \end{pgfonlayer}
    \begin{pgfonlayer}{edgelayer}
      \draw (2.center) to (4);
      \draw [style=dashed] (0.center) to (4);
      \ifnum #1 >0 {\draw [style=dashed] (4) to (1)=;}\else {\draw (4) to (1);}\fi
      \ifnum #2 >0 {\draw [style=dashed] (1) arc (-90:-210:.5);}\else {\draw (1) arc (-90:-210:.5);}\fi
      \ifnum #3 >0 {\draw [style=dashed] (1) arc (-90:30:.5);}\else {\draw (1) arc (-90:30:.5);}\fi
    \end{pgfonlayer}
  \end{tikzpicture}
}}
\newcommand{\FFb}[3]
{\scalebox{\width}[\height]{
  \begin{tikzpicture}[baseline={([yshift=-.5ex]current bounding box.center)},resize]
    \begin{pgfonlayer}{nodelayer}
      \node [style=none] (0) at (2, 0) {};
      \node [style=none] (1) at (-2, 0) {};
      \node [style=A] (2) at (0, 0) {};
      \node [style=A] (3) at (-1, 0) {};
      \node [style=BETA] (4) at (1, 0) {};
    \end{pgfonlayer}
    \begin{pgfonlayer}{edgelayer}
      \draw [style=dashed] (0.center) to (4);
      \draw (1.center) to (3);
      \ifnum #1 >0 {\draw [style=dashed] (2) arc (0:180:.5);}\else {\draw (2) arc (0:180:.5);}\fi
      \ifnum #2 >0 {\draw [style=dashed] (2) arc (0:-180:.5);}\else {\draw (2) arc (0:-180:.5);}\fi
      \ifnum #3 >0 {\draw [style=dashed](2) to (4);}\else {\draw (2) to (4);}\fi
    \end{pgfonlayer}
  \end{tikzpicture}
}}
\newcommand{\FFc}[3]
{\scalebox{\width}[\height]{
  \begin{tikzpicture}[baseline={([yshift=-2.1ex]current bounding box.center)},resize]
    \begin{pgfonlayer}{nodelayer}
      \node [style=none] (0) at (2, 0) {};
      \node [style=A] (1) at (1, 0) {};
      \node [style=none] (2) at (-2, 0) {};
      \node [style=BETA] (3) at (0, 1) {};
      \node [style=A] (4) at (-1, 0) {};
    \end{pgfonlayer}
    \begin{pgfonlayer}{edgelayer}
      \draw [style=dashed] (0.center) to (1);
      \draw (2.center) to (4);
      \ifnum #1 >0 {\draw [style=dashed] (1) arc (0:-180:1);}\else {\draw (1) arc (0:-180:1);}\fi
      \ifnum #2 >0 {\draw [style=dashed] (1) arc (0:61:1);}\else {\draw (1) arc (0:61:1);}\fi
      \ifnum #3 >0 {\draw [style=dashed] (4) arc (0:-61:-1);}\else {\draw (4) arc (0:-61:-1);}\fi
    \end{pgfonlayer}
  \end{tikzpicture}
}}
\newcommand{\DDFF}[4]
{\scalebox{\width}[\height]{
  \begin{tikzpicture}[baseline={([yshift=-.5ex]current bounding box.center)},small]
    \begin{pgfonlayer}{nodelayer}
      \node [style=A] (0) at (-2, 0) {};
      \node [style=A] (1) at (2, 0) {};
    \end{pgfonlayer}
    \begin{pgfonlayer}{edgelayer}
      \ifnum #1 >0 {\draw [style=dashed] (1) arc (0:180:2);}\else {\draw (1) arc (0:180:2);}\fi
      \ifnum #2 >0 {\draw [style=dashed,bend left=35] (0) to (1);}\else {\draw [bend left=35] (0) to (1);}\fi
      \ifnum #3 >0 {\draw [style=dashed,bend right=35] (0) to (1);}\else {\draw [bend right=35] (0) to (1);}\fi
      \ifnum #4 >0 {\draw [style=dashderder] (1.south) arc (-1:-179:2);}\else {\draw [style=derder] (1.south) arc (-1:-179:2);}\fi
    \end{pgfonlayer}
  \end{tikzpicture}
}}
\newcommand{\FFDDl}[4]
{\scalebox{\width}[\height]{
  \begin{tikzpicture}[baseline={([yshift=-.5ex]current bounding box.center)},small]
    \begin{pgfonlayer}{nodelayer}
      \node [style=A] (0) at (0, 0) {};
      \node [style=A] (1) at (2, 0) {};
    \end{pgfonlayer}
    \begin{pgfonlayer}{edgelayer}
      \ifnum #1 >0 {\draw [style=dashderloopback] (0) arc (0:360:1);}\else {\draw [style=derloopback] (0) arc (0:360:1);}\fi
      \ifnum #2 >0 {\draw [style=dashed] (1) arc (0:180:1);}\else {\draw (1) arc (0:180:1);}\fi
      \ifnum #3 >0 {\draw [style=dashder] (1) to (0);}\else {\draw [style=der] (1) to (0);}\fi
      \ifnum #4 >0 {\draw [style=dashed] (1) arc (0:-180:1);}\else {\draw (1) arc (0:-180:1);}\fi
    \end{pgfonlayer}
  \end{tikzpicture}
}}
\newcommand{\FFDDlprime}[4]
{\scalebox{\width}[\height]{
  \begin{tikzpicture}[baseline={([yshift=-.5ex]current bounding box.center)},small]
    \begin{pgfonlayer}{nodelayer}
      \node [style=A] (0) at (0, 0) {};
      \node [style=A] (1) at (2, 0) {};
    \end{pgfonlayer}
    \begin{pgfonlayer}{edgelayer}
      \ifnum #1 >0 {\draw [style=dashderloopback] (0) arc (0:-360:1);}\else {\draw [style=derloopback] (0) arc (0:-360:1);}\fi
      \ifnum #2 >0 {\draw [style=dashed] (1) arc (0:180:1);}\else {\draw (1) arc (0:180:1);}\fi
      \ifnum #3 >0 {\draw [style=dashder] (1) to (0);}\else {\draw [style=der] (1) to (0);}\fi
      \ifnum #4 >0 {\draw [style=dashed] (1) arc (0:-180:1);}\else {\draw (1) arc (0:-180:1);}\fi
    \end{pgfonlayer}
  \end{tikzpicture}
}}
\newcommand{\FFDlDl}[4]
{\scalebox{\width}[\height]{
  \begin{tikzpicture}[baseline={([yshift=-.5ex]current bounding box.center)},small]
    \begin{pgfonlayer}{nodelayer}
      \node [style=A] (0) at (-1, 0) {};
      \node [style=A] (1) at (1, 0) {};
    \end{pgfonlayer}
    \begin{pgfonlayer}{edgelayer}
      \ifnum #1 >0 {\draw [style=dashderloopback] (0) arc (0:360:.5);}\else {\draw [style=derloopback] (0) arc (0:360:.5);}\fi
      \ifnum #2 >0 {\draw [style=dashed] (1) arc (0:180:1);}\else {\draw (1) arc (0:180:1);}\fi
      \ifnum #3 >0 {\draw [style=dashed] (1) arc (0:-180:1);}\else {\draw (1) arc (0:-180:1);}\fi
      \ifnum #4 >0 {\draw [style=dashderloopback] (1) arc (0:360:-.5);}\else {\draw [style=derloopback] (1) arc (0:360:-.5);}\fi
    \end{pgfonlayer}
  \end{tikzpicture}
}}
\newcommand{\FFFD}[5]
{\scalebox{\width}[\height]{
  \begin{tikzpicture}[baseline={([yshift=-.5ex]current bounding box.center)},small]
    \begin{pgfonlayer}{nodelayer}
      \node [style=A] (0) at (-2, 0) {};
      \node [style=A] (1) at (2, 0) {};
      \node [style=A] (2) at (0, 2) {};
    \end{pgfonlayer}
    \begin{pgfonlayer}{edgelayer}
      \ifnum #1 >0 {\draw [style=dashderback] (2) arc (90:178:2);}\else {\draw [style=derback] (2) arc (90:178:2);}\fi
      \ifnum #2 >0 {\draw [style=dashed] (0) to (2);}\else {\draw (0) to (2);}\fi
      \ifnum #3 >0 {\draw [style=dashed] (1) arc (0:90:2);}\else {\draw (1) arc (0:90:2);}\fi
      \ifnum #4 >0 {\draw [style=dashed] (0) to (1);}\else {\draw (0) to (1);}\fi
      \ifnum #5 >0 {\draw [style=dashed] (1) arc (0:-180:2);}\else {\draw (1) arc (0:-180:2);}\fi
    \end{pgfonlayer}
  \end{tikzpicture}
}}
\newcommand{\FFFADa}[4]
{\scalebox{\width}[\height]{
  \begin{tikzpicture}[baseline={([yshift=-.5ex]current bounding box.center)},small]
     \begin{pgfonlayer}{nodelayer}
	\node [style=A] (0) at (-2, 0) {};
	\node [style=A] (1) at (2, 0) {};
	\node [style=FA] (2) at (0, 2) {};
     \end{pgfonlayer}
     \begin{pgfonlayer}{edgelayer}
	\ifnum #1 >0 {\draw [style=dashderback] (0) arc (0:-88:-2);}\else {\draw [style=derback] (0) arc (0:-88:-2);}\fi
	\ifnum #2 >0 {\draw [style=dashed] (1) arc (0:90:2);}\else {\draw (2) arc (0:90:2);}\fi
	\ifnum #3 >0 {\draw [style=dashed] (0) to (1);}\else {\draw (0) to (1);}\fi
	\ifnum #4 >0 {\draw [style=dashed] (1) arc (0:-180:2);}\else {\draw (1) arc (0:-180:2);}\fi
	\end{pgfonlayer}
  \end{tikzpicture}
}}
\newcommand{\FFFADb}[4]
{\scalebox{\width}[\height]{
  \begin{tikzpicture}[baseline={([yshift=-.5ex]current bounding box.center)},small]
    \begin{pgfonlayer}{nodelayer}
      \node [style=A] (0) at (-1, 0) {};
      \node [style=A] (1) at (2, 0) {};
      \node [style=FA] (2) at (-2, 0) {};
    \end{pgfonlayer}
    \begin{pgfonlayer}{edgelayer}
      \ifnum #1 >0 {\draw [style=dashed] (0) to (2);}\else {\draw (0) to (2);}\fi
      \ifnum #2 >0 {\draw [style=dashed] (1) arc (0:180:1.5);}\else {\draw (1) arc (0:180:1.5);}\fi
      \ifnum #3 >0 {\draw [style=dashder] (1) to (0);}\else {\draw [style=der] (1) to (0);}\fi
      \ifnum #4 >0 {\draw [style=dashed] (1) arc (0:-180:1.5);}\else {\draw (1) arc (0:-180:1.5);}\fi
    \end{pgfonlayer}
  \end{tikzpicture}
}}
\newcommand{\FFFADl}[4]
{\scalebox{\width}[\height]{
  \begin{tikzpicture}[baseline={([yshift=-.5ex]current bounding box.center)},small]
    \begin{pgfonlayer}{nodelayer}
      \node [style=A] (0) at (-1, 0) {};
      \node [style=A] (1) at (1, 0) {};
      \node [style=FA] (2) at (-2, 0) {};
    \end{pgfonlayer}
    \begin{pgfonlayer}{edgelayer}
      \ifnum #1 >0 {\draw [style=dashed] (0) to (2);}\else {\draw (0) to (2);}\fi
      \ifnum #2 >0 {\draw [style=dashed] (1) arc (0:180:1);}\else {\draw (1) arc (0:180:1);}\fi
      \ifnum #3 >0 {\draw [style=dashed] (1) arc (0:-180:1);}\else {\draw(1) arc (0:-180:1);}\fi
      \ifnum #4 >0 {\draw [style=dashderloopback] (1) arc (0:360:-.5);}\else {\draw [style=derloopback] (1) arc (0:360:-.5);}\fi
    \end{pgfonlayer}
  \end{tikzpicture}
}}
\newcommand{\FFFAFA}[4]
{\scalebox{\width}[\height]{
  \begin{tikzpicture}[baseline={([yshift=-.5ex]current bounding box.center)},small]
    \begin{pgfonlayer}{nodelayer}
      \node [style=A] (0) at (-1, 0) {};
      \node [style=A] (1) at (1, 0) {};
      \node [style=FA] (2) at (-2, 0) {};
      \node [style=FA] (3) at (2, 0) {};
    \end{pgfonlayer}
    \begin{pgfonlayer}{edgelayer}
      \ifnum #1 >0 {\draw [style=dashed] (0) to (2);}\else {\draw (0) to (2);}\fi
      \ifnum #2 >0 {\draw [style=dashed] (1) arc (0:180:1);}\else {\draw (1) arc (0:180:1);}\fi
      \ifnum #3 >0 {\draw [style=dashed] (1) arc (0:-180:1);}\else {\draw (1) arc (0:-180:1);}\fi
      \ifnum #4 >0 {\draw [style=dashed] (1) to (3);}\else {\draw (1) to (3);}\fi
    \end{pgfonlayer}
  \end{tikzpicture}
}}
\newcommand{\FFFFa}[6]
{\scalebox{\width}[\height]{
  \begin{tikzpicture}[baseline={([yshift=-.5ex]current bounding box.center)},small]
    \begin{pgfonlayer}{nodelayer}
      \node [style=A] (0) at (-1.75, 0) {};
      \node [style=A] (1) at (1.75, 0) {};
      \node [style=A,anchor=north] (2) at (-0.75, 1.65) {};
      \node [style=A,anchor=north] (3) at (0.75, 1.65) {};
    \end{pgfonlayer}
    \begin{pgfonlayer}{edgelayer}
      \ifnum #1 >0 {\draw [style=dashed] (0) arc (0:-atan2(1,.5):-1.75);}\else {\draw (0) arc (0:-atan2(1,.5):-1.75);}\fi
      \ifnum #2 >0 {\draw [style=dashed] (3) arc (0:180:.75);}\else {\draw (3) arc (0:180:.75);}\fi
      \ifnum #3 >0 {\draw [style=dashed] (3) arc (0:-180:.75);}\else {\draw (3) arc (0:-180:.75);}\fi
      \ifnum #4 >0 {\draw [style=dashed] (1) arc (0:atan2(1,.5):1.75);}\else {\draw (1) arc (0:atan2(1,.5):1.75);}\fi
      \ifnum #5 >0 {\draw [style=dashed] (0) to (1);}\else {\draw (0) to (1);}\fi
      \ifnum #6 >0 {\draw [style=dashed] (1) (1) arc (0:-180:1.75);}\else {\draw (1) (1) arc (0:-180:1.75);}\fi
    \end{pgfonlayer}
  \end{tikzpicture}
}}
\newcommand{\FFFFaT}
{\scalebox{\width}[\height]{
  \begin{tikzpicture}[baseline={([yshift=-.5ex]current bounding box.center)},small]
    \begin{pgfonlayer}{nodelayer}
      \node [style=A] (0) at (-1.75, 0) {};
      \node [style=A] (1) at (1.75, 0) {};
      \node [style=A,anchor=north] (2) at (-.75, 1.65) {};
      \node [style=A,anchor=north] (3) at (.75, 1.65) {};
    \end{pgfonlayer}
    \begin{pgfonlayer}{edgelayer}
      \draw (0) arc (0:-atan2(1,.5):-1.75);
      \draw (3) arc (0:180:.75);
      \draw (3) arc (0:-180:.75);
      \draw (1) arc (0:atan2(1,.5):1.75);
      \draw (0) to (1);
      \draw (1) arc (0:-180:1.75);
    \end{pgfonlayer}
  \end{tikzpicture}
}}
\newcommand{\FFFFb}[6]
{\scalebox{\width}[\height]{
  \begin{tikzpicture}[baseline={([yshift=-.5ex]current bounding box.center)},small]
    \begin{pgfonlayer}{nodelayer}
      \node [style=A] (0) at (-2, 0) {};
      \node [style=A] (1) at (0, 0) {};
      \node [style=A] (2) at (2, 0) {};
      \node [style=A] (3) at (0, -2) {};
    \end{pgfonlayer}
    \begin{pgfonlayer}{edgelayer}
      \ifnum #1 >0 {\draw [style=dashed](2) arc (0:180:2);}\else {\draw (2) arc (0:180:2);}\fi
      \ifnum #2 >0 {\draw [style=dashed](0) to (1);}\else {\draw (0) to (1);}\fi
      \ifnum #3 >0 {\draw [style=dashed](1) to (2);}\else {\draw (1) to (2);}\fi
      \ifnum #4 >0 {\draw [style=dashed](3) arc (-90:-180:2);}\else {\draw (3) arc (-90:-180:2);}\fi
      \ifnum #5 >0 {\draw [style=dashed](1) to (3);}\else {\draw (1) to (3);}\fi
      \ifnum #6 >0 {\draw [style=dashed](3) arc (-90:90:2);}\else {\draw (3) arc (-90:90:2);}\fi
    \end{pgfonlayer}
  \end{tikzpicture}
}}
\newcommand{\FFFFbT}
{\scalebox{\width}[\height]{
  \begin{tikzpicture}[baseline={([yshift=-.5ex]current bounding box.center)},small]
    \begin{pgfonlayer}{nodelayer}
      \node [style=A] (0) at (-2, 0) {};
      \node [style=A] (1) at (0, 0) {};
      \node [style=A] (2) at (2, 0) {};
      \node [style=A] (3) at (0, -2) {};
    \end{pgfonlayer}
    \begin{pgfonlayer}{edgelayer}
      \draw (2) arc (0:180:2);
      \draw (0) to (1);
      \draw (1) to (2);
      \draw (3) arc (-90:-180:2);
      \draw (1) to (3);
      \draw (3) arc (-90:90:2);
    \end{pgfonlayer}
  \end{tikzpicture}
}}
\newcommand{\dFAFFFaT}
{\scalebox{\width}[\height]{
  \begin{tikzpicture}[baseline={([yshift=-.5ex-\yshift]current bounding box.center)},small]
	\begin{pgfonlayer}{nodelayer}
		\node [style=none] (0) at (2, 0) {};
		\node [style=none] (1) at (-2, 0) {};
		\node [style=A] (2) at (0, 0) {};
		\node [style=A] (3) at (0, 1) {};
		\node [style=A] (4) at (0, 2) {};
		\node [style=FA] (5) at (1.75, 2) {};
	\end{pgfonlayer}
	\begin{pgfonlayer}{edgelayer}
		\draw [style=none] (0.center) to (2);
		\draw (1.center) to (2);
		\draw (2) to (3);
		\draw (3) arc (-90:-270:.5);
		\draw (3) arc (-90:90:.5);
		\draw  (4) to (5);
	\end{pgfonlayer}
  \end{tikzpicture}
}}
\newcommand{\dFAFFFa}[4]
{\scalebox{\width}[\height]{
  \begin{tikzpicture}[baseline={([yshift=-.5ex-\yshift]current bounding box.center)},small]
	\begin{pgfonlayer}{nodelayer}
		\node [style=none] (0) at (2, 0) {};
		\node [style=none] (1) at (-2, 0) {};
		\node [style=A] (2) at (0, 0) {};
		\node [style=A] (3) at (0, 1) {};
		\node [style=A] (4) at (0, 2) {};
		\node [style=FA] (5) at (1.75, 2) {};
	\end{pgfonlayer}
	\begin{pgfonlayer}{edgelayer}
		\draw [style=dashed] (0.center) to (2);
		\draw (1.center) to (2);
		\ifnum #1 >0 {\draw [style=dashed](2) to (3);}\else {\draw (2) to (3);}\fi
		\ifnum #2 >0 {\draw [style=dashed] (3) arc (-90:-270:.5);}\else {\draw (3) arc (-90:-270:.5);}\fi
		\ifnum #3 >0 {\draw [style=dashed] (3) arc (-90:90:.5);}\else {\draw (3) arc (-90:90:.5);}\fi
		\ifnum #4 >0 {\draw [style=dashed] (4) to (5);}\else {\draw  (4) to (5);}\fi
	\end{pgfonlayer}
  \end{tikzpicture}
}}
\newcommand{\dFAFFFbprime}[4]
{\scalebox{\width}[\height]{
  \begin{tikzpicture}[baseline={([yshift=.5ex-\yshift]current bounding box.center)},small]
	\begin{pgfonlayer}{nodelayer}
		\node [style=none] (0) at (2,0) {};
		\node [style=none] (1) at (-2,0) {};
		\node [style=A] (2) at (0,0) {};
		\node [style=A] (4) at (-1,0) {};
		\node [style=FA] (5) at (1,1.75) {};
		\node [style=A] (6) at (1,0) {};
	\end{pgfonlayer}
	\begin{pgfonlayer}{edgelayer}
		\draw (1.center) to (4);
		\draw [style=dashed] (0.center) to (6);
		\ifnum #1 >0 {\draw [style=dashed] (6) to (5);}\else {\draw (6) to (5);}\fi
		\ifnum #2 >0 {\draw [style=dashed] (6) to (2);}\else {\draw (6) to (2);}\fi
		\ifnum #3 >0 {\draw [style=dashed] (2) arc (0:180:.5);}\else {\draw (2) arc (0:180:.5);}\fi
		\ifnum #4 >0 {\draw [style=dashed] (2) arc (0:-180:.5);}\else {\draw (2) arc (0:-180:.5);}\fi
	\end{pgfonlayer}
  \end{tikzpicture}
}}
\newcommand{\dFAFFFb}[4]
{\scalebox{\width}[\height]{
  \begin{tikzpicture}[baseline={([yshift=.5ex-\yshift]current bounding box.center)},small]
	\begin{pgfonlayer}{nodelayer}
		\node [style=none] (0) at (2,0) {};
		\node [style=none] (1) at (-2,0) {};
		\node [style=A] (2) at (1,0) {};
		\node [style=A] (3) at (-1,0) {};
		\node [style=A] (4) at (0,0) {};
		\node [style=FA] (5) at (-1,1.75) {};
	\end{pgfonlayer}
	\begin{pgfonlayer}{edgelayer}
		\draw [style=dashed] (0.center) to (2);
		\draw (1.center) to (3);
		\ifnum #1 >0 {\draw [style=dashed] (2) arc (0:-180:.5);}\else {\draw (2) arc (0:-180:.5);}\fi
		\ifnum #2 >0 {\draw [style=dashed] (2) arc (0:180:.5);}\else {\draw (2) arc (0:180:.5);}\fi
		\ifnum #3 >0 {\draw [style=dashed](4) to (3);}\else {\draw (4) to (3);}\fi
		\ifnum #4 >0 {\draw [style=dashed] (3) to (5);}\else {\draw (3) to (5);}\fi
	\end{pgfonlayer}
  \end{tikzpicture}
}}
\newcommand{\dFAFFFbT}
{\scalebox{\width}[\height]{
  \begin{tikzpicture}[baseline={([yshift=-.5ex]current bounding box.center)},small]
	\begin{pgfonlayer}{nodelayer}
		\node [style=none] (0) at (2,0) {};
		\node [style=none] (1) at (-2,0) {};
		\node [style=A] (2) at (1,0) {};
		\node [style=A] (3) at (-1,0) {};
		\node [style=A] (4) at (0,0) {};
		\node [style=FA] (5) at (-1,1.75) {};
	\end{pgfonlayer}
	\begin{pgfonlayer}{edgelayer}
		\draw [style=none] (0.center) to (2);
		\draw (1.center) to (3);
		\draw (2) arc (0:-180:.5);
		\draw (2) arc (0:180:.5);
		\draw (4) to (3);
		\draw (3) to (5);
	\end{pgfonlayer}
  \end{tikzpicture}
}}
\newcommand{\dFAFFFc}[4]
{\scalebox{\width}[\height]{
  \begin{tikzpicture}[baseline={([yshift=-.75ex-\yshift/2]current bounding box.center)},small]
	\begin{pgfonlayer}{nodelayer}
		\node [style=none] (0) at (-2, 0) {};
		\node [style=none] (1) at (2, 0) {};
		\node [style=A] (2) at (-1, 0) {};
		\node [style=A] (3) at (1, 0) {};
		\node [style=A] (4) at (0, 1) {};
		\node [style=FA] (5) at (0, 2) {};
	\end{pgfonlayer}
	\begin{pgfonlayer}{edgelayer}
		\draw [style=dashed] (1) to (3);
		\draw (0) to (2);
		\ifnum #1 >0 {\draw [style=dashed] (3) arc (0:-180:1);}\else {\draw  (3) arc (0:-180:1);}\fi
		\ifnum #2 >0 {\draw [style=dashed] (3) arc (0:90:1);}\else {\draw  (3) arc (0:90:1);}\fi
		\ifnum #3 >0 {\draw [style=dashed] (4) arc (90:180:1);}\else {\draw  (4) arc (90:180:1);}\fi
		\ifnum #4 >0 {\draw [style=dashed] (4) to (5);}\else {\draw  (4) to (5);}\fi
	\end{pgfonlayer}
  \end{tikzpicture}
}}
\newcommand{\dFAFFFcT}
{\scalebox{\width}[\height]{
  \begin{tikzpicture}[baseline={([yshift=-.5ex]current bounding box.center)},small]
	\begin{pgfonlayer}{nodelayer}
		\node [style=none] (0) at (-2, 0) {};
		\node [style=none] (1) at (2, 0) {};
		\node [style=A] (2) at (-1, 0) {};
		\node [style=A] (3) at (1, 0) {};
		\node [style=A] (4) at (0, 1) {};
		\node [style=FA] (5) at (0,2) {};
	\end{pgfonlayer}
	\begin{pgfonlayer}{edgelayer}
		\draw [style=none] (1) to (3);
		\draw (0) to (2);
		\draw (3) arc (0:-180:1);
		\draw  (3) arc (0:90:1);
		\draw  (4) arc (90:180:1);
		\draw  (4) to (5);
	\end{pgfonlayer}
  \end{tikzpicture}
}}

\title{\boldmath O($D$,$D$)-covariant two-loop $\beta$-functions and Poisson-Lie T-duality}
\preprint{MI-TH-2031}

\author[a]{Falk Hassler}
\author[b]{and Thomas B. Rochais}

\emailAdd{falk@fhassler.de}
\emailAdd{thb@sas.upenn.edu}

\affiliation[a]{George P. \& Cynthia Woods Mitchell Institute for Fundamental Physics and Astronomy,\\ Texas A\&M University, College Station, TX 77843, USA}
\affiliation[b]{Department of Physics and Astronomy, University of Pennsylvania, Philadelphia, PA 19104, USA}

\abstract{We show that the one- and two-loop $\beta$-functions of the closed, bosonic string can be written in a manifestly O($D$,$D$)-covariant form. Based on this result, we prove that
\begin{enumerate}[leftmargin=1.5em,itemindent=1em,itemsep=0cm,label*=\arabic*)]
  \item Poisson-Lie symmetric $\sigma$-models are two-loop renormalisable and
  \item their $\beta$-functions are invariant under Poisson-Lie T-duality.
\end{enumerate}
Moreover, we identify a distinguished scheme in which Poisson-Lie symmetry is manifest. It simplifies the calculation of two-loop $\beta$-functions significantly and thereby provides a powerful new tool to advance into the quantum regime of integrable $\sigma$-models and generalised T-dualities. As an illustrating example, we present the two-loop $\beta$-functions of the integrable $\lambda$- and $\eta$-deformation.}

\begin{document}

\maketitle

\section{Introduction}
Two seemingly completely different theories, for example, one strongly coupled and the other one weakly coupled, may still exhibit the same physics. This remarkable phenomenon is governed by dualities and even if it is not generic, it can provide deep insights into the theories involved. A genuine duality is not restricted to the classical level but still applies after quantisation. Unfortunately, the dualities that are understood best, only apply to a very limited class of theories. A prominent example is abelian T-duality in string theory. It is restricted to target spaces with abelian isometries which are of course by no means generic. Yet, it provides many crucial insights into string theory. Therefore, it is unarguably an important challenge to advance our knowledge about dualities and their properties. In this process, one encounters the problem that the notion of duality outlined above is very strong. But often only certain properties of a theory are relevant to solve a problem. In this case, it is sufficient to ask: Is it possible to find two different theories that share at least these properties? This approach has the considerable advantage that it is much less constraining. A remarkable example along these lines is Poisson-Lie (PL) T-duality \cite{Klimcik:1995ux}. 

In fact the term PL T-duality is slightly ambiguous because it is sometimes used as a synonym for a whole family of different dualities. All started with non-abelian T-duality \cite{delaOssa:1992vci}. It is based on the observation that the Buscher procedure \cite{Buscher:1987sk}, which mediates abelian T-duality on the closed string $\sigma$-model, can be extended to non-abelian isometries. There are however two major problems one encounters in this generalisation \cite{Giveon:1993ai,*Alvarez:1993qi}:
\begin{enumerate}[leftmargin=1.5em,itemindent=1em,itemsep=0cm,label*=\arabic*)]
\item\label{item:prob1} The Buscher procedure employs a Lagrange multiplier that enforces a flat connection on the worldsheet. However, the connection might still have non-trivial monodromies around non-contractible cycles on the worldsheet. In the abelian case, this problem is resolved by using a periodic Lagrange multiplier \cite{Rocek:1991ps}. Physically this choice leads to the celebrated momentum winding exchange under abelian T-duality and allows for the identification of the topology of the dual target space. Unfortunately, this idea does not work for non-abelian isometries. Therefore, the global properties of non-abelian T-duality are not fully understood and a topic of active research. 
  \item\label{item:prob2} A second problem is that the resulting, dual target space geometry has in general a smaller isometry group which seemingly prohibits the duality to be inverted. This is particularly severe because by definition a duality has to be invertible.
\end{enumerate}
PL T-duality solves problem~\ref{item:prob2} by the seminal observation \cite{Klimcik:1995ux} that both, the original and the dual, $\sigma$-models originate from the same underlying structure, a Drinfeld double. Drinfeld doubles are in one-to-one correspondence with PL groups, which actually form the corresponding target spaces and give the duality its name. Remarkably, non-abelian T-duality is based on a further refined class of Drinfeld doubles with an abelian, maximally isotropic subgroup. Therefore, PL T-duality, which works for arbitrary Drinfeld doubles, not just shows that non-abelian T-duality is invertible but additionally gives rise to a broader family of dualities that do not need isometries at all. Intriguingly, this already rich notion of duality can be even pushed beyond Drinfeld doubles by relaxing the Poisson structure of the PL group to a quasi-Poisson structure \cite{Klimcik:2001vg,*Severa:2001qm}. Physically, this leads to a Wess-Zumino-Witten (WZW) term and describes $H$-flux in a non-trivial cohomology class. Eventually, the duality was extended from groups to cosets by the dressing coset construction \cite{Klimcik:1996np}. Thus, the term ``PL T-duality'' may refer to any member of the family
\begin{center}
  dressing cosets $\supset$ PL with WZW term $\supset$ PL $\supset$ non-abelian T-duality\,.
\end{center}
In this paper, we use it for all of them except for dressing cosets, which we hope to address in the future based on \cite{Demulder:2019vvh}.

Problem \ref{item:prob1} is still an issue since it prohibits discussions of PL T-duality on higher genus Riemann surfaces that appear in the $g_\mathrm{s}$ expansion of the string path integral. Moreover, beyond non-abelian T-duality we do not know a gauging procedure comparable to Buscher's original approach which could be used to check if the path integrals of dual theories match. Hence, PL T-duality is deemed to not be a genuine symmetry of string theory but at most a map between different conformal field theories (CFTs). However, quantum corrections to the classical string are not exclusively controlled by $g_\mathrm{s}$. Additionally, the $\alpha'$-expansion incorporates quantum effects for fixed worldsheet topologies. Fortunately, $\alpha'$-corrections are accessible even without solving problem \ref{item:prob1} and at the leading, one-loop order in this expansion, it is known that \cite{\refonelooprenorm}
\begin{enumerate}[leftmargin=1.5em,itemindent=1em,itemsep=0cm,label*=\arabic*)]
\item PL symmetric\footnote{PL symmetric refers to the properties a target space geometry must have to permit PL T-duality. We give an exact definition in section~\ref{sec:genframes}.} $\sigma$-models are renormalisable.
  \item The RG flows of two PL T-dual $\sigma$-models are identical because they share the same $\beta$-functions.
\end{enumerate}
These two points are important hints that PL T-duality is not just a classical phenomenon but captures quantum effects as well. An immediate question is if they continue to hold at higher loop orders. We will answer it in the affirmative at two loops in this paper by explicitly computing the one- and two-loop $\beta$-functions of the bosonic string. For string theory, most relevant are points in the moduli space where these functions vanish, and CFTs at fixed points of the RG flow emerge. In this case, it is instructive to expand the $\beta$-functions in the couplings $\lambda^a$. As we discuss in much more detail below, the resulting expansion is scheme dependent. However, there exists a particular scheme in which it reads \cite{Gaberdiel:2008fn,Codello:2017hhh}
\begin{equation}
  \beta^a = \mu \frac{\dd \lambda^a}{\dd \log\mu} = - ( 2 - \Delta_a ) \lambda^a + \sum\limits_{b,c} C^a{}_{bc} \lambda^b \lambda^c
  + \dots\,,
\end{equation}
where $\Delta_a$ and $C^c{}_{ab}$ denote the anomalous dimensions and coupling constants which appear in the OPE
\begin{equation}
  \left\langle \mathcal{O}_a(x) \mathcal{O}_b(y) \dots \right\rangle = 
    \sum_c \frac1{|x-y|^{\Delta_a + \Delta_b - \Delta_c}} C^c{}_{ab} \left\langle \mathcal{O}_c(x) \dots \right\rangle
\end{equation}
of the classically marginal operators $\mathcal{O}_a$ that correspond to the couplings $\lambda^a$. There are other primary fields in the CFT, too. Hence, the $\beta$-functions do not capture the CFT data completely. Still, as PL T-duality does not affect $\beta$-functions (at least up to two loops), the two CFTs it connects are clearly not unrelated and share at least a common subsector formed by the operators $\mathcal{O}_a$.

Hence, we conclude: A quantum version of PL T-duality is not out of reach and definitely worth studying. Especially, since this duality is tightly linked to integrable deformations of two dimensional $\sigma$-models\footnote{Recently, \cite{Lacroix:2020flf} constructed $\mathcal{E}$-models \cite{Klimcik:1995dy} for a large class of integrable $\sigma$-models and thereby makes their PL symmetry manifest.}. Prominent examples include Yang-Baxter deformations \cite{Klimcik:2002zj}, which are either governed by the homogeneous or inhomogeneous, classical Yang-Baxter equation, and $\lambda$-deformations \cite{Sfetsos:2013wia}. While all of them were discovered independently, they are actually linked by a web of PL T-dualities (and analytic continuations) \cite{\reflambdaetaPL}. Because the S-matrix of integrable models is strongly constrained it only depends on a small number of free parameters. Ultimately, these parameters originate from couplings in the underlying $\sigma$-model. Of course, this relation is extremely complicated but it suggests that if integrability is not broken by quantum effects, only these couplings are affected by RG flows. Motivated by this observation, it was possible to show that $\eta$- and $\lambda$-deformation are indeed two-loop renormalisable \cite{Hoare:2019ark,Hoare:2019mcc,Georgiou:2019nbz}. This is an important clue that PL symmetric $\sigma$-models, might be renormalisable beyond one-loop. Moreover, insights from double field theory (DFT) \cite{Siegel:1993th,*Siegel:1993xq,Hull:2009mi} were used to show that PL T-duality with adapted transformation rules maps CFTs to CFTs \cite{\refPLalphaprime}. Motivated by these findings we will use DFT techniques to compute $\beta$-functions for PL $\sigma$-models and show that they are renormalisable. Because the framework we are using is independent of the chosen duality frame, our results automatically imply that all $\beta$-functions are preserved under PL T-duality.

Because the computations which we present are technically challenging, we split their presentation into two parts. In section~\ref{sec:results}, we summarise our results and demonstrate them for the $\lambda$- and $\eta$-deformation. All required tools are reviewed, but no derivations are given. For readers who are mainly interested in computing the $\beta$-functions of particular PL $\sigma$-models, for example integrable deformations, reading this section should be sufficient. Detailed derivations are discussed in section~\ref{sec:derivation}. In particular, we exploit that the $\beta$-functions we are dealing with are governed by a gradient flow \cite{Metsaev:1987zx,Hull:1987yi,Friedan:2009ik}. We show how this flow arises in the conventional $\sigma$-model and then rewrite all its constituents in an O($D$,$D$)-covariant form. After capturing the target space geometry of a PL $\sigma$-model in terms of a generalised frame field and the corresponding generalised fluxes \cite{\reflongDFTPLlist}, this manifestly covariant form permits us to directly read off the results presented in section~\ref{sec:results}. However, the O($D$,$D$)-covariant $\beta$-functions, which we derive, are completely general and hold for arbitrary target space geometries. Section~\ref{sec:conclusion} concludes the paper with several still open questions and ideas for future research.

\section{One and two-loop \texorpdfstring{$\beta$}{beta}-functions}\label{sec:results}
In the following, we present a summary of the main result of this paper, the two-loop $\beta$-functions for a bosonic, PL symmetric $\sigma$-model of the form \cite{Fradkin:1984pq}
\begin{equation}\label{eqn:sigmamodel}
  S = \frac1{4\pi\alpha'} \int_\Sigma \dd z \dd \overline{z} ( \sqrt{h} h^{ab} g_{ij} \partial_a X^i \partial_b X^j + \ci \epsilon^{ab} B_{ij} \partial_a X^i \partial_b X^j + \alpha' \sqrt{h} R^{(2)} \phi )\,.
\end{equation}
The couplings of this model are the target space metric $g_{ij}$, $B$-field $B_{ij}$, and dilaton $\phi$. As we explain in section~\ref{sec:genframes}, PL symmetry constraints them significantly. After imposing it, only a finite number of couplings survive. We discuss their $\beta$-functions first at one-loop and eventually at two loops in sections~\ref{sec:oneloop} and \ref{sec:twoloops}, respectively. Along the way we introduce all required DFT techniques and apply them to the $\lambda$- and $\eta$-deformation on a Lie group $G$ \cite{Sfetsos:2013wia} to have an explicit example. Poisson-Lie T-duality is completely manifest in our framework and preserves the $\beta$-functions. This will allow us to deduce the RG flow of the $\eta$-deformation \cite{Klimcik:2002zj} directly from the results of the $\lambda$-deformation since both are related by PL T-duality and analytic continuation \cite{\reflambdaetaPL}.

\subsection{PL symmetry and generalised frame fields}\label{sec:genframes}
A very powerful way to describe PL symmetric target space geometries is in terms of a generalised frame field $E_A{}^I$ on the generalised tangent bundle $T M \oplus T^* M$ of the target space manifold $M$. Each element of this bundle has a vector and a one-form component. A generalised frame $E_A = E_A{}^i \partial_i + E_{A i} d x^i$ consists of $A=1, \dots, 2D$  such elements, where $D$ denotes the dimension of the target space. They are linearly independent and defined on every point $M$. We distinguish two different sets of indices: $A$ to $H$ are called flat and from $I$ on they are called curved. While the latter are naturally associated to the generalised tangent space, the former are valued in a doubled Lie algebra $\mfd$ with generators $T_A$ and the commutator relations
\begin{equation}
  [T_A, T_B] = F_{AB}{}^C t_C\,.
\end{equation}
Additionally, $\mfd$ is equipped with a ($D$,$D$)-signature pairing 
\begin{equation}
  \langle T_A, T_B \rangle = \eta_{AB}\,,
\end{equation}
which is invariant under the adjoint action of $\mfd$. As a direct consequence $F_{ABC} = F_{AB}{}^D \eta_{DC}$ is totally anti-symmetric. We follow the standard convention in DFT and lower/raise indices with $\eta_{AB}$/its inverse $\eta^{AB}$. Without loss of generality, they can always be brought into the form
\begin{equation}
  \eta_{AB} = \begin{pmatrix} \eta_{ab} & 0 \\
    0 & - \eta_{\bar a\bar b} \end{pmatrix} \qquad 
  \text{and} \qquad
  \eta^{AB} = \begin{pmatrix} \eta^{ab} & 0 \\
    0 & - \eta^{\bar a\bar b} \end{pmatrix}\,,
\end{equation}
where lowercase indices run only from $1$ to $D$ and $\eta_{ab}=\eta_{\bar a\bar b}=\eta^{ab}=\eta^{\bar a\bar b}$ is the invariant metric of the target space's Lorentz group. The generalised frame field translates between the structure on $\mfd$ and the generalised tangent space. More specifically, it relates $\eta^{AB}$ to the canonical pairing
\begin{equation}
  \eta^{IJ} = E_A{}^I E_B{}^J \eta^{AB} = \begin{pmatrix} 0 & \delta_i^j \\ \delta^i_j & 0 \end{pmatrix}
\end{equation}
on $T M \oplus T^*M$.

In this framework, PL symmetry is encoded by the partial differential equation \cite{Hassler:2017yza}
\begin{equation}\label{eqn:framealgebra}
  \mathcal{L}_{E_A} E_B{}^I = F_{AB}{}^C E_C{}^I\,,
\end{equation}
where $\mathcal{L}$ denotes the generalised Lie derivative
\begin{equation}
  \mathcal{L}_{E_A} E_B{}^I = E_A{}^J \partial_J E_B{}^I + \big( \partial^I E_{AJ} - \partial_J E_A{}^I \big) E_B{}^J
  \,.
\end{equation}
As its name suggests, it serves the same purpose as the Lie derivative in conventional geometry. But due to the structure of the generalised tangent space, it not only captures diffeomorphisms on $M$ but also $B$-field transformations. There is a slight subtlety concerning the partial derivatives $\partial_I$ in this expression. In DFT, they in general incorporate not only the $D$ coordinates of the target space but also $D$ additional coordinates on an auxiliary space. But in this setup, the generalised Lie derivative does not close into an algebra automatically. It only does if additional constraints are satisfied. The most restrictive one is the section condition, or strong constraint. It requires that arbitrary combinations of fields, denoted by $\cdot$, are annihilated by $\partial_I \cdot \partial^I \cdot = 0$. A trivial solution to this constraint is given by $\partial_I = \begin{pmatrix} 0 & \partial_i \end{pmatrix}$. It renders DFT equivalent to generalised geometry and we will use it for the rest of the paper. It is interesting to note that in the framework of generalised geometry, PL symmetric backgrounds mimic the structure of group manifolds in conventional geometry. More precisely, $E_A{}^I$ corresponds to $D$ vector fields that are dual to the left-invariant Maurer-Cartan form while the generalised Lie derivative is replaced by the standard Lie derivative.

Each generalised frame field, even when it is not a solution of \eqref{eqn:framealgebra}, can be brought into the form
\begin{equation}\label{eqn:decompframe}
  E_A{}^I = \frac{1}{\sqrt{2}} \begin{pmatrix}
    \delta_a{}^b & 0 \\
    0 & \Lambda_{\bar a}{}^{\bar b}
  \end{pmatrix}\begin{pmatrix}
    e_{bi} + e_b{}^j B_{ji} &  e_b{}^i \\
    - e_{\bar bi} + e_{\bar b}{}^j B_{ji} & e_{\bar b}{}^i
  \end{pmatrix} := \Lambda_A{}^B \Eh_B{}^I \,,
\end{equation}
where $B_{ij}$ is the $B$-field on the target space and $e_a{}^i = e_{\bar a}{}^i$ denotes a conventional frame field. The latter encodes the metric $g^{ij} = e_a{}^i \eta^{ab} e_b{}^j$ and a Lorentz frame. Note that we use the standard convention that lowercase, curved indices, like $i$, are lowered and raised by this metric and its inverse. Additionally, the generalised frame field incorporates a double Lorentz transformation $\Lambda_{\bar a}{}^{\bar b}$ with the defining property $\Lambda_{\bar a}{}^{\bar c} \Lambda_{\bar b}{}^{\bar d} \eta_{\bar c\bar d} = \eta_{\bar a\bar b}$. At a first glance, it seems irrelevant because it does not affect the target space geometry encoded by the metric and the $B$-field. However, except for a few special cases, it is crucial to solving the constraint \eqref{eqn:framealgebra} for PL symmetry. Moreover, we will see later that it plays a central role beyond one-loop. If the doubled Lie group $\DD$ associated to $\mfd$ has a maximally isotropic subgroup $H$, it is always possible to explicitly construct $E_A{}^I$ on the coset $H \backslash \DD$ \cite{\reflongDFTPLlist}. This construction has become standard and we will not repeat it here. Frequently, the explicit target space geometry is convoluted and while it can always be constructed, it is more elegant to extract as much information as possible directly from the doubled formalism. We will do exactly this for the one- and two-loop $\beta$-functions in the next subsections. A considerable advantage of this approach is that PL T-duality only affects the generalised frame field but not the structure coefficients $F_{ABC}$ and the pairing $\eta_{AB}$. Hence all quantities which can be exclusively written in terms of the latter are manifestly invariant under PL T-duality. Different dual target space geometries arise if $\DD$ has different maximally isotropic subgroups $H_i$. For each of them a different frame field on a different target space $M_i = H_i \backslash \DD$ can be constructed. 

The dilaton $\phi$ is encoded in the generalised dilaton
\begin{equation}\label{eqn:defgend}
  d = \phi - \frac14 \log \det g\,.
\end{equation}
Its condition for PL symmetry can be written in full analogy with the generalised frame field as
\begin{equation}\label{eqn:defFA}
  \mathcal{L}_{E_A} e^{-2 d} = - F_A e^{-2 d}\,, \qquad F_A = \text{const.}\,,
\end{equation}
where $e^{-2 d}$ transforms as a weight $+1$ density under the generalised Lie derivative, namely
\begin{equation}
  \mathcal{L}_{E_A} e^{-2 d} = E_A{}^I \partial_I e^{-2d} + e^{-2d} \partial_I E_A{}^I\,.
\end{equation}
$F_A$ is in one-to-one correspondence with the Lie algebra element $F^A T_A = F \in \mfd$. This element has to be in the center of $\mfd$, meaning that it is constrained by $[ T_A , F ] = 0$ for all generators $T_A$. Moreover, it has to be isotropic and therefore satisfy $\langle F, F \rangle = 0$. We find these two conditions directly from the closure of the generalised Lie derivatives \cite{Geissbuhler:2013uka}. 

\subsubsection{\texorpdfstring{$\lambda$}{Lambda}- and \texorpdfstring{$\eta$}{Eta}-deformation}\label{sec:genframelambda}
The $\lambda$-deformation on a semisimple group manifold $G$ \cite{Sfetsos:2013wia} is a good example to demonstrate this structure explicitly. It is governed by the doubled group $\DD = G \times G$ with the maximally isotropic subgroup $H=G_{\mathrm{diag}}$ \cite{Klimcik:2015gba} that is used to construct the generalised frame field $E_A{}^I$. The frame field $e_a{}^i$ in \eqref{eqn:decompframe} is written in terms of the inverse transpose of the left- and right invariant Maurer-Cartan forms
\begin{equation}\label{eqn:leftrightinv}
  t_a l^a{}_i d x^i = \sqrt{\frac{k}{2}} g^{-1} \dd g\,, \qquad
  t_a r^a{}_i d x^i = \sqrt{\frac{k}{2}} \dd g g^{-1}\,, \qquad
  [t_a, t_b] = f_{ab}{}^c t_c\,,
\end{equation}
($l_a{}^i l^b{}_i = \delta_a^b$, $r_a{}^i r^b{}_i = \delta_a^b$) and reads
\begin{equation}\label{eqn:eailambda}
  e_a{}^i = \frac{\kappa + 1}{2\sqrt{\kappa}} l_a{}^i + \frac{\kappa - 1}{2\sqrt{\kappa}} r_a{}^i\,,
\end{equation}
where $k$ and $\kappa$ are free parameters. To construct the $B$-field, a locally defined two-form, $B_0$, whose exterior derivative results in the three-form
\begin{equation}
  H_0 = -\frac1{3 \sqrt{2 k}} f_{abc} l^a \wedge l^b \wedge l^c = \dd B_0
\end{equation}
is required. It gives rise to
\begin{equation}
  B = B_0 - \frac{\kappa + 1}{2\sqrt{\kappa}} l^a{}_i e_{aj} d x^i \wedge d x^j
\end{equation}
and completes, together with
\begin{equation}\label{eqn:lambdalambdadef}
  \Lambda_{\bar a}{}^{\bar b} = \frac{(\kappa + 1) \delta_{\bar a}{}^{\bar b} - 2 \sqrt{\kappa} e_{\bar a}{}^i l^{\bar b}{}_i}{\kappa - 1}\,,
\end{equation}
the constituents of the generalised frame field \eqref{eqn:decompframe}. Apparently, these expressions look rather complicated and they turn out to become even more involved once a parameterisation for the group element $g$ is fixed. This is because the standard target fields obscure the underlying structure of the $\lambda$-deformation. The structure coefficients of $\mfd$ encode the same information but in a much more streamlined form. They arise from \eqref{eqn:framealgebra} and read
\begin{equation}\label{eqn:genfluxlambda}
  F_{abc} = \frac{\kappa^2 + 3}{4 \sqrt{\kappa k}} f_{abc} \,, \quad
  F_{ab\bar c} = \frac{\kappa^2 - 1}{4 \sqrt{\kappa k}} f_{ab\bar c} \,, \quad
  F_{a\bar b\bar c} = \frac{\kappa^2 - 1}{4 \sqrt{\kappa k}} f_{a\bar b\bar c} \,, \quad
  F_{\bar a\bar b\bar c} = \frac{\kappa^2 + 3}{4 \sqrt{\kappa k}} f_{\bar a\bar b\bar c}\,.
\end{equation}
Note that the remaining components are fixed by the total antisymmetry of $F_{ABC}$. Furthermore, in this form the symmetry $\kappa \rightarrow - \kappa$ and $k \rightarrow -k$ \cite{Itsios:2014lca} of the $\lambda$-deformation is immediately manifest. The semisimple doubled Lie algebra $\mfd=\mathfrak{g}\times\mathfrak{g}$ has no center and thus $F_A=0$. Starting from \eqref{eqn:defgend} and \eqref{eqn:defFA}, one can use this fact to extract the derivative of the dilaton
\begin{equation}
  \partial_i \phi = \frac12 \omega^a{}_{ai}\,,
\end{equation}
where $\omega_{ia}{}^b$ denotes the spin connection corresponding to the frame field \eqref{eqn:eailambda}.

PL T-duality relates the $\lambda$-deformation to the $\eta$-deformation up to an analytic continuation \cite{\reflambdaetaPL}. A generalised frame field for the latter can be easily constructed \cite{Demulder:2018lmj}. The detailed expressions for the metric and $B$-field are not relevant for our discussion. All information we rely on is contained in the structure coefficients $F_{ABC}$ and thus it is not surprising that the $\lambda$- and $\eta$-deformation are both captured by \eqref{eqn:genfluxlambda} after the identification
\begin{equation}\label{eqn:paramlambdaeta}
  \begin{aligned}
    \text{$\lambda$-deformation:} & &
      \kappa&=\frac{1 - \lambda}{1 + \lambda} &\qquad
      k&=k \\
    \text{$\eta$-deformation:} & &
      \kappa&= - \ci \eta &
      k&= \frac{\ci}{4 \eta t}\,.
  \end{aligned}
\end{equation}
Both form two different branches on the space of structure coefficients $F_{ABC}$, representing $\DD=G\times G$ and $\DD=G^{\mathbb{C}}$, respectively. There is a one-dimensional subspace where both meet. It is defined by the limit $\kappa \to 0$ and $k \to \infty$. In this case, we have $\lambda=1$ and $\eta=0$, whereas $t$ remains a free parameter with $\kappa = h/(4k)$. The corresponding model is the principal chiral model (PCM) on the group manifold $G$, and $\DD$ is contracted to $T^* G$.

\subsection{One-loop}\label{sec:oneloop}
A $\sigma$-model has an infinite number of couplings that are encoded in the metric $g_{ij}$, the $B$-field $B_{ij}$ and the dilaton $\phi$. As some of them are redundant, we first note that infinitesimal diffeomorphisms and gauge transformations,
\begin{equation}
  \dg_{ij} = 2 \nabla_{(i} \xi_{j)}\,, \qquad
  \dB_{ij} = H_{ijk} \xi^k + 2 \nabla_{[i} \chi_{j]}\,, \qquad \text{and} \qquad
  \dphi = \xi^i \nabla_i \phi\,,
\end{equation}
that are generated by the vector $\xi^i$ and the one-form $\chi_i$, do not affect any local observables. Thus, it is useful to define equivalence classes of $\beta$-functions which only differ by those transformations. Each class has a canonical representative for which the $\beta$-functions do not generate any diffeomorphisms or gauge transformations. We denote it with a bar and define an arbitrary member of its equivalence class by
\begin{equation}\label{eqn:defbh}
  \bh^E_{ij} = \bb^E_{ij} + 2 \nabla_{(i} \xi_{j)} + H_{ijk} \xi^k + 2 \nabla_{[i} \chi_{j]} \,,
    \qquad
  \bh^\phi = \bb^\phi + \xi^i \nabla_i \phi\,,
\end{equation}
where $E_{ij} = g_{ij} + B_{ij}$ unifies the metric and $B$-field into a single object. Furthermore we use the standard convention where the RG flow is governed by
\begin{equation}\label{eqn:RGflownormal}
  \mu \frac{\dd E_{ij}}{\dd\log\mu} = \beta^E_{ij} \qquad \text{and} \qquad
  \mu \frac{\dd \phi}{\dd\log\mu} = \beta^\phi\,.
\end{equation}

At one-loop $\bb^E_{ij}$ reads \cite{Friedan:1980jf,*Curtright:1984dz}
\begin{equation}\label{eqn:bb1E}
  \bb^{(1)E}_{ij} = R_{ij} - \frac14 H^2_{ij} - \frac12 \nabla_k H^k{}_{ij}
    \qquad \text{with} \qquad 
    H^2_{ij} = H_{imn} H_j{}^{mn}\,.
\end{equation}
It is the first non-vanishing term in the expansion
\begin{equation}
  \bb^{E}_{ij} = \alpha' \bb^{(1)E}_{ij}  + \alpha'^2 \bb^{(2)E}_{ij}  + \dots\,.
\end{equation}
We adopt the same notation for all other quantities that admit an $\alpha$-expansion, too. Saying that a quantity$^{(n)}$ comes with a factor of $\alpha'^n$. Because derivatives contribute a factor of $\sqrt{\alpha}$, we can alternatively conclude that quantities at the level $n$ normally contain $2n$ derivatives. The notable exceptions are the vector $\xi^i$ and the one form $\chi_i$ in \eqref{eqn:defbh}. They contain $2n - 1$ derivatives. Computing \eqref{eqn:bb1E} directly is cumbersome and one might ask if there is an easier way to obtain the RG flow. At this point working with doubled quantities, as they naturally appear in DFT, is very convenient. As already demonstrated in the last section, they are particularly powerful to describe PL symmetric target space whose flows we ultimately want to address. The doubled version of the first equation in \eqref{eqn:RGflownormal} becomes
\begin{equation}\label{eqn:flowdoubled}
  \mu \frac{\dd \Eh_A{}^I}{\dd \log\mu} \Eh_{BI} = \Bh^{(1)E}_{AB}
\end{equation}
in the framework of DFT. In this equation we prefer the partially double Lorentz fixed generalised frame field $\Eh_A{}^I$ over $E_A{}^I$ because its remaining, unfixed symmetries coincide with the diffeomorphisms, $B$-field and Lorentz transformations that are manifest symmetries of \eqref{eqn:bb1E}. The doubled $\beta$-function on the right hand side is based on $\bh^E_{ij}$ that arises from \eqref{eqn:defbh} with 
\begin{equation}\label{eqn:diffgauge1}
  \xi^{(1)i} = \nabla^i \phi\,, \qquad \chi^{(1)}_i = 0\,.
\end{equation}
More specifically, its off-diagonal contributions
\begin{equation}\label{eqn:betaEdoubled}
  \Bh^E_{AB} = \frac12 \begin{pmatrix} 
    0 & \bh^E_{a\bar b} \\
    - \bh^E_{b\bar a} & 0
  \end{pmatrix}
\end{equation}
are formed by $\bh^E_{ij}$ in flat indices. This embedding is motivated by the observation that all physical information is contained in the off-diagonal blocks while the diagonal blocks only generate double Lorentz transformations. Therefore, we set them to zero. In order to extract the physically relevant blocks, the projectors
\begin{equation}
  P_A{}^B = \begin{pmatrix} \delta_a{}^b & 0 \\
    0 & 0 \end{pmatrix}
  \qquad \text{and} \qquad
  \Pb_A{}^B = \begin{pmatrix} 0 & 0 \\
    0 & \delta_{\bar a}{}^{\bar b} \end{pmatrix}
\end{equation}
are required. Note the factor of $1/2$ in the definition \eqref{eqn:betaEdoubled}. It appears because $\bh^E_{ij}$ governs the flow of the metric and $B$-field directly, whereas $\Bh^E_{AB}$ captures the flow of a (generalised) frame field. The former is the square of the latter and of course the derivative of a square always introduces a factor of $2$. Due to this factor, we have to carefully distinguish between $\bh^E_{a\bar b}$ and $\Bh^E_{a\bar b}$.

Our primary objective is to find an expression for $\Bh^{(1)E}_{AB}$ that reproduces \eqref{eqn:bb1E} and can be written exclusively in terms of the doubled quantities we encountered so far, namely $\Fh_{ABC}$, $\Fh_A$, $\Dh_A = \Eh_A{}^I \partial_I$, $P^{AB}$, $\Pb^{AB}$. Hats over the $F$'s indicate that they are still computed by \eqref{eqn:framealgebra} and \eqref{eqn:defFA} but for $\Eh_A{}^I$ instead of $E_A{}^I$. Therefore, neither $\Fh_{ABC}$ nor $\Fh_A$ is constant. Using the parameterisation given in \eqref{eqn:decompframe}, we obtain the generalised fluxes
\begin{equation}
  \begin{aligned}
    \Fh_{abc} &= \frac1{2\sqrt{2}} ( H_{abc} - 6 \omega_{[abc]} ) & \qquad
    \Fh_{\bar a b c} &= \frac1{2\sqrt{2}} ( H_{\bar abc} - 2 \omega_{\bar a bc} ) \\
    \Fh_{a\bar b\bar c} &= \frac1{2\sqrt{2}} ( H_{a\bar b\bar c} + 2 \omega_{a\bar b\bar c} ) &
    \Fh_{\bar a\bar b\bar c} &= \frac1{2\sqrt{2}} ( H_{\bar a\bar b\bar c} + 6 \omega_{\bar a\bar b\bar c} )
  \end{aligned}
\end{equation}
and
\begin{equation}\label{eqn:Fha}
  \Fh_a = \Fh_{\bar a} = \sqrt{2} \Dh_a \phi - \frac1{\sqrt{2}} \omega^b{}_{ba}
    \qquad \text{with} \qquad
    \Dh_a = e_a{}^i \partial_i
\end{equation}
written in terms of the spin connection $\omega_{abc}$ and the $H$-flux. Eventually, one is able to come up with the one-loop, doubled $\beta$-function 
\begin{equation}\label{eqn:beta(1)Edoubled}
  \Bh^{(1)E}_{AB} = 2 P_{[A}{}^C \Pb_{B]}{}^D \left( \Fh_{CEG} \Fh_{DFH} P^{EF} \Pb^{GH} + \Fh_{CDE} \Fh_F P^{EF} + \Dh_D \Fh_C - \Dh_E \Fh_{CDF} P^{EF} \right)\,,
\end{equation}
which agrees with the starting point \eqref{eqn:bb1E}. A detailed derivation of this equation is given in section~\ref{sec:oneloopdetail}.

We will encounter more equations like this one. To see their structure more clearly, one might represent them in diagrammatic form. To this end, we identify the projectors $P_{AB}$ and $\Pb_{AB}$ with two different propagators
\begin{align}
  P_{AB} &= \tikz[baseline=(A.base)]{\draw (0,0) node[anchor=east] (A) {$A$} -- (1,0) node[anchor=west] {$B$};} &      
    \text{and} & &
    \Pb_{AB} &= \tikz[baseline=(A.base)]{\draw[dashed] (0,0) node[anchor=east] (A) {$A$} -- (1,0) node[anchor=west] {$B$};}   
  \,,
\intertext{while the fluxes become the vertices}
  \Fh_{ABC} &= \tikz[baseline=(B.base)] {
    \draw (0:0) node[A] {} -- (0:0.4) node[anchor=west] (B) {$B$};
    \draw (0:0) -- (120:0.4) node[anchor=east] {$A$};
    \draw (0:0) -- (240:0.4) node[anchor=east] {$C$};
  } & \text{and} & &
  \Fh_A &= \tikz[baseline=(A.base)] {
    \draw (0:0) node[FA] {} -- (0:0.4) node[anchor=west] (A) {$A$};
  }\,.
\end{align}
Finally, we denote a derivative with an arrow, for example
\begin{equation}
  \Dh_A \Fh_B = \tikz[baseline=(A.base)] { 
    \draw[der] (0:0) node[anchor=east] (A) {$A$} -- (0:0.4) node[FA,anchor=west] {};
    \draw (0:0.4) -- (0:0.8) node [anchor=west] {$B$};}\,.
\end{equation}
Dummy indices are suppressed in these diagrams and, if unambiguous, also external indices can be dropped. Making use of these conventions, \eqref{eqn:beta(1)Edoubled} can be written as
\begin{equation}\label{eqn:diagbeta1}
  \bh^{(1)E}_{a\bar b} = - 2 \dFsquare01 - 2\dFFA0 + 2\dDAFA + 2\dFADA0\,.
\end{equation}
For the $\beta$-function of the generalised dilaton, the same argument applies and one can check that (again all the details are given in section~\ref{sec:oneloopdetail})
\begin{equation}\label{eqn:diagbeta1d}
  \begin{aligned}
    \bh^{(1)d} &= \mu\frac{\dd d}{\dd\log\mu} = \bh^{(1)\phi} - \frac14 g^{ij} \bh^{(1)E}_{ij} = 
      -\frac14 R + \frac1{48} H^2 + (\nabla \phi)^2 - \nabla^2 \phi \\
      &= \frac{1}{12} \Fsquare000 + \frac14\Fsquare010 + \frac12\FAFA{0} - \FADA{0}
  \end{aligned}
\end{equation}
holds.

\subsubsection{Double Lorentz transformation}
Instead of $\Fh_{ABC}$ and $\Fh_A$, we would rather use $F_{ABC}$ and $F_A$ as they are the natural objects for a PL symmetric $\sigma$-model. They are connected to each other by the double Lorentz rotation $\Lambda_A{}^B$ defined in \eqref{eqn:decompframe}. Although the generalised fluxes and their derivatives transform anomalous under this rotation, the particular combination in which they enter the $\beta$-function cancels all anomalous contributions. This is a standard result in the flux formulation of DFT \cite{Hohm:2010xe,Geissbuhler:2013uka}, but since double Lorentz rotations become much more subtle beyond one-loop, we want to review how it arises: The finite transformation $\Lambda_A{}^B$ is a composition of infinitesimal transformations, namely $\Lambda_A{}^B = \exp ( \lambda_A{}^B )$ with $\lambda_{AB} = - \lambda_{BA}$. Covariant quantities, like $\B^{(1)E}_{AB}$, transform as
\begin{equation}\label{eqn:deltalambdabeta1}
  \delta_\lambda^{(0)} \Bh^{(1)E}_{AB} = \lambda_A{}^C \Bh^{(1)E}_{CB} + \lambda_B{}^C \Bh^{(1)E}_{AC}
\end{equation}
under the infinitesimal action $\delta_\lambda^{(0)}$. Note that this relation only holds to leading order in $\alpha'$, indicated by the superscript ${}^{(0)}$ on the action. As already mentioned, there are also  non-covariant quantities, like the generalised fluxed $\Fh_{ABC}$. To treat them in a methodical way, we introduce the ``anomalous'' contribution to  the transformation
\begin{equation}
  \Al = \delta_\lambda - \lambda \cdot\,. 
\end{equation}
With $\lambda \cdot$, we denote the standard action of $\lambda$ on every flat index. For example, the generalised fluxes have the leading order anomalous transformation
\begin{equation}\label{eqn:anFABC}
  \Al^{(0)} \Fh_{ABC} = \delta_\lambda^{(0)} \Fh_{ABC} - 3 \lambda_{[A}{}^D \Fh_{BC]D} = 3 \Dh_{[A} \lambda_{BC]}\,.
\end{equation}
Let us see in more detail how the right hand side of this equation arises. Because $\Al$ is a linear operator ($\Al (a + b) = \Al a + \Al b$) that acts as a derivative ($\Al ( a b ) = \Al a b + a \Al b$), all we need to evaluate \eqref{eqn:anFABC} from the definition $\Fh_{ABC} = 3 \Dh_{[A} \Eh_B{}^I \Eh_{C]I}$ is $\Al^{(0)} \Eh_A{}^I = 0$ and the commutator of $\Al^{(0)}$ and $\Dh_A$. The later is given by
\begin{equation}
  [ \Al^{(0)}, \Dh_A ] \Eh_B{}^I = \Dh_A \lambda_B{}^C \Eh_C{}^I\,.
\end{equation}
In the same vein one obtains $\Al^{(0)} \Fh_A = \Dh_B \lambda^B{}_A$ (after taking into account $\Al d=0$) and $\Al P^{AB} = - \Al \Pb^{AB} = 0$. Eventually, we can directly evaluate $\Al^{(0)} \Bh^{(1)E}_{AB}$ from \eqref{eqn:diagbeta1} and find that it vanishes. Hence, we come full circle and arrive again at \eqref{eqn:deltalambdabeta1}. 

A finite double Lorentz transformation arises from the exponential map
\begin{equation}\label{eqn:expdeltalambda}
  e^{\delta_\lambda} = e^{\Al + \lambda\cdot} = e^\lambda e^{\Al} = \Lambda \cdot e^{\Al} \,.
\end{equation}
$\Lambda \cdot$ denotes the group action of $\Lambda_A{}^B$ on every free index. Applying this relation to $\Bh^{(1)E}$, we eventually obtain 
\begin{equation}\label{eqn:BtoBh}
  \B^{(1)E}_{AB} = \Lambda_A{}^C \Lambda_B{}^D \Bh^{(1)E}_{CD}
\end{equation}
and prove that it is valid to drop all the hats in \eqref{eqn:beta(1)Edoubled} and use the rotated $\beta$-function $\beta^{(1)E}$ instead of $\Bh^{(1)E}$. It is important to stress that both only are written in different double Lorentz frames, but still describe exactly the same physics. However, the latter is much better adapted to PL symmetric target space geometries because all quantities are just constant. Hence all terms that contain derivatives $D_A$ drop out. Double Lorentz transformations do not affect the $\beta$-function of the generalised dilaton in \eqref{eqn:diagbeta1d} and we thus identify
\begin{equation}
  \bh^{(1)d} = \beta^{(1)d}\,.
\end{equation}

\subsubsection{Renormalisable \texorpdfstring{$\sigma$}{sigma}-models}\label{sec:renormalisable}
All information about the $\sigma$-model of the bosonic string \eqref{eqn:sigmamodel} is condensed in $F_{ABC}$ and $F_A$. We might take these two objects as being parameterised by $N$ coupling constants $c^\mu$ where $\mu=1, \dots, N$. The $\beta$-functions for these couplings arise directly from $\B^E_{AB}$ and $\beta^d$ through the relations
\begin{equation}\label{eqn:betamu}
  \begin{aligned}
    \beta^\mu \partial_\mu F_{ABC} &= 3 D_{[A} \B^E_{BC]} + 3 \B^E_{[A}{}^D F_{BC]D} \\
    \beta^\mu \partial_\mu F_A &= D^B \B^E_{BA} + \B^E_A{}^B F_B + 2 D_A \beta^d\,.
  \end{aligned}
\end{equation}
For general target space geometries, neither $F_{ABC}$ nor $F_A$ is constant. They rather have different values on every point of the target space manifold $M$. Hence, one needs infinitely many coupling constants $c^\mu$ to accommodate this information. In contrast, PL $\sigma$-models have by definition constant $F_{ABC}$'s and $F_A$'s. Therefore, PL symmetry just permits a finite number of couplings. If this property is preserved under RG flow, it renders the PL $\sigma$-model renormalisable. From \eqref{eqn:betamu} it follows that this is the case if
\begin{equation}\label{eqn:renormalisable}
  D_A \B^E_{BC} = 0 \qquad \text{and} \qquad D_A \beta^d = 0
\end{equation}
holds, which is clearly the case for the one-loop $\beta$-functions presented in \eqref{eqn:diagbeta1} and \eqref{eqn:diagbeta1d}. Hence, we conclude that PL $\sigma$-models are one-loop renormalisable. This observation is by now well established \cite{\refonelooprenorm}. However, all previous works we are aware of only incorporate the metric and the $B$-field but not the dilaton.

Another advantage of encoding all $\sigma$-model couplings in terms of $F_{ABC}$ and $F_A$ is that their transformation under infinitesimal generalised diffeomorphisms, which unify diffeomorphisms and $B$-field transformations, is very simple, namely
\begin{equation}
  \delta F_{ABC} = \Xi^I \partial_I F_{ABC} \qquad \text{and} \qquad
  \delta F_A = \Xi^I \partial_I F_A\,.
\end{equation}
Here $\Xi^I = \begin{pmatrix} \chi_i & \xi^i \end{pmatrix}$ contains the parameters introduced in \eqref{eqn:defbh}. PL symmetric backgrounds are invariant under such transformations because $F_A$ and $F_{ABC}$ are constant.

Remarkably, PL $\sigma$-models are just a particular example of a more general scheme: At one-loop, all target space geometries which admit a consistent truncation result in renormalisable $\sigma$-models. Both notions are related because the one-loop $\beta$-functions \eqref{eqn:diagbeta1} and \eqref{eqn:diagbeta1d} are equivalent to the field equations of the bosonic string's two-derivative target space effective action. One might understand field equations of a classical field theory as describing an infinite number of coupled degrees of freedom. Consistent truncations are based on the observation that it is sometimes possible to decouple a finite number of them from the rest, which then can be safely truncated. This technology is extremely useful to simplify the hard task of finding solutions to the field equations. Here, we see that it also has a natural interpretation in terms of two-dimensional, renormalisable field theories.

\subsubsection{\texorpdfstring{$\lambda$}{Lambda}- and \texorpdfstring{$\eta$}{Eta}-deformation}\label{sec:lambda1}
We have now all we need to compute the one-loop $\beta$-function of the coupling $\kappa$ and $k$ in the $\lambda$- and $\eta$-deformation. Only the first diagram in \eqref{eqn:diagbeta1} contributes to
\begin{equation}
  \beta^{(1)E}_{a\bar b} = - 2 \dFsquare01 = 2 F_{\bar c d a} F^{d \bar c}{}_{\bar b} = - \frac{(\kappa^2 - 1)^2}{8 k \kappa} c_G \eta_{a\bar b}\,.
\end{equation}
Here we use the normalisation $f_{ac}{}^d f_{bd}{}^c = - c_G \eta_{ab}$ for the structure coefficients of $G$'s Lie algebra with the dual Coxeter number $c_G$. From \eqref{eqn:betamu}, we extract
\begin{equation}\label{eqn:beta1kappa}
  \beta^k = 0 \qquad \text{and} \qquad \beta^\kappa = \frac{c_G}{8 k} (\kappa^2 - 1)^2\,.
\end{equation}
$\kappa$ is related to $\lambda$ and $\eta$ by \eqref{eqn:paramlambdaeta}, which eventually gives rise to
\begin{equation}
  \beta^\lambda = - \frac{\lambda^2 c_G}{k(\lambda + 1)^2}
    \qquad \text{and} \qquad
  \beta^\eta = \frac{\eta t c_G}2 ( 1 + \eta^2 )^2\,.
\end{equation}
These results match with the ones provided in the literature \cite{Itsios:2014lca,Sfetsos:2015nya}. We also compute the $\beta$-function for the generalised dilaton
\begin{equation}\label{eqn:lambdadefbeta1d}
  \beta^{(1)d} = \frac{1}{12} \Fsquare000 + \frac14\Fsquare010 = \frac{\kappa^4 - 6 \kappa^2 - 3}{96 k \kappa} c_G \dim G \,.
\end{equation}
At $\lambda=0$ the RG-flow has a fixed point, the WZW-model on the group $G$.

\subsection{Two loops}\label{sec:twoloops}
Beyond one-loop the $\beta$-functions become scheme dependent. Therefore, we first have to fix a particular scheme in which we present our results. As we will see, making a good choice is crucial because only in a distinguished scheme PL symmetry becomes manifest and the computations manageable. Different schemes arise from an ambiguity in choosing counter terms during the renormalisation of the $\sigma$-model. An alternative perspective is that different schemes are related by field redefinitions, which are diffeomorphisms on the space of couplings. Naively, choosing a scheme is the same as committing to a particular set of coordinates in general relativity. Obviously when dealing with a problem with rotational symmetry, it is a good idea to choose spherical coordinates instead of Cartesian coordinates. We know that the final, physical observables do not depend on this choice. But it is much easier to extract them in adapted coordinates.

\subsubsection{Scheme transformation}
There is one aspect of scheme transformations for $\sigma$-models which makes them slightly more complicated than the standard diffeomorphisms that we are used to from general relativity. Because a $\sigma$-model has an infinite number of coupling constants, one has to deal with diffeomorphisms on an infinite dimensional manifold. The tangent space of this manifold is spanned by the vectors $\delta_\Psi$ with $\Psi = \begin{pmatrix} \delta E_{AB} & \delta d \end{pmatrix}$. In working with them, it is very helpful to remember what happens after a projection onto a finite dimensional submanifold (this is exactly what PL symmetry will allow us to do later). In this case, $\Psi$ reduces to a column vector $\Psi^\mu$ and $\delta_\Psi$ becomes $\Psi^\mu \partial_\mu$. The derivative $\delta_\Psi$ is defined by its action on
\begin{equation}\label{eqn:variations}
  \begin{aligned}
   \delta_\Psi F_{ABC} & = 3 D_{[A} \delta E_{BC]} + 3 \delta E_{[A}{}^D F_{BC]D} & \qquad \delta_\Psi P_{AB} &= 0\\
   \delta_\Psi F_A &= D^B \delta E_{BA} + \delta E_A{}^B F_B + 2 D_A \delta d & \delta_\Psi \Pb_{AB} &= 0 \\
   \delta_\Psi D_A &= \delta E_A{}^B D_B + D_A \delta_\Psi\,.
  \end{aligned}
\end{equation}
Note that these relations allow us to rewrite \eqref{eqn:betamu} in the cleaner form
\begin{equation}
  \beta^\mu \partial_\mu = \delta_{\begin{pmatrix} \B^E & \beta^d \end{pmatrix}}
\end{equation}
and we see that above we actually restricted the infinite dimensional coupling space to the finite dimensional space of couplings which are compatible with PL symmetry.

An infinitesimal scheme transformation of the $\beta$-functions $\B = \begin{pmatrix} \B^E & \beta^d \end{pmatrix}$ with the parameter $\Psi$ is mediated by the Lie derivative
\begin{equation}
  L_\Psi \B = \delta_\Psi \B - \delta_\B \Psi - T(\Psi, \B) \,.
\end{equation}
The last term takes into account that the derivative $\delta_\Psi$ in general has torsion, which is defined by
\begin{equation}
  \delta_\Psi \delta_{\Psi'} - \delta_{\Psi'} \delta_{\Psi} = \delta_{T(\Psi,\Psi')}\,.
\end{equation}
From the definition \eqref{eqn:variations}, one indeed obtains the non-vanishing torsion
\begin{equation}
  T(\Psi, \Psi') = \begin{pmatrix} 2 \delta E^C{}_{[A} \delta{E}'_{B]C} & 0 \end{pmatrix}\,.
\end{equation}
Since $\delta E_A{}^B$ generates an O($D$,$D$) transformation, the non-trivial part of the torsion tensor may be written as $[\delta E, \delta E']$. This rewriting shows that the torsion we encounter originates from the O($D$,$D$) structure of the generalised tangent space. At a first glance, our choice of derivative might seem peculiar because it clearly differs from the canonical, torsion-free variation with respect to $g_{ij}$, $B_{ij}$, and $\phi$. In the end, one can check that both give rise to the same results. However, using this $\delta_\Psi$ simplifies the computations considerably and therefore we prefer it.

Infinitesimal scheme transformations are sufficient for our purpose because we are just concerned with contributions to the $\beta$-functions up to the order $\alpha'^2$, and for all $\Psi$ which we consider, $\Psi^{(0)}$ always vanishes. Consequentially $\B^{(1)}$ is not affected and $\B^{(2)}$ is corrected by
\begin{equation}\label{eqn:schemetrbeta}
  \B^{(2)} \rightarrow \B^{(2)} + L_{\Psi^{(1)}} \B^{(1)}\,.
\end{equation}
In principal, one could apply more general transformations with a non-trivial $\Psi^{(0)}$. But they would spoil the manifest symmetries of the one-loop results obtained in the last subsection. Hence, we are restricted to transformations that start with $\Psi^{(1)}$ and \eqref{eqn:schemetrbeta} applies.

\subsubsection{\texorpdfstring{$\beta$}{Beta}-functions}
Like in the last subsection, we again start with the known result for the two-loop $\beta$-functions of the metric, $B$-field, and dilaton in the Metsaev-Tseytlin (MT) scheme \cite{Metsaev:1987zx}. The reason why we preferred this scheme over other popular options, like the Hull-Townsend (HT) scheme, is purely technical and will be explained in section~\ref{sec:twoloopsdetails}. Compared to the discussion at one-loop, the most striking difference is that the two-loop $\beta$-functions, which are given in \eqref{eqn:betag2}-\eqref{eqn:betaphi2}, are considerably more complicated. However, we can still relate them a member in their equivalence class, which is suited to be written exclusively in terms of $\Fh_{ABC}$, $\Fh_A$, $\Dh_A$, $P^{AB}$ and $\Pb^{AB}$, by the infinitesimal diffeomorphism and gauge transformation
\begin{equation}\label{eqn:requireddiff+gauge}
  \xi^{(2)i} = - \frac1{48} \nabla^i H^2 + \frac14 H^{ijk} \bh^{(1)E}_{jk} \,, \qquad 
  \chi^{(2)}_i = \omega_i{}^{ab} \bh^{(1)E}_{ab}\,.
\end{equation}
Still, this is not sufficient and we furthermore have to change the scheme by \eqref{eqn:schemetrbeta} with $\Psi^{(1)} = \begin{pmatrix} \Delta^{(1)} E_{ij} & 0 \end{pmatrix}$ with $\Delta E_{ij} = \Delta g_{ij} + \Delta B_{ij}$ and
\begin{equation}\label{eqn:schemetr}
  \Delta^{(1)} g_{ij} = - \frac12 \omega_{ia}{}^b \omega_{jb}{}^a + \frac38 H^2_{ij}\,, \qquad 
  \Delta^{(1)} B_{ij} = - \frac12 H_{[i|a}{}^b \omega_{|j]b}{}^a \,.
\end{equation}

After a cumbersome computation, that we detail in the next section, one finds that $\bh^{(2)E}$ has in total 342 terms(=diagrams). They are invariant under the $\mathbb{Z}_2$ action $Z$ that swaps the projectors $P$ and $\Pb$. To illustrate how $Z$ acts on the level of diagrams, take for example
\begin{equation}\label{eqn:Z2}
  Z( \HCNa00001 ) = - \HCNa01111\,.
\end{equation}
Here, we first swap solid and dashed lines ($P \leftrightarrow \Pb$) and then bring the external solid line to the left and the dashed one to the right. This swapping of the external lines corresponds to $a \rightarrow \bar b$ and $\bar b \rightarrow a$, or equivalently $A\leftrightarrow B$, of the antisymmetric $\Bh^{(2)E}_{AB}$ and therefore introduces a minus sign. Because the two-loop $\beta$-functions of the bosonic string satisfies
\begin{equation}
  Z(\bh^{(2)E}_{a\bar b}) = \bh^{(2)E}_{a\bar b}\,,
\end{equation}
we actually only have to cope with 172 different diagrams for $\bh^{(2)E}$ while the others are fixed by the $\mathbb{Z}_2$ symmetry. It is not very illuminated to present this bulky result here. Fortunately, for PL symmetric target spaces it can be simplified considerably. But to benefit from the structure introduced in section~\ref{sec:genframes}, we again have to switch to unhatted quantities by applying the double Lorentz transformation $\Lambda_A{}^B$.

\subsubsection{Double Lorentz transformation}
At this point we encounter another important subtlety that we need to handle beyond one-loop: Double Lorentz transformations of the generalised frame field pick up the anomalous contribution
\begin{equation}\label{eqn:genGStr}
  \Al^{(1)} \Eh_A{}^I \Eh_{BI} = - P_{[A}{}^C \Pb_{B]}{}^D \Dh_C \lambda_{EF} \Fh_{DGH} P^{EG} P^{FH} - P \leftrightarrow \Pb\,.
\end{equation}
It originates from the non-Lorentz-covariant scheme transformation \eqref{eqn:schemetr} and was dubbed generalised Green-Schwarz transformation (gGS) \cite{Marques:2015vua}. The name is motivated by the observation that the $B$-field of the heterotic string receives a non-Lorentz-covariant contribution to its transformation at the subleading order of $\alpha'$. This correction is captured by the first term on the left hand side of \eqref{eqn:genGStr} and gives rise to the celebrated Green-Schwarz anomaly cancellation mechanism \cite{Green:1984sg}. Moreover, gGS transformations play a central role in constructing $\alpha'$-corrections in DFT, where the one-loop and two-loop effective target space actions, $\Sh^{(1)}$ and $\Sh^{(2)}$, are related by \cite{Marques:2015vua}
\begin{equation}\label{eqn:DLS2}
  \Al^{(0)} \Sh^{(2)} = \delta_{\AlE} \Sh^{(1)} = L_{\AlE} \Sh^{(1)}\,.
\end{equation}
Actually, this relation is so strong that it fixes $S^{(2)}$ completely. Note that the generalised dilaton is not affected and $\Al d = 0$ still holds. Following the steps that we demonstrated at one-loop, one obtains the anomalous transformation of the two-loop $\beta$-functions,
\begin{equation}\label{eqn:DLbeta2}
  \Al^{(0)} \Bh^{(2)} = L_{\AlE} \Bh^{(1)}\,,
\end{equation}
which is of the same form as \eqref{eqn:DLS2}. We present the derivation of this important relation in section~\ref{sec:twoloopsdetails}. For the moment, we are rather interested in a finite version of the left hand side. Because $\Al$ acts as a derivative on the Lie derivative\footnote{One can show that for two arbitrary vectors $X$ and $Y$,
\begin{equation}
  \Al (L_X Y) = L_{\Al X} Y + L_X (\Al Y)
\end{equation}
holds.} one obtains
\begin{equation}\label{eqn:beta2DL}
  \B^{(2)} = \Lambda \cdot \Bigl( \Bh^{(2)} + L_{\begin{pmatrix} \DL^{(1)} \Eh & 0 \end{pmatrix}} \bh^{(1)} \Bigr)\,.
\end{equation}
According to our convention, $\DL^{(1)} \Eh$ denotes the term in the finite gGS transformation
\begin{equation}
  \DL \Eh = \left( e^{\Al} - 1 \right) \Eh
\end{equation}
of the generalised frame field $\Eh_A{}^I$ at the leading order in $\alpha'$.

Hence, we conclude that to go from hatted to unhatted quantities at the two-loop level, not only a rotation by $\Lambda_A{}^B$, but also a scheme transformation is required. Fortunately, neither affects any observables of the theory.  Consequentially, we can drop the hats in the expression for $\bh^{(2)E}$ as we did already at one-loop. PL symmetry removes all terms with flat derivatives $D_A$ and we are left with 20 diagrams contributing to
\begin{equation}\label{eqn:diagbeta2}
  \begin{split}
    \beta^{(2)E}_{a\bar b} = 
      &\HCNa00001 \co{+}  \HCNa00101 \co{+2} \HCNb00011 \co{+2} \HCNb00100 \co{+4} \HCNb01001 \\
            \co{-4} &\HCNb01010 \co{+2} \HCNb11000 \co{-2} \HCNc00010 \co{-4} \HCNc00011 \co{+} \HCNc00110 \\
            \co{+4} &\HCNc01010 \co{+} \HCNc10001 \co{-2} \HCNc10010 \co{+} \dFAFFFa1000 \co{-2} \dFAFFFc0000 \\
            \co{-2} &\dFAFFFc0010 \co{+} \dFAFFFbprime0000 \co{+2} \dFAFFFbprime0010 \co{+} \dFAFFFbprime0100 {-2} \dFAFFFb0100 
      + P \leftrightarrow \Pb\,.
  \end{split}\raisetag{6em}
\end{equation}
To avoid problems with the translation from diagrams to a tensor expression, we give here the explicit result
\newcommand{\refpoint}[1]{\tikz[remember picture,overlay]{\node (#1) {};}}
\begin{align}
  \B^{(2)E}_{AB} =&
 -F_{IJM}F_{CDF}F_{EGH}F_{KLN}P_{[A}{}^{C}P^{DE}P^{FG}P^{HI}P^{JK}\Pb_{B]}{}^{L}\Pb^{MN} \refpoint{HCNa1} \quad\refpoint{HRZ}\qquad\qquad\qquad \nonumber \\&+
 F_{ILN}F_{CDF}F_{EGK}F_{HJM}P_{[A}{}^{C}P^{DE}P^{FG}P^{HI}\Pb_{B]}{}^{J}\Pb^{KL}\Pb^{MN} \refpoint{HCNa2} \nonumber \\&+
 2F_{CDF}F_{EHK}F_{GIM}F_{JLN}P_{[A}{}^{C}P^{DE}P^{FG}P^{HI}\Pb_{B]}{}^{J}\Pb^{KL}\Pb^{MN} \refpoint{HCNb1} \nonumber \\&+
 2F_{IKL}F_{CDF}F_{EHM}F_{GJN}P_{[A}{}^{C}P^{DE}P^{FG}P^{HI}P^{JK}\Pb_{B]}{}^{L}\Pb^{MN} \nonumber \\&- 
 4F_{ILN}F_{CDK}F_{EFH}F_{GJM}P_{[A}{}^{C}P^{DE}P^{FG}P^{HI}\Pb_{B]}{}^{J}\Pb^{KL}\Pb^{MN} \nonumber \\&-
 4F_{IJN}F_{CDK}F_{EFM}F_{GHL}P_{[A}{}^{C}P^{DE}P^{FG}P^{HI}\Pb_{B]}{}^{J}\Pb^{KL}\Pb^{MN}   \nonumber \\&+ 
 2F_{CKM}F_{DFJ}F_{EHL}F_{GIN}P_{[A}{}^{C}P^{DE}P^{FG}P^{HI}\Pb_{B]}{}^{J}\Pb^{KL}\Pb^{MN} \refpoint{HCNb2} \nonumber
  \tikz[remember picture,overlay,br/.style={decorate,decoration={brace}}] {
    \draw[br] ($(HRZ|-HCNa1)+(0,1em)$) -- ($(HRZ|-HCNa2)-(0,.5em)$) node[midway, anchor=west] {\HCNaT};
    \draw[br] ($(HRZ|-HCNb1)+(0,1em)$) -- ($(HRZ|-HCNb2)-(0,.5em)$) node[midway, anchor=west] {\HCNbT}; } 
  \displaybreak \\&-
 2F_{IKN}F_{CDF}F_{EHL}F_{GJM}P_{[A}{}^{C}P^{DE}P^{FG}P^{HI}P^{JK}\Pb_{B]}{}^{L}\Pb^{MN}  \refpoint{HCNc1} \nonumber \\&-
 4F_{ILN}F_{CDF}F_{EHK}F_{GJM}P_{[A}{}^{C}P^{DE}P^{FG}P^{HI}\Pb_{B]}{}^{J}\Pb^{KL}\Pb^{MN}  \nonumber \\&+
 F_{ILN}F_{CDF}F_{EHJ}F_{GKM}P_{[A}{}^{C}P^{DE}P^{FG}P^{HI}\Pb_{B]}{}^{J}\Pb^{KL}\Pb^{MN}   \nonumber \\&+
 4F_{IJL}F_{CDK}F_{EFM}F_{GHN}P_{[A}{}^{C}P^{DE}P^{FG}P^{HI}\Pb_{B]}{}^{J}\Pb^{KL}\Pb^{MN} \refpoint{HCNc2} \nonumber \\&-
 F_{CDK}F_{EJM}F_{FHL}F_{GIN}P_{[A}{}^{C}P^{DE}P^{FG}P^{HI}\Pb_{B]}{}^{J}\Pb^{KL}\Pb^{MN}   \nonumber \\&+
 2F_{ILN}F_{CDK}F_{EFJ}F_{GHM}P_{[A}{}^{C}P^{DE}P^{FG}P^{HI}\Pb_{B]}{}^{J}\Pb^{KL}\Pb^{MN}  \refpoint{HCNc2} \nonumber \\&+
 F_{DJK}F_{EFH}F_{GIL}P_{[A}{}^{D}P^{CE}P^{FG}P^{HI}\Pb_{B]}{}^{J}\Pb^{KL}F_{C}   \refpoint{FAa}  \nonumber \\&+ 
 2F_{IKL}F_{DFH}F_{EGJ}P_{[A}{}^{D}P^{CE}P^{FG}P^{HI}P^{JK}\Pb_{B]}{}^{L}F_{C}  \refpoint{FAb1} \nonumber \\&+
 2F_{DFK}F_{EHL}F_{GIJ}P_{[A}{}^{D}P^{CE}P^{FG}P^{HI}\Pb_{B]}{}^{J}\Pb^{KL}F_{C}  \refpoint{FAb2} \nonumber \\&+ 
 F_{DFH}F_{EJL}F_{GIK}P_{[A}{}^{D}P^{CE}P^{FG}P^{HI}P^{JK}\Pb_{B]}{}^{L}F_{C}  \refpoint{FAc1} \nonumber \\&-
 2F_{DFK}F_{EHJ}F_{GIL}P_{[A}{}^{D}P^{CE}P^{FG}P^{HI}\Pb_{B]}{}^{J}\Pb^{KL}F_{C}  \nonumber \\&- 
 F_{DFH}F_{EJK}F_{GIL}P_{[A}{}^{D}P^{CE}P^{FG}P^{HI}\Pb_{B]}{}^{J}\Pb^{KL}F_{C}  \nonumber \\&-
 2F_{IJL}F_{DEF}F_{GHK}P_{[A}{}^{D}P^{CE}P^{FG}P^{HI}\Pb_{B]}{}^{J}\Pb^{KL}F_{C} \refpoint{FAc2} \nonumber \\&
+ P \leftrightarrow \Pb\,.
  \tikz[remember picture,overlay,br/.style={decorate,decoration={brace}}] {
    \draw[br] ($(HRZ|-HCNc1)+(0,1em)$) -- ($(HRZ|-HCNc2)-(0,.5em)$) node[midway, anchor=west] {\HCNcT};
    \draw[br] ($(HRZ|-FAa) +(0,1em)$) --  ($(HRZ|-FAa)-(0,.5em)$) node[pos=-.2, , anchor=west] {\dFAFFFaT};
    \draw[br] ($(HRZ|-FAb1)+(0,1em)$) --  ($(HRZ|-FAb2)-(0,.5em)$) node[midway, anchor=west] {\dFAFFFcT};
    \draw[br] ($(HRZ|-FAc1)+(0,1em)$) --  ($(HRZ|-FAc2)-(0,.5em)$) node[midway, anchor=west] {\dFAFFFbT};}
\end{align}
For the reader's convenience also a maschine readable version of $\beta^{(2)E}_{a\bar b}$ is available in the Mathematica notebook \appnotebook.

The finite gGS transformation $\Delta_\Lambda \widehat E$ in \eqref{eqn:beta2DL} is a pivotal ingredient the $\alpha'$-corrected PL T-duality transformation rules \cite{\refPLalphaprime}. Thus, it is not surprising that it appears here. In the next subsection, we explain how it is used to extract the metric, $B$-field, and dilaton in the MT scheme. Besides this technical point, it is important to remember that our discussion started from known results for ${\Bb}$ and eventually identified them with the PL duality invariant $\B$ by following the steps
\begin{equation}
  \tikz[baseline=(betabar.center)]{
    \node (betabar) {$\Bb$};
    \node[at=(betabar.east),anchor=west,xshift=6cm] (betahat) {$\Bh$};
    \node[at=(betahat.east),anchor=west,xshift=6cm]  (beta) {$\B$\,.};
    \draw[<->] (betabar.east) -- (betahat.west) node[above,midway] {gauge \& scheme transformation};
    \draw[<->] (betahat.east) -- (beta.west) node[above,midway] {finite gGS scheme transformation};
  }
\end{equation}
Let us stress again that all $\beta$-functions in this diagram capture the same physics. Hence, for practical purposes one might start directly with $\B$ and, if required, reconstruct the much more complicated $\Bb$ by inverting the transformations we found. We will do exactly this for the $\lambda$-deformation below. For completeness, let us just state the results for the two-loop $\beta$-function of the generalised dilaton, either in terms of diagrams
\begin{equation}\label{eqn:diagbeta2d} 
  \b^{(2)d} = \frac14 \FFFFa001011 - \frac14 \FFFFa010010 + \frac14 \FFFFb001100 + 
  \frac13 \FFFFb101001 + \frac14 \FFFAFA0000 + P \leftrightarrow \Pb\,
\end{equation}
or as the tensor expression
\begin{align}
  \b^{(2)d} &= \frac14  F_{ACG} F_{BEH} F_{DIK} F_{FJL} P^{AB} P^{CD} P^{EF} \Pb^{GH} \Pb^{IJ} \Pb^{KL} \nonumber \\
  &-\frac14 F_{ACI} F_{BEJ} F_{DGK} F_{FHL} P^{AB} P^{CD} P^{EF} P^{GH} \Pb^{IJ} \Pb^{KL} \nonumber \\
  &+\frac14  F_{ACI} F_{BEK} F_{DGL} F_{FHJ} P^{AB} P^{CD} P^{EF} P^{GH} \Pb^{IJ} \Pb^{KL} \nonumber \\
  &+\frac13  F_{ACE} F_{BGI} F_{DHK} F_{FJL} P^{AB} P^{CD} P^{EF} \Pb^{GH} \Pb^{IJ} \Pb^{KL} \nonumber\\
  &-\frac14 F_{CEG} F_{DFH} P^{CD} P^{EF} P^{AG} P^{BH} F_{A} F_{B}
  + P \leftrightarrow \Pb \,.
\end{align}

\subsubsection{Renormalisable PL \texorpdfstring{$\sigma$}{sigma}-models}
Again, the argument from section~\ref{sec:renormalisable} applies: Because both $\B^{(2)E}$ and $\beta^{(2)d}$ satisfy \eqref{eqn:renormalisable}, PL $\sigma$-models are two-loop renormalisable and PL T-duality leaves RG-flows invariant. It would be interesting to see if this result can be extended to more general backgrounds by extending the currently available tools for consistent truncations to include $\alpha'$-corrections. We comment more on this point in the conclusion in section~\ref{sec:conclusion}.

\subsubsection{\texorpdfstring{$\lambda$}{Lambda}- and \texorpdfstring{$\eta$}{Eta}-deformation}\label{sec:lambda2}
Using \eqref{eqn:diagbeta2} and \eqref{eqn:diagbeta2d}, it is straightforward to calculate the two-loop $\beta$-functions of the $\lambda$- and $\eta$-deformation. A considerable simplification arises because all four components of the generalised fluxes $F_{ABC}$ in \eqref{eqn:genfluxlambda} just differ by a prefactor and $F_A=0$. Therefore, the remaining diagrams in $\beta^{(2)E}_{a\bar b}$ decompose into two contributions: A topological piece, which is independent of the particular projectors involved, and a coefficient capturing the projector structure. We encounter three different diagram topologies. They are denoted by 
\begin{equation}
  A \sim \HCNaT \,, \quad
  B \sim \HCNbT \,, \quad \text{and} \quad
  C \sim \HCNcT \,.
\end{equation}
From this structure it follows that the two-loop $\beta$-function has to have the form
\begin{equation}
  \beta^{(2)E}_{a\bar b} = \frac{2A + B + 2C}{2 ( 16 \kappa k )^2} c_G^2 \eta_{a\bar b}\,,
    \quad \text{or} \quad
  \beta^{(2)k} = 0 \quad \text{and} \quad 
  \beta^{(2)\kappa} = - \frac{2 A + B + 2 C}{2 (16 k)^2 \kappa} c_G^2
\end{equation}
after taking into account \eqref{eqn:betamu}. The coefficients
\begin{equation}\label{eqn:ABC}
  \begin{aligned}
    A &= 2 \left[ ( -1 ) x^2 y^2 + ( +1 ) x y^3 \right]\\
    B &= 2 \left[ ( +2 ) x^2 y^2 + ( -2 +4 ) x y^3 + ( -4 + 2 ) y^4 \right]\\
    C &= 2 \left[ ( +2 -4 +1 )xy^3 + ( +4 +1 -2 )y^4 \right]
  \end{aligned}
\end{equation}
follow directly from the rules: For each vertex in a diagram of \eqref{eqn:diagbeta2} with no dashed propagators (no $\Pb$s) or all dashed propagators (three $P$'s) put a $x=\kappa^2 + 3$, for all other vertices put a $y=\kappa^2 - 1$. Furthermore, every internal $\Pb$ contributes with a minus sign. Note that swapping $P\leftrightarrow \Pb$ neither changes the topology nor the contributing powers of $x$ and $y$ for a diagram. Thus, we just can introduce an overall factor of two on the left hand side of each line in \eqref{eqn:ABC} and restrict the discussion to the 13 diagrams printed in \eqref{eqn:diagbeta2}. Remarkably, this is sufficient to obtain the two-loop $\beta$-function
\begin{align}
  \beta^{(2)k} &= 0 & &\text{and} &
  \beta^{(2)\kappa} &= - \frac{(3\kappa^2 + 1)(\kappa^2 - 1)^3 c_G^2}{128 k^2 \kappa}\,,
\intertext{or equivalently}
  \beta^{(2)\lambda} &= - \frac{\lambda^3 ( 1 - \lambda + \lambda^2 ) c_G^2}{(1 - \lambda)(1 + \lambda)^5 k^2} 
    & &\text{or} &
  \beta^{(2)\eta} &= - \frac{(1-3\eta^2)(1+\eta^2)}{16\eta} (\eta t c_G)^2
  \,.
\end{align}
Our result matches with the one presented in equation (3.9) of \cite{Georgiou:2019nbz} for the $\lambda$-deformation.

For the dilaton the two relevant topologies are
\begin{equation}
  A \sim \FFFFaT \,, \quad \text{and} \quad B \sim \FFFFbT
  \quad \text{with} \quad \beta^{(2)d} = \frac{2 A + B}{2 (16 \kappa k)^2} c_G^2 \dim G\,.
\end{equation}
By applying the same rules as for $\beta^{(2)E}_{a\bar b}$, we obtain
\begin{equation}\label{eqn:lambdadefbeta2d}
  A = - y^4\,, \quad B = \frac12 y^4 - \frac23 x y^3\,, \quad \text{and} \quad
  \beta^{(2)d} = \frac{(1 - \kappa^2)^3 (3 +13 \kappa^2)}{3072 (\kappa k)^2} c_G^2 \dim G\,.
\end{equation}
Fixed points of the RG-flow give rise to CFTs. Their central charge is related to the value of $\beta^d$ as \cite{Metsaev:1987zx}
\begin{equation}\label{eqn:cfrombetad}
  c = 6 \beta^d \,.
\end{equation}
Taking into account that $\beta^{(0)d} = D/6$, where $D$ denotes the dimension of the target space, we extract for the fixed point at $\lambda = 0$ the central charge
\begin{equation}
  c = \dim G \left( 1 - (2 k)^{-1} c_G \right) + \mathcal{O}(k^{-3})
\end{equation}
by combining \eqref{eqn:lambdadefbeta1d} and \eqref{eqn:lambdadefbeta2d}. Matching it with the central charge of the level $\kh$ WZW-model on the Lie group $G$ \cite{Knizhnik:1984nr},
\begin{equation}
  c = \frac{2 \kh \dim G}{2 \kh + c_G}\,,
\end{equation}
we see that $k = \kh + 1/2 c_G$. Again this observation is in agreement with equation (3.8) of \cite{Georgiou:2019nbz}.

\subsection{Finite generalised Green-Schwarz transformation}\label{sec:finitegGS}
From a conceptual point of view, finite gGS transformations are straightforward. They just exponentiate the infinitesimal version $\delta_\lambda$. Formally, this was already done in \eqref{eqn:expdeltalambda} but at the end of the day, one needs an explicit prescription how this transformation acts on the metric, $B$-field, and dilaton. This is more complicated than one might initially think because it requires an infinite tower of $\alpha'$-corrections. As we restrict our discussion to $\beta$-functions up to two loops, we can fortunately circumvent this problem and just need to compute the first contribution. More precisely, we consider
\begin{equation}
  e^{\delta_\lambda} \Eh_A{}^I \Eh_{BI} = \Lambda_A{}^C \Lambda_B{}^D ( \Lambda_{CD} + \DL \Eh_{CD} )
\end{equation}
with
\begin{equation}
  \DL \Eh_{AB} = \frac12 \begin{pmatrix} 0 & \DL g_{a\bar b} + \DL B_{a\bar b} \\
    - \DL g_{b \bar a} + \DL B_{b \bar a} & 0
  \end{pmatrix}\,.
\end{equation}
To evaluate the scheme transformation \eqref{eqn:beta2DL} that links $\Bh$ with $\B$, one has to compute $\DL^{(1)} g_{a\bar b}$ and $\DL^{(1)} B_{a\bar b}$, respectively. There are slightly different ways how one can do this \cite{Borsato:2020bqo,Hassler:2020tvz}. Of course all of them lead to the same result \cite{Borsato:2020bqo}
\begin{equation}\label{eqn:finitegGStr}
  \begin{aligned}
    \DL^{(1)} g_{ij} &= - \frac12 \Theta_{(i| \bar a}{}^{\bar b} \omega^{(-)}_{|j)\bar b}{}^{\bar a} + 
      \frac14 \Theta_{i\bar a}{}^{\bar b} \Theta_{j\bar b}{}^{\bar a}\\
    \DL^{(1)} B_{ij} &= - \frac12 \Theta_{[i| \bar a}{}^{\bar b} \omega^{(-)}_{|j]\bar b}{}^{\bar a} + B^\Theta_{ij}
  \end{aligned}
\end{equation}
with
\begin{equation}
  \Theta_{i\bar a}{}^{\bar b} = \partial_i \Lambda^{\bar c}{}_{\bar a} \Lambda_{\bar c}{}^{\bar b}\,, \qquad
  \omega^{(-)}_{i\bar a}{}^{\bar b} = \omega_{i\bar a}{}^{\bar b} - \frac12 H_{i\bar a}{}^{\bar b}\,, 
\end{equation}
and
\begin{equation}
  \dd B^\Theta = -\frac1{12} \Theta_{i\bar a}{}^{\bar b} \Theta_{j\bar b}{}^{\bar c} \Theta_{k\bar c}{}^{\bar a} \dd x^i \wedge \dd x^j \wedge \dd x^k\,.
\end{equation}
In order to keep these equations as simple as possible, we frequently switch between flat and curved indices by contracting with the frame $e_a{}^i = e_{\bar a}{}^i$ from \eqref{eqn:decompframe} or its inverse transpose $e^a{}_i = e^{\bar a}{}_i$. Not surprisingly, the resulting field redefinitions are still quite complicated and cumbersome even for simple, low dimensional examples. Hence, one should rather perform all calculations in the adopted scheme of $\B$. Let us revisit the simplest $\lambda$-deformation on SU(2) \cite{Sfetsos:2013wia} to emphasise this claim.

\subsubsection{SU(2) \texorpdfstring{$\lambda$}{lambda}-deformation}
Generators in the fundamental representation of the Lie algebra $\mathfrak{su}(2)$ can be conveniently written in terms of the three Pauli matrices $\sigma_a$ as 
\begin{equation}\label{eqn:su(2)gen}
  t_a = - \frac{i}{\sqrt{2}} \sigma_a\,.
\end{equation}
Note that we use an exotic normalisation that results in $c_{\mathrm{SU(2)}} = 4$ rather than the standard value of $2$. It will become obvious shortly that this choice is required to match with the results in \cite{Hoare:2019mcc}. With the group element
\begin{equation}
  g = \begin{pmatrix}
    \sqrt{1-\alpha^2} - i \alpha \cos\gamma\sin\beta & -\alpha ( \cos\beta - i \sin\beta \sin\gamma) \\
    \alpha (\cos\beta + i \sin\beta\sin\gamma) & \sqrt{1-\alpha^2} + i \alpha \cos\gamma\sin\beta
  \end{pmatrix}\,,
\end{equation}
we obtain the leading order metric, $H$-flux and dilaton,
\begin{equation}
  \begin{aligned}
    \dd s^{2(0)} &= \frac{k}{\kappa (1 - \alpha^2)}\dd \alpha^2 + \frac{\alpha^2 \kappa k}{\Delta} 
      \dd s^2(S^2) \\
    H^{(0)} &= \frac{k \alpha^2 \left[ 2 \kappa^2 + (1-\kappa^2) \Delta \right]}{%
      \sqrt{1-\alpha^2}\Delta^2} \dd \alpha \wedge \mathrm{vol}(S^2)\\
    \phi^{(0)} &= - \frac12 \log \Delta & 
    \quad \text{with} \quad \Delta &=\kappa^2 + \alpha^2 (1-\kappa^2)\,,
  \end{aligned}
\end{equation}
after implementing the discussion in section~\ref{sec:genframelambda}. They match equation (3.10) in \cite{Hoare:2019mcc} and fix the normalisation \eqref{eqn:su(2)gen} we use for the generators $t_a$. In order to make the expression more readable, we use the round-two sphere $S^2$ with the metric $\dd s^2(S^2) = \dd \beta^2 + \sin\beta \dd \gamma^2$ and the volume form $\mathrm{vol}(S^2) = \sin\beta \dd \beta \wedge \dd \gamma$ as a reference. Evaluating \eqref{eqn:defbh} for \eqref{eqn:bb1E} and \eqref{eqn:diffgauge1}, one obtains the one-loop $\beta$-function for the metric and $B$-field, which can be written as
\begin{equation}
  \bh^{(1)E}_{ij} = - c_G \frac{(\kappa^2 - 1)^2}{8 k \kappa} \Lambda_{ji}\,,
\end{equation}
where $\Lambda_{ij}$ is the curved version of $\Lambda_{\bar a\bar b}$ in \eqref{eqn:lambdalambdadef}. As expected, this equation agrees with \eqref{eqn:BtoBh}. Another remarkable property, which is not directly obvious at the level of the target space fields, is
\begin{equation}
  \frac{\dd\left(g_{ij}^{(0)} + B_{ij}^{(0)}\right)}{\dd \kappa} = - \frac1{\kappa} \Lambda_{ji}\,.
\end{equation}
It can be used to verify the $\beta$-function for $\kappa$ in \eqref{eqn:beta1kappa} and emphasise that already the one-loop computations involving $\bh$ are more opaque than the ones for $\beta$. In the same vein, one checks the $\beta$-function of the generalised dilaton.

Using \eqref{eqn:finitegGStr}, we evaluate the corrections to the metric and $B$-field,
\begin{equation}
  \begin{aligned}
    \DL^{(1)} \dd s^2 &= \frac{(1-\kappa^2)\Delta + 2 \kappa^2}{\Delta^2} \left[
      (1-\alpha^2)^{-1} \dd \alpha^2 + \alpha^2 \dd s^2 ( S^2 ) \right]\\
    \DL^{(1)} B &= \frac{\alpha^2 \kappa \sin\beta}{\Delta^2 \sqrt{1 - \alpha^2}} \left[
      4 \gamma \dd\alpha\wedge\dd\beta + 2 \alpha (\alpha^2 - 1) (\kappa^2 - 1) 
      \dd\beta\wedge\dd\gamma \right]\,,
  \end{aligned}
\end{equation}
which originate from the finite gGS transformation with the parameter $\Lambda_{\bar a\bar b}$. Combining them with the scheme transformation \eqref{eqn:schemetr}, we obtain the $\alpha'$-corrections
\begin{equation}
  \begin{aligned}
    \dd s^{2(1)} &=  - \frac{8 \kappa^4 - 8 \kappa^2(1 - \kappa^2)\Delta - 3 (1 - \kappa^2)^2 \Delta^2}{
        4 \kappa k \Delta^2} d s^{2 (0)}\\
    H^{(1)} &= \frac{\kappa^4 \left[12 + \Delta(3\Delta-14)\right] + 2 \kappa^2(3-2\Delta)\Delta +
          \Delta^2}{k \left[ \kappa^2 (2 - \Delta) + \Delta \right] \Delta^2} \kappa H^{(0)}\\
    d^{(1)} &= 0\,.
  \end{aligned}
\end{equation}
As a check, one can evaluate the two-loop $\beta$-function \eqref{eqn:bh2dpure}\footnote{Equation~\eqref{eqn:bh2dpure} does not include the infinitesimal diffeomorphism $\xi^{(1) i}(\bh^{(1)B})$ from \eqref{eqn:difftodoubleK} which generates the second term in \eqref{eqn:requireddiff+gauge}. Thus, we add it to get the $\bh^{(2)}$ in \eqref{eqn:bh2dexample}.} for the generalised dilaton for this corrected target space geometry. With the help of the xCoba Mathematica package, we find
\begin{equation}\label{eqn:bh2dexample}
  \bh^{(2)d} = \frac{(1 - \kappa^2)^3 (3 +13 \kappa^2)}{3072 (\kappa k)^2} c_G^2 \dim G \,,
\end{equation}
which matches \eqref{eqn:lambdadefbeta2d}. We could continue to compute the $\beta$-functions of the metric and $B$-field. For them, performing the scheme transformation is more involved. Moreover, one has to account for a further correction from a partial double Lorentz frame fixing, as explained in section~\ref{sec:twoloopsdetails}. Because it will not provided any further insights, we will not present this complicated calculation here.

\section{Doubled gradient flow}\label{sec:derivation}
The results in the last section are self-contained and can be used without additional insights into how they were obtained. Still, it is of course interesting to see how we systematically derive expressions like \eqref{eqn:diagbeta1} and \eqref{eqn:diagbeta2}. Thus, we will go step by step through the derivation in the following.

A crucial observation is that it is in general highly complicated to compute O($D$,$D$)-covariant $\beta$-functions directly. To avoid this problem, we exploit the fact that they alternatively arise from a gradient flow,
\begin{equation}\label{eqn:gradientflow}
  \delta_\Psi S = \intvol \Psi \cdot K ( \B )\,,
\end{equation}
where $K(\B)$ is an invertible rank two tensor on the coupling space. In order to obtain $\B$, it is sufficient to know $S$ and $K$. Splitting \eqref{eqn:gradientflow} order by order in $\alpha'$, one finds
\begin{equation}\label{eqn:SandK}
  \begin{aligned}
    \delta_\Psi S^{(1)} &= \intvol \Psi \cdot K^{(0)} (\B^{(1)}) \\
    \delta_\Psi S^{(2)} &= \intvol \Psi \cdot \left[ K^{(0)}(\B^{(2)}) + K^{(1)}(\B^{(1)}) \right]\,.
  \end{aligned}
\end{equation}
Because $K^{(0)}$ does not contain any derivatives, it is just a matrix and can be inverted easily. With the inverse, which is fixed completely by a one-loop computation, it is straightforward to extract $\B^{(1)}$ and $\B^{(2)}$. At a first glance, this route might seem more complicated than just trying to directly rewrite the known results for one and two-loop $\beta$-functions of the bosonic string in a doubled, O($D$,$D$)-covariant way. However, we will see that it is much easier. Especially, since the covariant expressions for $S^{(1)}$ and $S^{(2)}$ are already known \cite{Marques:2015vua}. Furthermore, $K^{(0)}$ follows nearly immediately from known results in DFT. Hence, the remaining challenge is to find $K^{(1)}$ and bring it in an O($D$,$D$)-covariant form. In doing so, a considerable advantage is that $K^{(1)}$ just contains two derivatives and dealing with Bianchi identities simplifies significantly compared to $S^{(2)}$ or $\B^{(2)}$.

We start with the one-loop computation in the next subsection. It contains all the major ingredients of the gradient flow \eqref{eqn:gradientflow} in a simple setting. After introducing all required quantities, we demonstrate how the $\beta$-functions from section~\ref{sec:oneloop} arise. Subsequently, we address the two-loop $\beta$-functions in section~\ref{sec:twoloopsdetails}. They require to additionally discuss scheme transformations, partial double Lorentz gauge fixing, and gGS transformations. Manifest PL symmetry does not only simplifies the $\beta$-functions considerably but also $K$, which governs the gradient flow. Hence, we explain in section~\ref{sec:cfunction} how one computes the $c$-function and the corresponding gradient flow metric of PL $\sigma$-model. In the spirit of section~\ref{sec:results}, we discuss the $\lambda$-deformation as an explicit example.

\subsection{One-loop}\label{sec:oneloopdetail}
The starting point of our derivation is the one-loop $\beta$-functions $\bh^{(1)E}_{ij}$ and $\bh^{(1)\phi}$ from section~\ref{sec:oneloop}. For convenience, we decompose the former into its metric and $B$-field contribution, $\bh^{E}_{ij} = \bh^{g}_{ij} + \bh^{B}_{ij}$. Hence, the three $\beta$-functions
\begin{align}
    \bh^{(1)g}_{ij} &= R_{ij} - \frac14 H^2_{ij} + 2 \nabla_i \nabla_j \phi \,, \nonumber \\
    \bh^{(1)B}_{ij} &= -\frac12 \nabla_l H^l{}_{ij} + \nabla_l \phi H^l{}_{ij} \,, \nonumber \\  
    \label{eqn:betah1}
    \bh^{(1)\phi}  &= - \frac12 \nabla^2 \phi - \frac1{24} H^2 + (\nabla \phi)^2
\end{align}
form the basis of our discussion. Each line contains an infinitesimal diffeomorphism with $\xi^{(1)i}=\nabla^i \phi$, which relates the respective $\bh$-function to $\bb^{(1)g}_{ij}$, $\bb^{(1)B}_{ij}$ and $\bb^{(1)\phi}$. 

In order to understand how these $\beta$-functions arise from a gradient flow, we vary the one-loop effective target space action, 
\begin{equation}\label{eqn:action1}
  \Sh^{(1)} = \int \dd^D x \sqrt{g} e^{-2\phi} \left( R - \frac1{12} H^2 - 4 (\nabla \phi)^2 + 4 \nabla^2 \phi \right)\,,
\end{equation}
of the bosonic string. By comparing the result
\begin{equation}\label{eqn:varaction1}
  \delta_\Psi \Sh^{(1)} = \intvol \left( - \delta g^{ij} \bh^{(1)g}_{ij} - \delta B^{ij} \bh^{(1)B}_{ij} + 8 \delta d \bh^{(1)d} \right)
\end{equation}
with the first equation in \eqref{eqn:SandK}, we verify that the $\beta$-functions \eqref{eqn:betah1} indeed arise from a gradient flow. Like before, we use the $\beta$-function for the generalised dilaton \eqref{eqn:diagbeta1d} instead of $\bh^\phi$. Moreover, we can easily read off $K^{(0)}(\Bh)$. For the following discussion it is crucial that the first two terms in the integral come both with a minus sign. For the metric, this sign is subtle as we can either vary with respect to the metric or its inverse (both differ by a sign). All metric variations we perform are with respect to the metric $g_{ij}$. Thus, the natural index position for $\delta g$ is $\delta g_{ij}$ and the corresponding $\beta$-function has both indices raised. Due to the superscripts $\bh^{(1)g}$ is carrying, it is usually more convenient to use exactly the opposite notation, like in \eqref{eqn:varaction1}. This is perfectly fine, as long as we keep in mind that the variation is still with respect to the metric and not its inverse.

Our next objective is to rewrite \eqref{eqn:varaction1} in terms of the O($D$,$D$)-covariant quantities from section~\ref{sec:genframes}. To this end, we first obtain the variation of the generalised frame $\Eh_A{}^I$,
\begin{equation}\label{eqn:deltahatE}
  \dEh_A{}^I \Eh_{BI} = \dEh_{AB} + \dEgf_{AB}
\end{equation}
with
\begin{equation}
  \dEh_{AB} = \frac12 \begin{pmatrix}
    0 & \dg_{a\bar b} + \dB_{a\bar b} \\
    -\dg_{b \bar a} + \dB_{b \bar a} & 0
  \end{pmatrix} \qquad \text{and} \qquad
  \dEgf_{AB} = \frac12 \begin{pmatrix}
    \dB_{ab} & 0 \\
    0 & \dB_{\bar a\bar b}
  \end{pmatrix}\,.
\end{equation}
$\delta g_{ab}$ and $\delta B_{ab}$ denote the flattened variations of the metric and $B$-field ($\delta g_{ab} = e_a{}^i e_b{}^j \delta g_{ij}$ and $\delta B_{ab}=e_a{}^i e_b{}^j \delta B_{ij}$). All fluctuations of the generalised frame field in \eqref{eqn:deltahatE} split into two parts because $\Eh_A{}^I$ is partially gauge fixed to a distinguished double Lorentz frame. If we would only apply $\dEh_{AB}$, whose form is identical to the $\Bh_{AB}$ in \eqref{eqn:betaEdoubled}, we would destroy this gauge fixing. Hence, we have to additionally apply the compensating gauge transformation $\dEgf_{AB}$. The same pattern applies to the $\beta$-function and we readily define
\begin{equation}
  \BhEgf = \frac12 \begin{pmatrix}
    \bh^{B}_{ab} & 0 \\
    0 & \bh^{B}_{\bar a\bar b}
  \end{pmatrix}\,.
\end{equation}
Because the one-loop action $\Sh^{(1)}$ is invariant under double Lorentz transformations, gauge fixing terms drop out from the doubled version,
\begin{equation}\label{eqn:varaction1doubled}
  \delta_\Psi \Sh^{(1)} = \intvol \left( \delta\Eh^{AB} K^{(0)}_{AB}{}^{CD} \Bh^{(1)E}_{AB} + 8 \delta d \bh^{(1)d} \right)\,,
\end{equation}
of \eqref{eqn:varaction1}. But they will become relevant at two loops, as we discuss in section~\ref{sec:PDLgf}. From \eqref{eqn:varaction1doubled}, we read off $K^{(0)}$ for the metric and the $B$-field. It takes the surprisingly simple form
\begin{equation}\label{eqn:K0doubled}
  K^{(0)}_{ABCD} = 2 \eta_{AC} \eta_{BD}\,.
\end{equation} 

In DFT, the action $\Sh^{(1)}$ is expressed in terms of the generalised Ricci scalar,
\begin{equation}
  \Sh^{(1)} = \intvol \scRh^{(1)}\,.
\end{equation}
There are two different ways to write $\scRh^{(1)}$, either in terms of a generalised metric or generalised fluxes. We adopt the latter, the flux formulation \cite{Siegel:1993th,*Siegel:1993xq,Hohm:2010xe,Geissbuhler:2013uka}, where it reads
\begin{equation}\label{eqn:genR}
  \widehat{\mathcal{R}}^{(1)} = P^{AB} P^{CD} \left( \overline{P}^{EF} + \frac13 P^{EF} \right) \Fh_{ACE} \Fh_{BDF}  + 2 P^{AB} ( 2 \Dh_A \Fh_B - \Fh_A \Fh_B )\,.
\end{equation}
Finally, we compute the variation\footnote{%
The variations are exactly the ones given in \eqref{eqn:variations}. Furthermore, performing integration by parts with
\begin{equation}
  \intvol \Dh_A \cdot = \intvol \Fh_A
\end{equation}
is required.%
} of this action with respect to $\delta\Eh_{AB}$ and $\delta d$, 
\begin{equation}
  \delta_\Psi \Sh^{(1)} = \intvol \left( \delta \Eh^{AB} \widehat{\mathcal{G}}_{AB}^{(1)} - 2 \delta d \widehat{\mathcal{R}}^{(1)} \right)
\end{equation}
with
\begin{equation}
  \mathcal{G}^{(1)}_{AB} = 4 P_{[A}{}^C \Pb_{B]}{}^D \left( \Fh_{CEG} \Fh_{DFH} P^{EF} \Pb^{GH} + \Fh_{CDE} \Fh_E P^{EF} + \Dh_D \Fh_C - \Dh_E \Fh_{CDF} \Pb^{EF} \right)\,,
\end{equation}
and compare the result with \eqref{eqn:varaction1doubled}. One directly reads off $\bh^{(1)E}_{a\bar b} = \widehat{\mathcal{G}}^{(1)}_{a\bar b}$, $\bh^{(1)d}=-\frac14\widehat{\mathcal{R}}^{(1)}$ and thereby obtains the results discussed in section~\ref{sec:oneloop}.

\subsection{Two loops}\label{sec:twoloopsdetails}
Beyond one-loop, $\beta$-functions become scheme dependent and we have to choose a scheme to start with. There are two popular options for the bosonic string, the Metsaev-Tseytlin \cite{Metsaev:1987zx} (MT) scheme and the Hull-Townsend (HT) scheme. Both are connected by a scheme transformation which is detailed in appendix~\ref{app:HTtoMT}. We found it a little easier to extract $K^{(1)}$ from the results presented in \cite{Metsaev:1987zx} and therefore we start from the two-loop $\beta$-functions in the MT scheme\footnote{Note that $H^2_{ab} = H_{acd}H_b{}^{cd}$, $H^4_{ab} = H_{acd}H^{cef}H_{eg}{}^d H^{g}{}_{bf}$ and that the signs of the last two terms in the first line of \eqref{eqn:betaB2} are flipped compared to \cite{Metsaev:1987zx}. It seems that there happened a misprint in \cite{Metsaev:1987zx}, because the combination of the signs in \eqref{eqn:betaB2} is the one which arises from the variation of the target space effective action in appendix~\ref{app:var2laction}. It is also required to obtain the $B$-field $\beta$-function in the HT scheme after the appropriate scheme transformation (see appendix~\ref{app:HTtoMT} for details).} \cite{Metsaev:1987zx}
\begin{align}\label{eqn:betag2}
  \bb^{(2)g}_{ab} &= 
    \frac{1}{2} \Big[ R_{acde}R_b{}^{cde} + \frac{1}{8} (H^4)_{ab} + 
    \frac{3}{4} \nabla_c H_{ade} \nabla^c H_b{}^{de} + \frac{1}{8} H^{cd}{}_a H_{db}{}^{e}{}(H^2)_{ce} 
    \\ & \qquad
    - R^{cdef} H_{acd} H_{bef} - \frac{5}{2} R_{(a}{}^{cde} H_{b)cf}{} H_{dbe}{}^{f} - 
    \frac{1}{2} R^c{}_{ab}{}^d (H^2)_{cd} + \frac{1}{12} \nabla_a H_{cde} \nabla_b H^{cde} \Big] \,, 
    \nonumber \\ \label{eqn:betaB2}
  \bb^{(2)B}_{ab} &= 
    \frac{1}{4} \Big[ 2R_{[a|cde} \nabla^c H_{|b]}{}^{de} + \nabla^c H^{}_{de[a} H_{b]}{}^{fd} H_{cf}{}^e + 
      2 \nabla_c (H^{2})_{d[a} H_{b]}{}^{dc} - \frac{1}{2} H^2_{cd} \nabla^c H^d{}_{ab} \Big]\,, \\
      \label{eqn:betaphi2}
  \bb^{(2)\phi} &= 
    \frac1{16} \Big[ R_{abcd} R^{abcd} + \frac5{24} H^4 + \frac{4}{3} \nabla_d H_{abc} \nabla^d H^{abc} +
    \frac38 H_{ab}^2(H^2)^{ab} 
    \\ & \qquad\qquad
    - \frac{11}2 R^{abcd} H_{abe} H_{cd}{}^e - 2 H^2_{ab} \nabla^a \nabla^b \phi \Big] \,. \nonumber
\end{align}
Note that we use flat indices because this is more in line with the objects we expect to find in the O($D$,$D$)-covariant rewriting we are looking for. But as the covariant derivative $\nabla_i$ annihilates by construction the frame field $e_a{}^i$, which is used to go from flat to curved indices, switching between the two becomes just a relabeling. Like we have seen in the last subsection, instead of $\Bb$, the gradient flow usually involves a different member in the same equivalence class, $\Bh$, which is obtained by an infinitesimal diffeomorphism and/or a $B$-field gauge transformation. More precisely, we take $\xi^{(2) i} = -\frac1{48}\nabla^i H^2$ and $\chi^{(2)}_i = 0$ in \eqref{eqn:defbh} to fix $\bh^{(2)E}_{ij}$ and $\bh^{(2)\phi}$ respectively.

After a cumbersome computation, which is summarised in appendix~\ref{app:var2laction}, we find that the variation of the two-loop target space effective action
\begin{equation}\label{eq:S2}
  \Sh^{(2)} = \intvol \frac14 \Big[ R_{abcd}R^{abcd} - \frac12 R^{abcd} H_{abe}     H_{cd}{}^{e} + \frac1{24} H^4 - \frac18 (H^2_{ab})^2 \Big]
\end{equation}
gives rise to
\begin{align}\label{eqn:varS2}
  \delta_\Psi \Sh^{(2)} = \intvol \Bigr[& - \delta g^{ab} \bh^{(2)g}_{ab} - \delta B^{ab} \bh^{(2)B}_{ab} + 8 \delta d \b^{(2)d} + \Kh^{(1)d}(\bh^{(1)B})\\
  & + \delta g^{ab} \Kh^{(1)g}_{ab} ( \bh^{(1)g}, \bh^{(1)B} ) + \delta B^{ab} \Kh^{(1)B}_{ab}( \bh^{(1)g}, \bh^{(1)B} )  \Bigl] \nonumber
\end{align}
with
\begin{align}\label{eqn:Kh(1)d}
  \Kh^{(1)d}(\beta^B) &= H^{abc} \nabla_a \beta^B_{bc}\,, \\ \label{eqn:Kh(1)g}
  \Kh^{(1)g}_{ab}(\beta^g,\beta^B) &= 
    -\nabla^2\beta^g_{ab} + \nabla^c \nabla^{\vphantom{g}}_{(a|} \beta^g_{c|b)} - 
    \frac34 H_{acd} H_{be}{}^{d} (\beta^g)^{ec} - \frac14 H^2_{(a|c} (\beta^g)_{|b)}{}^{c} 
    \\ &\quad 
    - 2\left(\nabla_{(a|}\beta^g_{cb)} - \nabla_c \beta^g_{ab}\right)\nabla^c\phi + 
    H^{cd}{}_{(a|} \nabla_c \beta^B_{d|b)} + \frac12 H^{cd}{}_{(a|} \nabla_{|b)} \beta^B_{cd}
    \nonumber \\ &\quad 
    - \frac12\nabla_{(a} H_{b)cd}\beta^{B cd} + \beta^B_{ca}\beta^{B c}{}_{b} \,,
    \nonumber \\ \label{eqn:Kh(1)B}
  \Kh^{(1)B}_{ab}(\beta^g,\beta^B) &=  
    -H_{[a|}{}^{cd} \nabla_c \beta^g_{d|b]} - \frac12 R_{ab}{}^{cd} \beta^B_{cd} - 
    \frac14 H_{abc} H^{dec} \beta^B_{de} + \frac14 H_{acd}H_{be}{}^{c} \beta^{B ed}
    \\ &\quad 
    -\frac12 H_{de[a|}H^{dec}\beta^B_{c|b]} \,. \nonumber
\end{align}
It is actually rather non-trivial to bring $\delta_\Psi \Sh^{(2)}$ into this form. That it is still possible demonstrates the power of the gradient flow equations \eqref{eqn:SandK}.

\subsubsection{Physically equivalent choices for \texorpdfstring{$\Kh^{(1)}$}{Khat(1)}}\label{sec:fixKh1}
The expressions \eqref{eqn:Kh(1)d} to \eqref{eqn:Kh(1)B}, we obtained for $\Kh^{(1)}$, cannot be brought into a doubled, O($D$,$D$)-covariant form as given. However, they can be modified to overcome this problem. More specifically there are at least four different ways to change an arbitrary $\Kh$ while keeping the physics it describes unchanged:
\begin{enumerate}[leftmargin=1.5em,itemindent=0cm,itemsep=0cm,label*=\arabic*)]
  \item Assume that the $\beta$-functions are shifted by a combination of an infinitesimal gauge transformation and diffeomorphism which is parameterised by $\Xi^I = \begin{pmatrix} \chi_i & \xi^i \end{pmatrix}$ and mediated by the generalised Lie derivative $\mathcal{L}$, 
    \begin{equation}
      \Bh \rightarrow \Bh + \mathcal{L}_{\Xi} \Bh\,.
    \end{equation}
    Moreover, take $\Xi$ to be a function of the one-loop $\beta$-functions, which contains one additional derivative. If we want to keep the second gradient flow equation in \eqref{eqn:SandK} invariant, we have to adapt $\Kh^{(1)}$ according to
    \begin{equation}
      \Kh^{(1)} \rightarrow \Kh^{(1)} - K^{(0)} \mathcal{L}_{\Xi^{(1)}} \,.
    \end{equation}
    We will do exactly this with
    \begin{equation}\label{eqn:difftodoubleK}
      \xi^{(1) i}(\bh^{(1)B}) = \frac14 H^{ijk} \bh^{(1)B}_{ij} \qquad \text{and} \qquad
      \chi^{(1)}_i(\bh^{(1)B}) =  \frac12 \omega_i{}^{ab} \bh^{(1)B}_{ab}\,.
    \end{equation}
  \item Additionally, the invariance of the action $\Sh$ under generalised diffeomorphisms gives rise to relations between $\beta$-functions. In particular, one can use
    \begin{equation}
      \delta_{\mathcal{L}_\Xi \begin{pmatrix} \Eh & d \end{pmatrix}} \Sh^{(1)} = 0 
    \end{equation}
    to obtain the identities
    \begin{equation}
      \begin{aligned}
        0 &= \bh^{(1)g}{}_{ab} \nabla^b \phi - \frac12 \nabla^b \bh^{(1)g}{}_{ab} +
          \frac14 H_a{}^{bc} \bh^{(1)B}_{bc} - \nabla_a \bh^{(1)d} \\
        0 &= \nabla^b \Bigl( e^{-2\phi} \bh^{(1)B}_{ab} \Bigr)\,.
      \end{aligned}
    \end{equation}
  \item Shifting $\Kh^{(1)B}_{ab}(\beta^g, \beta^B)$ by
    \begin{equation}
      \frac12 \left( \bh^{(0)B}_{c[a} \beta^g{}_{b]}{}^c - 
        \beta^{B c}{}_{[a} \bh^{(1)g}_{b]c} \right)
    \end{equation}
    does not affect \eqref{eqn:varS2}, because for $\beta^g=\bh^{(1)g}$ and $\beta^B=\bh^{(1)B}$ it vanishes.
  \item Eventually, we perform a scheme transformation from the MT scheme to the generalised Bergshoeff-de Roe scheme (gBdR). This transformation is required to bring the action $\Sh^{(2)}$ into an O($D$,$D$)-covariant form \cite{Marques:2015vua}. Thus, it is natural to apply it to $\Kh^{(1)}$, too. In our conventions, this tranformation is parameterised by
  \begin{equation}\label{eqn:MTtogBdR}
    \begin{aligned}
      \Delta^{(1)} g_{ij} &= - \frac12 \omega_{ia}{}^b \omega_{jb}{}^a + \frac38 H^2_{ij}\,,\\
      \Delta^{(1)} B_{ij} &= - \bh^{B(1)}_{ij} - \frac12 H_{[ia}{}^b \omega_{j]b}{}^a \,, \qquad &
      \Delta^{(1)} d &= 0
    \end{aligned}
  \end{equation}
  and implemented by the Lie derivative on the coupling space. We already discussed the latter for $\beta$-functions. Here, we extend it in the canonical way to $K^{(1)}$, namely
  \begin{equation}
    K^{(1)}(\Psi', \B) \rightarrow K^{(1)}(\Psi', \B) + L_{\Psi^{(1)}} K^{(0)}(\Psi', \B) 
  \end{equation}
  with
  \begin{equation}\label{eqn:LieK(0)}
    \begin{aligned}
      L_\Psi K^{(0)}(\Psi', \B) = &K^{(0)}(\delta_{\Psi'} \Psi, \B) + 
        K^{(0)}(\Psi', \delta_\B \Psi) + \\
        &K^{(0)}(T(\Psi',\Psi), \B) + K^{(0)}(\Psi',T(\B,\Psi))\,,
    \end{aligned}
  \end{equation}
  where we understand $K$ as a pairing between two vectors,
  \begin{equation}
    K(\Psi, \B) = \intvol \Psi\cdot K(\B)\,,
  \end{equation}
  on the infinite dimensional coupling space. Evaluating \eqref{eqn:LieK(0)} for \eqref{eqn:MTtogBdR} is cumbersome, especially because \eqref{eqn:MTtogBdR} contains Lorentz symmetry violating terms. We approach this challenge by writing the one-loop $\beta$-functions in terms of the spin connection $\omega_{abc}$, the flat derivative $\Dh_a$, $\Fh_a$ from \eqref{eqn:Fha} and the $H$-flux $H_{abc}$ with the following, non-vanishing, variations
  \begin{equation}
    \begin{aligned}
      \delta_\Psi \omega_{abc} &= D_{[c} \delta g_{b]a} + \delta g_{d[b} \omega_{c]a}{}^d + \delta g_{ad} \omega_{[cb]}{}^d - \frac12 \delta g_{ad} \omega^d{}_{bc}\,, \\
      \delta_\Psi H_{abc} &= - \frac32 \delta g_{[a}{}^d H_{bc]d} + 3 \nabla_{[a} \delta B_{bc]}\,, \\
      \delta_\Psi \Fh_{a} &= \sqrt{2} \Dh_a \delta d + \frac1{2\sqrt{2}} \Dh_b \delta g_a{}^b - \frac12 \Fh_b \delta g_a{}^b \qquad \text{and} \qquad & [ \delta_\Psi, \Dh_a ] = - \frac12 \delta g_a{}^b \Dh_b \,.
    \end{aligned}
  \end{equation}
\end{enumerate}

To keep the following discussion more tractable, we split $K$ into a symmetric and an antisymmetric part,
\begin{equation}
  K_\pm(\Psi, \Psi') = \frac12 \left[ K(\Psi, \Psi') \pm K(\Psi', \Psi) \right]\,.
\end{equation}
Most of $\Kh^{(1)}_+$ can be re-expressed in terms of O($D$,$D$)-covariant quantities. Unfortunately, the situation for the asymmetric part is much worse. Hence, one might hope that there is a way to get rid of $\Kh^{(1)}_-$ and, while doing so, also to obtain the missing terms that are required to complete the doubling of $\Kh^{(1)}_+$. Remarkably, this is indeed possible by applying a scheme transformation, which is linear in the one-loop $\beta$-functions, namely
\begin{equation}\label{eqn:2ndschemetr}
    \Delta^{(1)} g_{ij} = 0 \,, \qquad
    \Delta^{(1)} B_{ij} = \bh^{B(1)}_{ij}\,,\qquad 
    \Delta^{(1)} d = 0\,.
\end{equation}
But instead of applying it to all quantities in \eqref{eqn:SandK}, we only transform the $\beta$-functions
\begin{equation}
  \Bh'^{(2)} = \Bh^{(2)} + L_{\Psi'} \Bh^{(1)}\,.
\end{equation}
While the first equation of \eqref{eqn:SandK} is not affected by this transformation, the second one becomes
\begin{equation}
  \delta_\Psi \Sh^{(2)} = \intvol \Psi \left[ K^{(0)}( \Bh'^{(2)} ) + \Kh'^{(1)} ( \Bh^{(1)} ) \right]
\end{equation}
with
\begin{equation}
  \Kh'^{(1)}(\B^{(1)}) = \Kh^{(1)} (\B^{(1)}) - K^{(0)}(L_{\Psi'} \B^{(1)})\,.
\end{equation}
Here $\Psi'$ generates the scheme transformation \eqref{eqn:2ndschemetr} and, as intended, all terms of $K'^{(1)}_-$ vanish. For the sake of brevity, we drop the prime from now on.

\subsubsection{\texorpdfstring{O($D$,$D$)}{O(D,D)}-covariant rewriting of \texorpdfstring{$\Kh^{(1)}$}{Khat(1)}}\label{sec:matchingdiagrams}%
\begin{table}[!t]
  \centering
\begin{tabular}{ |c|c|c|c|c|c|c|c|c|c| } 
  \hline
  & & & & \multicolumn{3}{|c|}{\# of diagrams} &  \multicolumn{3}{|c|}{terms in $\Kh^{(1)}$ to match}\\
  \hline
  type & $\Fh_{ABC}$ & $\Fh_A$ & $\Dh_A$ & class $A$ & class $B$ & class $C$ &
    class $A$ & class $B$ & class $C$ \\ \hline
  I   & 0 & 0 & 2 & 0& 1& 1 & 0 & 2 & 2\\
  II  & 1 & 0 & 1 & 4& 4& 0 & 9 & 16 & 0\\
  III & 0 & 1 & 1 & 0& 4& 2 & 0 & 2 & 2\\
  IV  & 2 & 0 & 0 & 3& 3& 2 & 16 & 16 & 0\\
  V   & 1 & 1 & 0 & 2& 2& 0 & 3 & 8 & 0\\
  VI  & 0 & 2 & 0 & 0& 1& 1 & 0 & 0 & 0\\
 \hline
\end{tabular}
\caption{Different combinations of the three tensors $\Fh_{ABC}$, $\Fh_{A}$, and $\Dh_A$ with the number of different possible diagrams obtained by combining two of them. We reference each combination with a Roman numeral from I to VI and further specify one of three classes, $A$, $B$, or $C$.\label{tab:diagclasses}}
\end{table}%
Written in terms of the spin connection $\omega_{abc}$, the flat derivative $\Dh_a$, the $H$-flux $H_{abc}$, and $\Fh_a$, $\Kh^{(1)}(\B^{(1)})$ consists of 76 different terms which can be recast using exclusively $P^{AB}$, $\Pb^{AB}$, $\Fh_{ABC}$, $\Fh_A$ and $\Dh_A$. For this job, the diagrams, which we have introduced in section~\ref{sec:oneloop}, are a convenient tool because they make keeping track of all different terms which could possibly appear much easier. Hence, we first have to  determine all diagrams with
\begin{enumerate}[leftmargin=1.5em,itemindent=1em,itemsep=0cm,label*=\arabic*)]
  \item two external legs, one with a $P$ and the other one with a $\Pb$
  \item internally $\beta^{E}_{a\bar b} = \tikz[baseline={($(beta.west)+(0,-4pt)$)}]{
      \node[draw=gray!60!black,shape=circle,inner sep=7pt,{label={center:{\Large $\beta$}}}] (beta) {};
      \draw[dashed] (beta.east) -- +(1,0);
      \draw (beta.west) -- +(-1,0);
    }$, representing the argument of $\Kh^{(1)}(\B)$
  \item two derivatives\,.
\end{enumerate}
$\Fh_{ABC}$, $\Fh_A$ and $\Dh_A$ contribute one derivative each. Thus, only two of them can be present in a diagram at a time. This results in six different combinations that we call types and number by roman numerals. Moreover, there are three different classes of diagrams where $\beta^{(1)E}_{a\bar b}$ is connected to
\begin{enumerate}[leftmargin=1.5em,itemindent=1em,itemsep=0cm,label*=\Alph*)]
  \item no external leg
  \item one external leg
  \item both external legs\,.
\end{enumerate}
Finally, note that all diagrams have to come in pairs because $\Kh^{(1)}$ is even under the $\mathbb{Z}_2$ symmetry defined in \eqref{eqn:Z2}. Hence, we only draw one diagram of each pair and understand that it has to be complemented by its partner, which arises under the swapping $P \leftrightarrow \Pb$. The resulting number of admissible diagrams for all types and the corresponding classes is summarised in table \ref{tab:diagclasses}. Going through this list, we find the factors listed in table~\ref{tab:diagcoeffs} in front of the relevant diagrams by equating coefficients. At this point, we have to refine our prescription to construct the diagrams slightly because diagrams of type I have two derivatives acting on $\beta^{E}_{a\bar b}$. But these two derivatives do not commute and thus we have to decide which one comes first. Our convention is that we go from top to bottom and left to right. The order how we encounter derivatives is the order we write them down in the tensorial expression.
\begin{table}[!t]
  \centering
  \begin{tabular}{|l|l|l|l|l|l|}
    \hline
    \multicolumn{2}{|c|}{type} & \multicolumn{4}{|c|}{diagrams - $P\leftrightarrow \Pb$}  \\ \hline  
    I   & B & \multicolumn{4}{|l|}{$\co{+ 2} \DDl0 $}\\   \hline 
    I   & C & \multicolumn{4}{|l|}{$\co{- 2} \DADA0$} \\  \hline 
    II  & A & \multicolumn{4}{|l|}{$\co{- 2} \FDla10 \co{+ 0} \DFa01  \co{+ 2} \DFb01 \co{+ 2} \DFc01 $} \\  \hline 
    II  & B & \multicolumn{4}{|l|}{$\co{- 2} \FDlb00 \co{+ 2} \FDlb01 \co{+ 4} \DFd11 \co{+ 4} \DFd01 $} \\  \hline 
    III & B & \multicolumn{4}{|l|}{$\co{+ 0} \FADla0 \co{- 2} \DFAa0  \co{+ 0} \DFAb0 \co{+ 0} \DFAc0 $} \\  \hline 
    III & C & \multicolumn{4}{|l|}{$\co{+ 2} \DFAd0  \co{+ 0} \FADlb0 $} \\  \hline 
    IV  & A & \multicolumn{4}{|l|}{$\co{+ 0} \FFa010 \co{+ 2} \FFc001 \co{+ 4} \FFc010 $} \\  \hline 
    IV  & B & \multicolumn{4}{|l|}{$\co{- 1} \FFb000 \co{+ 2} \FFb010 \co{- 1} \FFb110$} \\   \hline 
    V   & A & \multicolumn{4}{|l|}{$\co{- 2} \FFAb10 \co{+ 0} \FFAa01$} \\   \hline 
    V   & B & \multicolumn{4}{|l|}{$\co{+ 2} \FFAc00 \co{- 2} \FFAc10$} \\   \hline
  \end{tabular}
  \caption{Diagrams which might contribute to the doubling of $\Kh^{(1)}$ and their respective coefficients. The equivalent tensor expression is given in \eqref{eqn:K1doubled}.}
  \label{tab:diagcoeffs}
\end{table}%
To avoid any confusion and since it is a main result of our work, the explicit tensor expression corresponding to the diagrams in table~\ref{tab:diagcoeffs} reads
\begin{align*}\label{eqn:K1doubled}
    \Kh^{(1)}_{AB}&(\B) = -2P_{[A}{}^{C}P^{DE}\Pb_{B]}{}^{F}D_{E}D_{C}\B_{DF}
 +2P_{[A}{}^{C}P^{DE}\Pb_{B]}{}^{F}D_{E}D_{D}\B_{CF} \\&
 +2F_{CFH}P_{[A}{}^{C}P^{DE}\Pb_{B]}{}^{F}\Pb^{GH}D_{E}\B_{DG}
 +2P_{[A}{}^{C}P^{DE}\Pb_{B]}{}^{F}\Pb^{GH}\B_{DG}D_{E}F_{CFH} \\&
 +2F_{EFH}P_{[A}{}^{C}P^{DE}\Pb_{B]}{}^{F}\Pb^{GH}D_{C}\B_{DG}
 -2P_{[A}{}^{C}P^{DE}P^{FG}\Pb_{B]}{}^{H}\B_{DH}D_{G}F_{CEF} \\&
 -2P_{[A}{}^{C}P^{DE}\Pb_{B]}{}^{F}\Pb^{GH}\B_{CG}D_{E}F_{DFH}
 -4F_{EFH}P_{[A}{}^{C}P^{DE}\Pb_{B]}{}^{F}\Pb^{GH}D_{D}\B_{CG} \\&
 -4F_{CEG}P_{[A}{}^{C}P^{DE}P^{FG}\Pb_{B]}{}^{H}D_{F}\B_{DH}
 +2P_{[A}{}^{D}P^{CE}\Pb_{B]}{}^{F}F_{C}D_{D}\B_{EF} \\&
 -2P_{[A}{}^{D}P^{CE}\Pb_{B]}{}^{F}F_{C}D_{E}\B_{DF}
 +2F_{CHI}F_{EFJ}P_{[A}{}^{C}P^{DE}\Pb^{IJ}\Pb_{B]}{}^{F}\Pb^{GH}\B_{DG} \\&
 -4F_{CEI}F_{FHJ}P_{[A}{}^{C}P^{DE}\Pb^{IJ}\Pb_{B]}{}^{F}\Pb^{GH}\B_{DG}
 +F_{DGI}F_{FHJ}P_{[A}{}^{C}\Pb^{IJ}\Pb_{B]}{}^{D}\Pb^{EF}\Pb^{GH}\B_{CE} \\&
 -2F_{DFI}F_{EHJ}P_{[A}{}^{C}P^{DE}\Pb^{IJ}\Pb_{B]}{}^{F}\Pb^{GH}\B_{CG}
 +F_{DFH}F_{EGJ}P_{[A}{}^{C}P^{DE}P^{FG}\Pb^{IJ}\Pb_{B]}{}^{H}\B_{CI} \\&
 -2F_{DFH}P_{[A}{}^{D}P^{CE}\Pb_{B]}{}^{F}\Pb^{GH}F_{C}\B_{EG}
 +4F_{EFH}P_{[A}{}^{D}P^{CE}\Pb_{B]}{}^{F}\Pb^{GH}F_{C}\B_{DG} \\&
 - P \leftrightarrow \Pb \,. \stepcounter{equation}\tag{\theequation}
\end{align*}

\subsubsection{Partial double Lorentz gauge fixing}\label{sec:PDLgf}
There are still a few terms in $\Kh^{(1)}(\B)$ which cannot be matched by the procedure above. However, we will now show that they are just an artifact of the partially double Lorentz fixed generalised frame field $\Eh_A{}^I$ used in the calculation. Its variation \eqref{eqn:deltahatE} contains the compensating double Lorentz transformation $\dEgf_{AB}$ and by restricting \eqref{eqn:SandK} to it, we find
\begin{equation}\label{eqn:S(2)DL}
  \delta_{\Psi^{\mathrm{gf}}} \Sh^{(2)} = \intvol \dEgf \cdot \Kh^{(1)}( \Bh^{(1)} )\,.
\end{equation}
Since $\Sh^{(2)}$ is not invariant under double Lorentz transformations, $\Kh^{(1)}$ has to have contributions which relate physical degrees of freedom with gauge transformations. We fix them by remembering that $\Sh^{(2)}$ has been constructed such that the relation \cite{Marques:2015vua}
\begin{equation}
  \delta_{\Psi^{\mathrm{gf}}} \Sh^{(2)} = \intvol ( \A^{(1)}_{\Psi^{\mathrm{gf}}} \Eh ) \bh^{(1)E}
\end{equation}
holds. An analogous mechanism governs gauge fixed, two-loop $\beta$-functions, too. More precisely, they split into the two contributions
\begin{equation}\label{eqn:GFdecompbetaE}
  \Bh^{E(2)} = \Bh'^{E(2)} + \A^{(1)}_{\Bh^{(1)E\mathrm{gf}}} \Eh\,,
\end{equation}
where $\Bh'^{(1)E}$ is not gauge fixed. In the final result, we neither want to include \eqref{eqn:S(2)DL} nor the second term on the left hand side of \eqref{eqn:GFdecompbetaE}. The reason is that we are looking for two-loop $\beta$-functions which do not depend on a particular gauge fixing. All terms that we therefore drop can be neatly combined in the symmetric, double Lorentz gauge fixing term
\begin{equation}
  \widehat{K}^{(1)\mathrm{gf}}(\delta\Eh,\Bh) = \intvol \left[ -\frac14 \left( \delta g^a{}_b H^{bcd} - 2 \delta B^a{}_b\omega^{bcd} \right) D_a \bh^B_{cd} + \left( \bh^E \leftrightarrow \delta E \right) \right]
\end{equation}
and we eventually find that $\Kh^{(1)}$ can be written as
\begin{equation}
  \Kh^{(1)}(\delta\Eh, \Bh) = \intvol \delta \Eh^{AB} \Kh^{(1)}_{AB}(\bh) + \Kh^{(1)\mathrm{gf}}( \delta\Eh, \bh )\,.
\end{equation}
Note that we have dropped the prime on the $\Bh'^E$ to avoid cluttering our notation. From now on all doubled $\beta$-functions are free of any gauge fixing.

\subsubsection{Extracting the \texorpdfstring{$\beta$}{beta}-functions}
Since, we have been successful in writing $\Kh^{(1)}_{AB}(\B)$ in the O($D$,$D$)-covariant form \eqref{eqn:K1doubled}, it is straightforward to compute the two-loop $\beta$-functions. The procedure goes along the same line as at one-loop in section~\ref{sec:oneloopdetail}: First, we rewrite the gradient flow \eqref{eqn:varS2} in terms of doubled quantities. More specifically, we take the components of
\begin{equation}
  \Kh^{AB}(\B) = \begin{pmatrix}
    0 & \Kh^{a\bar b}(\B) \\
    -\Kh^{b\bar a}(\B) & 0 
  \end{pmatrix}\,,
    \qquad \text{with} \qquad
    \Kh^{a\bar b}(\B) = \Kh^{g\, a\bar b}(\B) + \Kh^{B\, a\bar b}(\B)\,,
\end{equation}
to rewrite \eqref{eqn:varS2} as
\begin{equation}
  \delta_\Psi \Sh^{(2)} = \intvol \left[ \delta E^{AB} K^{(0)}_{AB}{}^{CD} \left( \bh^{(2)E}_{CD} - \frac12 \Kh^{(1)}_{CD}(\Bh) \right) + 8 \delta d  \bh^{(2)d} \right]\,.
\end{equation}
For the discussion in section~\ref{sec:fixKh1}, we know that the action $\Sh^{(2)}$ has to be in the gBdR scheme to be compatible with our $\Kh^{(1)}_{AB}$ from section~\ref{sec:matchingdiagrams}. In this scheme, it can be written in the O($D$,$D$)-covariant form \cite{Marques:2015vua,Baron:2017dvb}
\begin{equation}
  \Sh^{(2)} = \int d^D x e^{-2d} \scRh^{(2)} \qquad \text{with} 
  \qquad \scRh^{(2)} = - \scRh^+ - \scRh^- \,,
\end{equation}
where the explicit expression for $\scRh^\pm$ is given in (2.33) of \cite{Baron:2017dvb}. In terms of diagrams $\scR^{(2)}$ reads
\begin{equation}
  \begin{array}{lll}
    \scRh^{(2)} =
    &\co{-}  \FFFFa100110 \co{+} \FFFFa010010 &\co{-\frac43} \FFFFb101001 \co{-} \FFFFb001100 \co{-}\FFFAFA1001 \\[0.8cm]
      &\co{+4} \FFFD10010   \co{-} \FFFD01100 &\co{+}  \FFFD11100 \co{+2} \FFFADl1001 \co{+2}\FFFADb1010 \\[0.8cm]
      &\co{+2} \FFFADa1100  \co{+\frac12} \DDFF0010 &\co{-\frac12} \DDFF0011 \co{+} \FFDDl1010 \co{+} \FFDDlprime1010 \\[0.8cm]
      &\co{-} \FFDlDl1001 + P \leftrightarrow \Pb \,.
  \end{array}
\end{equation}
All that is left to be done is compute the variation of this action. It has the form
\begin{equation}
  \delta_\Psi \Sh^{(2)} = \intvol \left( \delta E^{AB} \scGh^{(2)}_{AB} - 2 \delta d \scRh^{(2)} \right)
\end{equation}
and immediately allows for the identification
\begin{equation}
  \bh^{(2)E}_{a\bar b} = \scGh^{(2)}_{a\bar b} + \widehat{K}^{(1)}_{a\bar b}( \Bh^{(1)} )\,,
    \qquad
  \bh^{(2)d} = - \frac14 \scRh^{(2)}\,.
\end{equation}
We already computed $\Kh^{(1)}_{a\bar b}$ and therefore we only need $\scGh^{(2)}_{a\bar b}$ to obtain the final result. We compute it with the xTensor package of the xAct suite and get the results presented in section~\ref{sec:twoloops}.

\subsubsection{Generalised Green-Schwarz transformation}
The last thing we have to do to make full contact with section~\ref{sec:twoloops} is to prove that \eqref{eqn:DLbeta2} holds. To this end, we take a closer look at the identity 
\begin{equation}\label{eqn:deltaPsiLchiS2}
    \delta_\Psi L_{\chi^{(1)}} \Sh^{(1)} = L_{\chi^{(1)}} ( \widehat{K}^{(0)} ) ( \Psi, \Bh^{(1)} ) +
      \widehat{K}^{(0)}( \Psi, L_{\chi^{(1)}} \Bh^{(1)} )\,,
\end{equation}
which arises if we apply $L_\chi$ to both sides of the first equation in \eqref{eqn:SandK}. We now identify $\chi^{(1)} = \Al^{(1)} \begin{pmatrix} \Eh & d \end{pmatrix}$ to further simply this relation by using
\begin{equation}
  L_{\chi^{(1)}} \Sh^{(1)} = \Al^{(0)} \Sh^{(2)} \qquad \text{and} \quad
  L_{\chi^{(1)}} \Bh^{(1)} = \Al^{(0)} \Bh^{(2)}\,,
\end{equation}
which are equivalent to \eqref{eqn:DLS2} and \eqref{eqn:DLbeta2}, respectively. Together with \eqref{eqn:deltaPsiLchiS2} they imply
\begin{equation}
  \delta_\Psi \Al^{(0)} \Sh^{(2)} = L_{\chi^{(1)}} (\widehat{K}^{(0)}) (\Psi, \Bh^{(1)} ) +
    \widehat{K}^{(0)} (\Psi, \Al^{(0)} \Bh^{(2)}) = \Al^{(0)} \delta_\Psi \Sh^{(2)}\,.
\end{equation}
Note that we are able to swap $\delta_\Psi$ and $\Al$ because the variation parameter $\Psi$ does not transform anomalously under double Lorentz transformations and therefore $\Al \Psi = 0$ holds. This equation can be alternatively obtained by applying $\Al^{(0)}$ to the left and right side of the second equation of \eqref{eqn:SandK}, if we further impose
\begin{equation}\label{eqn:DLK0andK1}
  L_{\Al^{(1)} \begin{pmatrix} \Eh & d \end{pmatrix}} \widehat{K}^{(0)} = \Al^{(0)} \widehat{K}^{(1)} \,.
\end{equation}
Equally, one might conclude that if this identity holds for $\widehat{K}^{(0)}$ and $\widehat{K}^{(1)}$, it implies \eqref{eqn:DLbeta2}. This result is not very surprising, because we expect $\Kh$, like the action and the $\beta$-functions, to transform covariantly under gGS transformation. Indeed one can check that the expressions we have presented in \eqref{eqn:K0doubled} and \eqref{eqn:K1doubled} satisfy \eqref{eqn:DLK0andK1}. This result provides an important consistency check. Moreover, it would be interesting to see if, similar to the action $\Sh^{(2)}$, $\widehat{K}^{(1)}$ can be completely fixed by just imposing its covariance under gGS transformations.

\subsection{\texorpdfstring{$c$}{c}-function and gradient flow metric}\label{sec:cfunction}
We argue in section~\ref{sec:renormalisable} that PL symmetry restricts the $\sigma$-model $\beta$-functions to a finite dimensional subspace of the coupling space. The same is true for $K_{AB}(\B)$, which looses all derivatives on a PL symmetric background and thus can be written as
\begin{equation}
  K(\Psi, \B)= \delta E_{AB} \beta^{E}_{CD} K^{ABCD} V\,,
    \qquad \text{with} \qquad
  V = \intvol\,.
\end{equation}
Here, $K^{ABCD}$ only depends on the couplings that enter through $F_{ABC}$ and $F_A$. In the same vein, we rewrite the low-energy effective target space action,
\begin{equation}\label{eqn:Scfunction}
  S = V \scR = -4 V \beta^d = -\frac23 V c\,,
\end{equation}
where the last identity originates from \eqref{eqn:cfrombetad}. Now, the gradient flow \eqref{eqn:gradientflow} takes a form that matches (14) in Zamolodchikov's famous paper \cite{Zamolodchikov:1986gt}, namely
\begin{equation}
  \partial_\nu c  = 12 G_{\mu\nu} \beta^\nu
\end{equation}
with the gradient flow metric
\begin{equation}\label{eqn:Zmetric}
  G_{\mu\nu} = - \frac18 J_\mu{}^{AB} J_\nu{}^{CD} K_{ABCD}\,,
\end{equation}
and the Jacobian
\begin{equation}
  J_\mu{}^{AB} = \partial_\mu E^{AI} E^B{}_I\,.
\end{equation}

Because $K^{(n)}_{ABCD}$ is symmetric under the exchange of the indices $AB\leftrightarrow CD$, $G^{(n)}_{\mu\nu}$ is a symmetric tensor, at least for $n=0\,, 1$. Hence, one might conclude that the latter is the Zamolodchikov metric \cite{Zamolodchikov:1986gt}. But the gradient flow away from the conformal point has a more general form and incorporates corrections \cite{Friedan:2009ik}. Therefore, we prefer the term gradient flow metric for $G_{\mu\nu}$. On the other hand, the action $S$ in \eqref{eqn:Scfunction} has the ``central charge'' form of \cite{Tseytlin:1987bz} and thus, what we call $c$ should match Zamolodchikov's definition. A thorough comparison between the quantities, we identified here, and results from the fixed point CFT and its conformal perturbation theory is required to settle these points completely. This analysis is beyond the scope of this paper. But as a first step, we discuss the $\lambda$-deformation in the following, which was extensively studied from a CFT perspective \cite{Itsios:2014lca,Georgiou:2016iom,*Georgiou:2016zyo}.

\subsubsection{\texorpdfstring{$\lambda$}{Lambda}-deformation}\label{sec:lambdagradient}
We already have computed $c$ of the $\lambda$- and $\eta$-deformation for one and two loops in the sections~\ref{sec:lambda1} and \ref{sec:lambda2}, respectively. For convenience, we repeat it here,
\begin{equation}
  c = D - \frac{1 + 2 \lambda + 2 \lambda^3 + \lambda^4}{2 k (1 - \lambda) (1 + \lambda)^3} c_G D + 
    \frac{\lambda^3 (4 - 5\lambda + 4\lambda^2 )}{2 k^2 (1-\lambda)^2 (1+\lambda)^6} c_G^2 D\,,
\end{equation}
in terms of $\lambda$ instead of $\kappa$. While the first two terms match (3.30) of \cite{Georgiou:2019nbz} perfectly, the last term deviates. A possible explanation is that our $c$ and theirs actually capture different quantities. The derivation of $c$ in \cite{Georgiou:2019nbz} starts from the Zamolodchikov metric, obtained by conformal perturbation theory. Combining the Zamolodchikov metric and the $\beta$-functions, $\partial_\mu c$ is calculated and then integrated to obtain $c$. As explained above, our $G_{\mu\nu}$ is expected to differ from the Zamolodchikov metric away from the conformal point.

Because there is only one coupling that flows, $G_{\mu\nu}$ is solely formed by $G_{\lambda\lambda}$. Evaluating \eqref{eqn:Zmetric} with
\begin{equation}
  J_\lambda{}^{AB} = \frac1{1-\lambda^2}\begin{pmatrix} 0 & \eta_{a\bar b} \\
    - \eta_{b\bar a} & 0 \end{pmatrix}
\end{equation}
works along the same line as for the $\beta$-functions. We will not repeat the details here but instead refer to the accompanying Mathematica notebook \appnotebook. The result 
\begin{equation}\label{eqn:Gll}
  G_{\lambda\lambda} = \frac{D}{2(1 - \lambda^2 )^2} \left( 1 + \frac{Q(\lambda)}{k (1-\lambda) (1+\lambda)^3} c_G \right)\,.
\end{equation}
matches (3.16) of \cite{Georgiou:2019nbz}. There it is argued that the function $Q$ have to have the form
\begin{equation}
  Q(\lambda) = c_0 + c_1 \lambda + c_2 \lambda^2 + c_1 \lambda^3 + c_0 \lambda^4
\end{equation}
to be compatible with the symmetry $\lambda \leftrightarrow \lambda^{-1}$, $k \leftrightarrow -k$. We find a $Q(\lambda)$ of this form, but instead of $c_0=c_1=c_2=0$ \cite{Georgiou:2019nbz}, we obtain
\begin{equation}
  c_0 = -1\,, \qquad c_1 = 2\,, \quad \text{and} \quad c_2 = -4\,.
\end{equation}
This is not very surprising because already our $c$-function is different from theirs.

It should be possible to better understand this discrepancy by using alternative techniques to obtain the values of these coefficients. In particular, $c_0$ is accessible from the level $\kh$ WZW-model on the group manifold $G$, which arises at the fixed point $\lambda=0$ of the RG flow. At this distinguished point, the marginal operator that triggers the flow is
\begin{equation}\label{eqn:Olambda}
  \mathcal{O}_\lambda(z,\zb) = \frac{\gamma}{k} \eta_{a\bar b} \, j^a(z) \jb^{\bar b}(\zb) \,,
\end{equation}
where $\gamma$ is a numerical factor. Most important is that $\mathcal{O}_\lambda$ is proportional to $k^{-1}$ and not $\kh$. This dependence enters through the left and right invariant forms \eqref{eqn:leftrightinv}. The Ka\v{c}-Moody currents, which $\mathcal{O}_\lambda$ is formed of, are governed by the OPE
\begin{equation}
  j^a(z) j^b(w) = \frac{\kh \eta^{ab}}{(z-w)^2} + \frac{f^{ab}{}_c}{z-w} j^c(w) + \dots \,.
\end{equation}
The anti-chiral currents $\jb^a(\zb)$ are governed by an analogous version. Moreover, they commute with all chiral currents $j^b(z)$. We now know everything we need to compute the Zamolodchikov metric \begin{equation}
  G_{\lambda\lambda}(0) = \lim_{z\rightarrow w} | z - w |^4 \left\langle \mathcal{O}_\lambda(z,\zb) \mathcal{O}_\lambda(w,\wb) \right\rangle = \gamma^2 D \frac{\kh^2}{k^2} = \gamma^2 D \left ( 1 - \frac{c_G}{k} + \frac{c_G^2}{4 k^2} \right)
\end{equation}
from (6c) in \cite{Zamolodchikov:1986gt}. Matching this result with \eqref{eqn:Gll}, we recover $c_0 = -1$ and furthermore fix $\gamma^2 = 1/2$. This is consistent with the observation that, at least at the fixed point, additional corrections \cite{Friedan:2009ik} vanish and therefore $G_{\lambda\lambda}(\lambda=0)$ becomes the Zamolodchikov metric. The difference to $c_0 = 0$ in \cite{Georgiou:2019nbz} originates from a different normalisation of $\mathcal{O}_\lambda$, since they use $\hat{k}$ instead of $k$ in \eqref{eqn:Olambda}\footnote{We thank the authors of \cite{Georgiou:2019nbz} for explaining to use their calculations.}. Clearly, more work is required to understand this difference and to try to reproduces the remaining two coefficients, $c_1$ and $c_2$, from conformal perturbation theory.

\section{Conclusion}\label{sec:conclusion}
In this paper, we have established three main results for the bosonic string:
\begin{enumerate}[leftmargin=1.5em,itemindent=1em,itemsep=0cm,label*=\arabic*)]
  \item In an appropriate scheme, the two-loop $\beta$-functions for the metric, $B$-field, and dilaton can be written in a manifestly O($D$,$D$)-covariant form.
  \item PL $\sigma$-models are one and two-loop renormalisable.
  \item The respective RG flows are invariant under PL T-duality.
\end{enumerate}
One might expect that the best way to obtain them is to start from a worldsheet theory with manifest, classical PL symmetry and apply the background field method like in \cite{Sfetsos:2009vt,*Severa:2016lwc,*Pulmann:2020omk}. However, this idea has not been implemented successfully yet. Therefore, we chose a different approach which heavily relies on previous insights in DFT and on the option to obtain the one and two-loop $\beta$-functions from a gradient flow. An important lesson learned is that it is crucial to work in the right scheme. The latter is tightly linked to the deformation of double Lorentz symmetry on the target space and the corresponding gGS transformations. So it might be promising to revisit the worldsheet approach with this knowledge.

The one-loop RG flow has a natural interpretation in terms of a generalised Ricci flow (see \cite{Garcia-Fernandez:2020ope} for a recent review), the generalised geometry version of the celebrated Ricci flow \cite{hamilton1982} used in Perelman's resolution of the Poincar\'e and Thurston geometrisation conjecture \cite{Perelman:2006un,*Perelman:2006up,*Perelman:2003uq}. Therefore, all involved quantities possess a (generalised) geometric origin. It is tempting to speculate that something similar might be true for the two-loop flow. Since fundamental symmetries of generalised geometry (like double Lorentz transformations) are deformed in its derivation, it is likely that also the underlying notion of geometry has to be adapted. PL symmetric target space geometries provide intriguing clues on the required modifications: Remember that a significant class of such target spaces is formed by PL groups. But PL groups are just the classical limit of a quantum group (see for example \cite{Chaichian:1996ah} for an introduction). Quantum groups can be approached from different angles. Most significant for us is that they give rise to non-commutative geometries. Hence, we conjecture that $\beta$-functions beyond one-loop might be governed by non-commutative geometry where the deformation parameter is related to the string length $\sim \sqrt{\alpha}'$. A related clue in this direction is that integrable deformations, like the $\lambda$- and $\eta$-deformation, which we discuss in section~\ref{sec:results}, possess a hidden quantum group symmetry \cite{Hollowood:2015dpa,Delduc:2016ihq}. The respective deformations parameters, $q=\exp(i \pi / k)$ and $q=\exp(4 \pi \eta t)$, are RG invariants at one and two loops. It is instructive to restore $\alpha'$ in these expressions. We know that $F_{ABC}$ comes with one derivative and therefore a factor of $\sqrt{\alpha'}$. Hence, we are actually dealing with $\qh=q^{\alpha'}$. In the semiclassical limit, $\alpha'\rightarrow 0$, a $\qh$ deformed quantum group transitions into a Poisson-Hopf algebra with the deformation parameter $q$. It is the latter which partially captures the global symmetries of the classical $\eta$-deformation \cite{Delduc:2016ihq}. Consequentially, we might understand $\alpha'$-corrections as the driving force from the classical Poisson-Hopf algebra to the associated quantum group. Of course, these speculations have to be supplemented with further quantitative evidence. But if we assume that they are justified, it would imply that we could extract all order $\beta$-functions and their generating low-energy, effective target space actions. Another reason to be optimistic that our results can be extended beyond two loops is that gGS transformations and the corresponding O($D$,$D$)-covariant action are in principle (even though they become extremely complicated) available to arbitrary order in $\alpha'$ \cite{Baron:2020xel}.

Two immediate applications for our results are integrable deformations and consistent truncations with higher derivative corrections. The former are motivated by the observation that nearly all currently known integrable $\sigma$-models possess PL symmetry. Already at one-loop, they have interesting RG flows (examples include \cite{\refoneLRGintegr}) with generic features like multiple fixed points \cite{Georgiou:2020eoo}. Recently, first efforts were made to push this analysis to two-loop \cite{Hoare:2019ark,Hoare:2019mcc,Georgiou:2019nbz}. At the level of the target space fields this is challenging, as we have demonstrated in section~\ref{sec:finitegGS} for the $\lambda$-deformation. But with the formalism we develop in this paper, it becomes a much simpler task. Moreover, PL T-dualities between different integrable deformations are manifest. Due to this fact, we could obtain the flows of the $\lambda$- and $\eta$-deformation from a single calculation. We furthermore noticed that at one-loop, renormalisable $\sigma$-models are in one-to-one correspondence with consistent truncations of the low energy effective theory in the target space. Due to their potential to produce new, sophisticated solutions in (gauged) supergravity they have been intensively studied (for an early work see for example \cite{Duff:1985jd}). But only recently, systematic constructions of such truncations have been discussed and the framework of generalised geometry/double/exceptional field theory is predestined for them \cite{Cassani:2019vcl}. All of the work in this direction, that we are aware of, is based on a two-derivative action and its field equations. Since PL $\sigma$-models are two-loop renormalisable, they result in a large class of consistent truncations involving up to four derivatives. Hence, one might use them as guiding examples to construct a higher derivative version of the current constructions. Another important step that is required to make contact with $\alpha'$-corrected half-maximal gauged supergravities, is to extend our results to the heterotic string.

\section*{Acknowledgements}
We would like to thank D. Butter, G. Georgiou, C. Klim\v{c}\'ik, W. Linch, D. Marqu\'es, G. Piccinini, C. Pope, P. \v{S}evera, E. Sagkrioti, K. Sfetsos, K. Siampos, D. Thompson and A. Tseytlin for inspiring discussions and very helpful correspondence. FH acknowledges the online seminar series ``Exceptional Geometry Seminar Series'' for the opportunity to present the main results of this paper. We are grateful that the team behind the xAct bundle of Mathematica packages has made them openly accessible. Without their software many of our computations would have been much harder to implement. In particular, we benefit from the packages xPerm, xTensor, xCoba, xPert and xTras \cite{xPerm,*xPert,*xTras}.

\appendix
\section{Two-loop \texorpdfstring{$\beta$}{beta}-functions}\label{app:var2laction}
In this appendix we demonstrate how the two-loop $\beta$-functions arise from the variation of the two-loop low-energy effective target space action in the MT scheme. We follow the presentation of \cite{Metsaev:1987zx} closely but keep the one-loop $\beta$-functions in all steps instead of setting them to zero.

\subsection{Metric}
We begin by varying the two-loop low-energy effective target space action \eqref{eq:S2} with respect to the metric. As explained in section \ref{sec:oneloopdetail}, it is important that one does not vary with respect to the inverse metric, as this would introduce a wrong sign. For instance, the variation of the Riemann tensor is given by
\begin{equation}
  \delta R^e{}_{abc} = \frac{1}{2} g^{de}\left(\nabla_b\delta\left(\partial_a g_{dc}+\partial_c g_{da}-\partial_l g_{ca}\right)
  -\nabla_c\delta\left(\partial_a g_{db}+\partial_b g_{da}- \partial_d g_{ba}\right)\right),
\end{equation}
so that the variation of the first term in \eqref{eq:S2} reads
\begin{equation}\label{eqn:vargfirstterm}
  \begin{split}
    \delta \int \dd^D x \sqrt{g} e^{-2 \phi} \frac14  R_{efcd}R^{efcd} &= -\int \dd^D x \sqrt{g} e^{-2 \phi} R^{abcd} \nabla_a \nabla_c \delta g_{bd}\\
    &= -\int \dd^D x \sqrt{g} \nabla_a \nabla_c \left( e^{-2 \phi}R^{abcd} \right) \delta g_{bd}\\
    &= -\int \dd^D x \sqrt{g} \left(2R^{acdb}\Big(\nabla_c\partial_d\phi - 2\partial_c\phi\partial_d\phi\right) \\
    &\hspace{5em}  +4\nabla_c R^{c(ab)d}\partial_d\phi -\nabla_c\nabla_d R^{acdb} \Big)\delta g_{ab},
  \end{split}
\end{equation}
where we integrated by parts twice in the second line. Following \cite{Metsaev:1987zx}, we break down the metric variation of \eqref{eq:S2} into the three terms
\begin{align}
  P_{ab} &\equiv e^{2\phi} \frac{\delta}{\delta g^{ab}}\int \frac{1}{4} e^{-2\phi}R_{cdef}R^{cdef}, \label{Pmunu}\\
  Q_{ab} &\equiv e^{2\phi} \frac{\delta}{\delta g^{ab}}\int \left(-\frac{1}{8}\right)e^{-2\phi}R^{cdef}H_{cdh}H_{ef}{}^{h}, \label{Qmunu} \\
  O_{ab} &\equiv \frac{\delta}{\delta g^{ab}} \int \frac{1}{4}\left(\frac{1}{24}H_{cde}H^d{}_{fg}H^{fhe}H_h{}^{cg}-\frac{1}{8}H_{cde}H_f{}^{de}H^{cgh}H^f{}_{gh} \right), \label{Omunu}
\end{align}
where we have already partially treated the first term in \eqref{eqn:vargfirstterm}. We have to apply Bianchi identities to further simplify these three contributions. In particular, starting from the second Bianchi identity of the Riemann tensor, we derive the identity
\begin{equation}
  \nabla^c\nabla^d R_{acdb} = - \nabla^2R_{ab} +\frac12 \nabla_a\nabla_b R + R_{acdb}R^{cd} + R_{ac}  R^c{}_{b}.
\end{equation}
Applying it to the first term, we obtain
\begin{equation}
  \begin{split}
    P_{ab} = &-\frac12R_{acde}R_b{}^{cde} - \nabla^2R_{ab} + \frac12 \nabla_a\nabla_bR + R_{acdb}R^{cd} + R_{ac}  R^c{}_{b} \\
    &- 4\left(\nabla_{(a} R_{b)c}-\nabla_c R_{ab}\right)\partial^c \phi -2R_{acdb}\left(\nabla^c\partial^d\phi - 2\partial^c\phi\partial^d\phi\right).
  \end{split}
\end{equation}
Similarly, the second and third terms reduce to
\begin{align}
  Q_{ab} &= \frac{1}{2}R^{cdef}H_{acd}H_{bef}+\frac{3}{2}R^{cde}{}_{(a}H_{b)ge}H_{cd}{}^{g} + e^{2\phi} \nabla^c\nabla^d\left(e^{-2\phi}H_{ace}H_{bd}{}^{e}\right), \\
  O_{ab} &=-\frac{1}{16} (H^4)_{ab} + \frac{1}{16} H^2_{ac}(H^2)_b{}^{c} + \frac{1}{8}(H^2)^{cd}H_{ace}H_{bd}{}^{e}.
\end{align}

As suggested by \cite{Metsaev:1987zx}, we use the one-loop $\beta$-functions to remove all $\phi$-dependence. Consequentially, the variation of the action decomposed into terms containing $\bh^{(1)g}$ or $\bh^{(1)b}$ and terms without them. The latter form the two-loop $\beta$-function for the metric, $\bh^{(2)g}_{ab}$, while the former give rise to $\Kh^{(1)g}(\bh^{(1)g},\bh^{(1)B})$. During the computation we use the identities
\begin{equation}
  \begin{aligned}  
    & R_{acdb}\nabla^c\phi = 2 \nabla_{[c} \nabla_{b]}\nabla_a \phi = \nabla_{[d} \left(\bh^g_{b]a}-R_{b]a} + \frac{1}{4}H^2_{b]a} \right), \\
    &\nabla^c\nabla_a R_{cb} = \frac{1}{2} \nabla_a \nabla_b R + R_{ac}R^{c}{}_{b} + R_{acdb}R^{cd}, \\
    &\nabla_b H_{acd} = \nabla_c H_{bda}+\nabla_d H_{bac}+\nabla_a H_{bcd}\,,    
  \end{aligned}
\end{equation}
which eventually yield
\begin{equation}
  \begin{split}
    P_{ab} =& -\frac12 R_{acde}R_b{}^{cde} - \frac12 \nabla_e H_{acd}\nabla^e H_{b}{}^{cd} - \frac{1}{16}(H^2)_{a}{}^{c} H^2_{cb} - \frac14 H_{cea}H_{bd}{}^{e}(H^2)^{cd}  \\
    & + \frac12 R^{cdef}H_{acd}H_{bef} + \frac14 R_{acdb}(H^2)^{cd} + R_{(a}{}^{cde}H_{b)cf}H_{de}{}^{f} +\frac{1}{24} \nabla_a\nabla_bH^2 \\
    & - \nabla^2\bh^{(1)g}_{ab} + \nabla^c \nabla_{(a} \bh^{(1)g}_{b)c} - H^{cd}{}_{a}H_{edb}(\bh^{(1)g})_c{}^{e} - \frac14 H^2_{ac} (\bh^{(1)g})_{b}{}^{c} - 2\left(\nabla_{(a}\bh^{(1)g}_{b)c}-\nabla_c\bh^{(1)g}_{ab}\right)\partial^c\phi \\
    & + 2H^{cd}{}_{a} \nabla_c \bh^{(1)b}_{db} + \frac12 H^{cd}{}_{a} \nabla_{b} \bh^{(1)b}_{cd} - \frac12 \nabla_{(a} H_{b)cd}(\bh^{(1)b})^{cd}, \\
    Q_{ab} =& \frac14 R_{(a}{}^{cde}H_{b)cf}H_{de}{}^{f}+\frac{1}{16}H_{ace}H_{bd}{}^{e}(H^2)^{cd}+\frac{1}{8}\nabla_e H_{acd}\nabla^e H_{b}{}^{cd}-\frac{1}{24}\nabla_a H_{cde}\nabla_b H^{cde} \\
    & +\frac14 H_{ace} H_{bd}{}^{e} (\bh^{(1)g})^{dc} - \nabla^c \bh_{d(a}^{(1)b} H_{b)c}{}^{d} + (\bh^{(1)b})_{ca}(\bh^{(1)b})^{c}{}_{b}, \\
    O_{ab} =& -\frac{1}{16} (H^4)_{ab} + \frac{1}{16} H^2_{ac}(H^2)_b{}^{c} + \frac{1}{8}(H^2)^{cd}H_{ace}H_{bd}{}^{e}\,.
  \end{split}
\end{equation}
Adding those terms back together gives rise to
\begin{equation}
  P_{ab} + Q_{ab} + O_{ab} = - \bh^{(2)g}_{ab} + \Kh^{(1)g}_{ab} ( \bh^{(1)g}, \bh^{(1)b} )\,.
\end{equation}
Now, we read off $\Kh^{(1)g}_{ab} ( \beta^g, \beta^B )$ in \eqref{eqn:Kh(1)g} and
\begin{equation}
  \bh^{(2)g}_{ab} =  \bb^{(2)g}_{ab} - \frac1{24} \nabla_a \nabla_b H^2\,.
\end{equation}

\subsection{\texorpdfstring{$B$}{B}-field}\label{app:varBfield}
Since the $B$-field only appears indirectly through $H=d B$ ($H_{abc} = 3\partial_{[a} B_{bc]}$),\footnote{Due to the total antisymmetrisation we can equivalently write $H_{abc} = 3\nabla_{[a} B_{bc]}$.}
we vary the action \eqref{eq:S2} with respect to $H$, apply the chain-rule and integrate by parts
\begin{equation}
  \delta \Sh^{(2)}= -3\int \dd^D x \sqrt{g} \nabla_c \frac{\delta \Lh^{(2)}}{\delta H_{cab}}\delta B_{ab}\,,
  \qquad \text{where} \qquad \Sh^{(2)} = \int \dd^D x \sqrt{g} \Lh^{(2)}\,.
\end{equation}
The result reads
\begin{equation}
  \begin{split}
  \frac{\delta \Sh^{(2)}}{\delta B^{ab}} =& \int \mathrm{d}^Dx\sqrt{g} \frac{1}{4} \nabla^e \left(e^{-2\phi}\left(R_{ab}{}^{cd}H_{cde}+R_{ea}{}^{cd}H_{cdb}+R_{be}{}^{cd}H_{cda}\right)\right) \\
       & -\frac{1}{8} \nabla^f  \left(e^{-2\phi}H_{acd}H_{be}{}^{c}H_f{}^{ed}\right) \\
  & +\frac{1}{8}\nabla^f \left(e^{-2\phi}\left(H_{abc}H_{def}H^{dec}+H_{fac}H_{deb}H^{dec}+H_{bfc}H_{dea}H^{dec}\right)\right) \\
  =& \int \mathrm{d}^Dx\sqrt{g} e^{-2\phi} \Big[
    -\frac{1}{2}\nabla^e \phi \left(R_{ab}{}^{cd}H_{cde}+R_{ea}{}^{cd}H_{cdb}+R_{be}{}^{cd}H_{cda}\right)
    +\frac{1}{4}\nabla^f \phi H_{acd}H_{be}{}^{c}H_f{}^{ed} \\
    & -\frac{1}{4}\nabla^f \phi \left(H_{abc}H_{def}H^{dec}+H_{fac}H_{deb}H^{dec}+H_{bfc}H_{dea}H^{dec}\right) \\
    & +\frac{1}{4} \nabla^e \left(R_{ab}{}^{cd}H_{cde}+R_{ea}{}^{cd}H_{cdb}+R_{be}{}^{cd}H_{cda}\right) -\frac{1}{8} \nabla^f  \left(H_{acd}H_{be}{}^{c}H_f{}^{ed}\right) \\
    &+\frac{1}{8} \nabla^f \left(H_{abc}H_{def}H^{dec}+H_{fac}H_{deb}H^{dec}+H_{bfc}H_{dea}H^{dec}\right)
    \Big]\,.
  \end{split}
\end{equation}
Again all terms containing the dilaton can be eliminated in favour of $\bh^{(1)g}$ and $\bh^{(1)B}$, yielding
\begin{equation}
  \begin{split}
    \frac{\delta \Sh^{(2)}}{\delta B^{ab}} &= \int \mathrm{d}^Dx\sqrt{g} e^{-2\phi} \Big[
    -\frac{1}{2} R^{}_{[aecd}\nabla^e H^{cd}{}_{b]} - \frac{1}{4} \nabla^f  H^{}_{cd[a}H_{b]e}{}^{c}H_f{}^{ed} + \frac{1}{2} \nabla_c H^2_{d[b} H_{a]}{}^{dc}\\
    &+\frac{1}{8} H^2_{ec} \nabla^e H^c{}_{ab} + \frac{1}{8} H_{abc} H_{fde}\nabla^f H^{dec} -H_{[a|}{}^{cd}\nabla_c \bh^{(1)g}_{d |b]}- \frac{1}{2} R_{ab}{}^{cd} \bh^{(1)B}_{cd} \\
    &- \frac{1}{4} H_{abc} H^{dec} \bh^{(1)B}_{de} + \frac{1}{4} H_{acd}H_{be}{}^{c} (\bh^{(1)B})^{ed} - \frac{1}{2} H_{de[a|}H^{dec}\bh^{(1)B}_{c|b]}
    \Big]\,,
  \end{split}
\end{equation}
from which we read off $\Kh^{(1)B}_{ab}(\beta^g,\beta^B)$ in \eqref{eqn:Kh(1)B} and
\begin{equation}
  \bh^{(2)B\,ab} = \bb^{(2)B} - \frac1{48} H_{ab}{}^c \nabla_c H^2\,.
\end{equation}

\subsection{Dilaton}
Finally, for the dilaton, we begin with the two-loop $\beta$-function in the MT scheme, given in equation (6.10) of \cite{Metsaev:1987zx}, namely
\begin{equation}
  \bh^{(2)\phi} = \bb^{(2)\phi} - \frac1{48}\nabla^c\phi \nabla_cH^2\,.
\end{equation}
Combined with the $\beta$-function of the metric, it gives rise to\footnote{We make use of the Bianchi identity $\nabla^2 H^2 = 6 R^{ab} (H^2)_{ab} - 6 RHH + 2 \nabla_d H_{abc} \nabla^d H^{abc} + 6 H_{abc}\nabla^c\nabla_l H^{lab}$ in the step before last.}
\begin{align}
    \bh^{(2)d} =& \bh^{(2)\phi} - \frac14 g^{ab} \bh^{(2)g}_{ab} \nonumber \\
      =& -\frac1{16} \Big( R_{abcd} R^{abcd} + \frac1{24} H^4 + \frac13 \nabla_d H_{abc} \nabla^d H^{abc} -\frac18 H_{ab}^2(H^2)^{ab} \nonumber \\
      & -\frac32 RHH + R^{ab}H^2_{ab}  -\frac16\nabla^2H^2 + \frac13 \nabla^c\phi \nabla_c H^2 + 2H^2_{ab} \nabla^a\nabla^b \phi \Big) \nonumber \\
      =& -\frac1{16} \Big( R_{abcd} R^{abcd} + \frac1{24} H^4  -\frac18 H_{ab}^2(H^2)^{ab} -\frac12 RHH  - H_{abc}\nabla^c\nabla_l H^l{}_{ab} \nonumber \\
      & +\frac13 \nabla^c\phi \nabla_c H^2 +2 H^2_{ab} \nabla^a\nabla^b \phi \Big) \nonumber \\ \label{eqn:bh2dpure}
      =& -\frac1{16} \Big( R_{abcd} R^{abcd} + \frac1{24} H^4  -\frac18 H_{ab}^2(H^2)^{ab} -\frac12 RHH  + 2H^{abc}\nabla_c \bh^{(1)B}_{ab} \Big)\,,
\end{align}
where in the last step, we absorbed the terms involving $\phi$ into the one-loop $\beta$-function of the $B$-field. Moreover, the variation of the action \eqref{eq:S2}, with respect to the dilaton is given
\begin{equation}
  \begin{aligned}
    \frac{\delta S^{(2)}}{\delta d} = -2 \int \dd^D x e^{-2 d} \frac14 \Big[ & R_{abcd}R^{abcd} - \frac12 R^{abcd} H_{abe} H_{cd}{}^{e} \\ & + \frac1{24} H_{abc} H^b{}_{de} H^{dfc} H_f{}^{ae} - \frac18 H_{abc} H_d{}^{bc} H^{aef} H^d{}_{ef} \Big] \,. 
  \end{aligned}
\end{equation}
Combining it with \eqref{eqn:varS2}, we read off the value of $\Kh^{(1)d}(\beta^B)$ given in \eqref{eqn:Kh(1)d}.

\section{Transformation from HT to the MT scheme}\label{app:HTtoMT}
Starting from the two-loop $\beta$-function of the $B$-field in the HT scheme, we show the details of the scheme transformations required to obtain the corresponding $\beta$-function in the MT scheme. Our main motivation for this calculation is to have a cross check for \eqref{eqn:betaB2}, because it deviates by two signs from \cite{Metsaev:1987zx}. $\beta$-functions in both schemes are in general related by
\begin{equation}
  \bh^{\mathrm{MT}}_{ij} = \bh^{\mathrm{HT}}_{ij} - \Delta \bh_{ij} \label{eq:HTtoMT}\,,
\end{equation}
and the metric is shifted by \cite{Metsaev:1987zx}
\begin{equation}
  \Delta g^{(1)}_{ij} = \frac12 H_{ij}^2\,, \label{eq:scheme}
\end{equation}
while the $B$-field and the dilaton are not affected. Accordingly, the $B$-field $\beta$-function is shifted by
\begin{equation}
  \Delta \bh^{(2)B} = \Delta g^{(1)} \cdot \frac{\delta}{\delta g} \bh^{(1)B}\,. 
\end{equation}
Explicitly calculating the variation with respect to the metric on the right hand side yields
\begin{equation}
  \delta_g \bh^{(1)B}_{ij} = \frac12 H_{ij}{}^k \nabla^l \delta g_{kl} + H_{[i}{}^{kl}\nabla_k \delta g_{j]l} - \frac14 g^{kl} H_{ij}{}^n\nabla_n \delta g_{kl} + \frac12 \delta g_{lk} \nabla^l H^k{}_{ij} - \delta g_{lk} \nabla^l\phi H^k{}_{ij}\,.
\end{equation}
and thus
\begin{equation}\label{eqn:Deltabh(2)B}
  \Delta \bh^{(2)B} = -\frac12 \nabla_k H^2_{l[j}H_{i]}{}^{kl} + \frac14 H^2_{lk}\nabla^l H^k{}_{ij} + H_{ij}{}^k \xi_k
\end{equation}
with
\begin{equation}
  \xi_k = \frac14\nabla^lH^2_{lk} -\frac18 \nabla_k H^2 - \frac12\nabla^l\phi H^2_{lk}\,.
\end{equation}
Since that last term in \eqref{eqn:Deltabh(2)B} just generates an infinitesimal diffeomorphism, we can drop it when computing $\bb^{(2)B\,\mathrm{MT}}_{ab}$ from the expression in the HT scheme,
\begin{equation}
  \bb^{(2)B\,\mathrm{HT}}_{ij} = \frac12\nabla^k H^{lm}{}_{[j}R_{i]klm} - \frac14 \nabla^l H^{km}{}_{[j}H_{i]kn}H_{lm}{}^n + \frac18 \nabla_k H_{lij} (H^2)^{kl}\,.
\end{equation}
The result
\begin{equation}
    \bb^{(2)B\,\mathrm{MT}}_{ij} = \frac12R_{[i|klm}\nabla^k H^{lm}{}_{|j]} +\frac14 \nabla^lH_{mn[i}H_{j]k}{}^mH_l{}^{kn} - \frac18H^2_{kl}\nabla^kH^l{}_{ij} + \frac12 H_{[i}{}^{kl}\nabla_k H^2_{j]l} \,.
\end{equation}
matches \eqref{eqn:betaB2} and confirms our result from appendix~\ref{app:varBfield}.

\section{Mathematica notebook}\label{app:notebook}
We include in the arXiv submission of this paper the Mathematica notebook \notebook. It contains equations \eqref{eqn:diagbeta1}, \eqref{eqn:diagbeta1d}, \eqref{eqn:diagbeta2}, \eqref{eqn:diagbeta2d}, \eqref{eqn:K0doubled} and \eqref{eqn:K1doubled} in a machine readable form. To demonstrate how to use them, this notebook further demonstrates detailed computations for the SU(2) $\lambda$- and $\eta$-deformation along the lines of sections~\ref{sec:lambda1}, \ref{sec:lambda2} and \ref{sec:lambdagradient}.

There are two ways to access \notebook: It can be downloaded directly from 
\begin{center}
  \href{https://fhassler.de/files/PLtwoloop.nb}{https://fhassler.de/files/PLtwoloop.nb}.
\end{center}
Alternatively, one can download the source of the arXiv submission from the arXiv website. It is typically a .tar.gz archive and has to be extracted with an appropriate program.

\bibliography{literatur}
   
\bibliographystyle{JHEP}
\end{document}